\documentclass[
  aps,
  prx,
  reprint,
  longbibliography,
  superscriptaddress,
  nofootinbib,
  nobalancelastpage
]{revtex4-2}

\usepackage[english]{babel}
\usepackage[utf8]{inputenc}
\usepackage[colorinlistoftodos, color=green!40, prependcaption]
{todonotes}
\usepackage{tabularx}
\usepackage{subcaption}
\usepackage{float}
\usepackage{multirow}
\usepackage[margin=10pt,font=small,labelfont=bf,
labelsep=endash]{caption}
\usepackage{ragged2e}
\usepackage{orcidlink}
\usepackage{amsthm}
\usepackage{mathtools}
\usepackage{physics}
\usepackage{xcolor}
\usepackage{graphicx}
\usepackage{adjustbox}
\usepackage{placeins}
\usepackage[T1]{fontenc}
\usepackage{lipsum}
\usepackage{csquotes}
\usepackage{amsmath}
\usepackage{amsfonts}
\usepackage{amssymb}
\usepackage{graphicx}
\usepackage{dsfont}
\usepackage{float}
\usepackage{subcaption}
\usepackage{tabularx}
\usepackage[toc,page]{appendix}
\usepackage{booktabs}

\newcommand\numberthis{\addtocounter{equation}{1}\tag{\theequation}}

\hypersetup{
    colorlinks=true,
    linkcolor=red!75!black,
    citecolor=red!75!black,
    urlcolor=blue,
    pdftitle={Bias-Preserving Entanglement of Kerr-Cat Qubits with an ATS Coupler},
    pdfauthor={Kerem Yurtseven, Shantanu R. Jha, Apoorva Limaye, Risav Pokhrel, Maral Mahmoudi Kamelabad, Muhammad Waqar Amin},
}
\begin{document}
\title{Bias-Preserving Entanglement of Kerr-Cat Qubits with an ATS Coupler}

\author{Kerem Yurtseven\,\orcidlink{0009-0007-6387-9795}}
\email[Correspondence email address: ]{keremyurtseven@tudelft.nl}
\affiliation{Department of Electrical and Electronics Engineering, Bogazici University, 34342, Istanbul, Turkey}
\affiliation{Department of Physics, Bogazici University, 34342, Istanbul, Turkey}
\affiliation{Department of Microelectronics, Delft University of Technology, 2628 CD, Delft, Netherlands}

\author{Shantanu R. Jha}
\email[Correspondence email address: ]{shanjha@mit.edu}
\affiliation{Research Laboratory of Electronics, Massachusetts Institute of Technology, 02139, Cambridge, Massachusetts, USA}
\affiliation{Department of Electrical Engineering and Computer Science, Massachusetts Institute of Technology, 02139, Cambridge, Massachusetts, USA}

\author{Apoorva Limaye}
\affiliation{Department of Microelectronics, Delft University of Technology, 2628 CD, Delft, Netherlands}

\author{Risav~Pokhrel}
\affiliation{Pulchowk Campus, Institute of Engineering, Tribhuvan University, 44600, Lalitpur, Nepal}

\author{Maral Mahmoudi Kamelabad}
\affiliation{Faculty of Electrical and Electronics Engineering, Istanbul Technical University, 34485, Istanbul, Turkey}

\author{Muhammad Waqar Amin}
\affiliation{Quantum Science and Technology Group, Department of Physics and Applied Mathematics, Pakistan Institute of Engineering and Applied Sciences, 45650, Islamabad, Pakistan.}

\date{\today}

\begin{abstract}
{Kerr-cat qubits offer a hardware-efficient route to fault-tolerant quantum computing by suppressing bit-flip errors exponentially, while phase-flip errors increase only linearly with the mean photon number. To maintain this protection, entangling operations must be bias-preserving, ensuring that phase-flip errors occurring during the gate do not propagate into bit-flip errors. In this paper, we propose an asymmetrically threaded SQUID (ATS) parametric coupler to implement high-fidelity, bias-preserving entangling gates between Kerr-cat qubits. By engineering a photon-number asymmetry between the control and target modes, we enable fast gate operations that suppress non-adiabatic leakage without complex pulse shaping. Through numerical simulations of the full circuit Hamiltonian, we demonstrate a bias-preserving CX gate achieving 99.69\% fidelity in the absence of photon loss for a gate duration of 50 ns. Under a single-photon loss rate of 100 kHz, our protocol yields a noise bias of 9922, exceeding the noise bias of a naive non-bias-preserving CX implementation by approximately a factor of 570. These results support the ATS coupler as a valuable building block for concatenated cat-qubit architectures.}
\end{abstract}

\keywords{Kerr-Cat Qubits, Asymmetrically Threaded SQUID, Bias-Preserving Gates, Bosonic Qubits}

\maketitle
\section{Introduction} \label{sec:outline}
Quantum systems are inherently susceptible to environmental noise that limits the accuracy of quantum computations \cite{Shor1996, Harrington2022, deVega2017, Breuer2016}. Quantum error correction (QEC) has nevertheless seen substantial progress, a prerequisite for achieving fault-tolerant quantum computation \cite{Campbell2017}. Standard discrete-variable QEC approaches have achieved experimental milestones, most notably demonstrating logical error suppression below the surface code threshold \cite{Google2024}. Despite these successes, reducing the physical-qubit overhead required to reach useful logical error rates remains a major challenge, which motivates qubits with biased noise. For example, biased noise, in which a single Pauli error channel dominates, can enable substantially higher thresholds with a lower qubit overhead \cite{Aliferis2008, Webster2015, Tuckett2018, Xu2023}.

Cat qubits are a leading biased-noise qubit. They encode the logical basis states approximately as coherent states $\lvert\pm\alpha\rangle$ of an oscillator, so bit-flip errors are suppressed exponentially in the mean photon number while phase-flip errors grow only linearly \cite{Cochrane1999, Mirrahimi2014, Puri2017, Puri2019}. The cat manifold can be stabilized by engineered two-photon dissipation \cite{Mirrahimi2014, Lescanne2020, Berdou2023, Reglade2024} or, as in this work, by a Kerr nonlinearity and a two-photon drive \cite{Puri2017, Grimm2020}. Kerr-cat qubits suit gate engineering because these two terms open a large energy gap that permits fast, coherent operations \cite{Frattini2024, Hajr2024, Bhandari2024, Ding2025}. The remaining phase flips are corrected by an outer discrete-variable code \cite{Guillaud2019, Chamberland2022, Darmawan2021, Ruiz2025, Tuckett2019}; this was recently demonstrated in a distance-5 repetition code with cat data qubits and transmon ancillas \cite{Ocelot2025}.

Realizing concatenated cat codes, in which both the data and the measurement qubits are cat qubits, requires entangling gates between cat qubits that preserve their biased-noise structure, and in particular a bias-preserving CX gate. With only CZ available, syndrome-extraction overhead erases the noise-bias advantage at realistic error rates, while a bias-preserving CX keeps the bias advantage and enables hardware-efficient magic-state preparation \cite{Ruiz2026, Ruiz2026Magic}. As we show in Section~\ref{sec:theory}, a CX gate implemented in the naive way is generated by a Hamiltonian that does not commute with the dominant phase-flip channel of the target, so that a phase flip during the gate can become a bit flip, thereby compromising the exponential suppression of bit-flip errors.

Effective-Hamiltonian and architecture studies established the bias-preserving CX and its dissipative and hybrid variants \cite{Puri2020, Guillaud2019, Chamberland2022, Xu2022, Gautier2022, Hann2025}. Circuit-level proposals for Kerr-cat qubits either implement ZZ-rotation ($R_{ZZ}$) gates without quantifying the logical noise bias \cite{Chono2022, Aoki2024, Aoki2025}, or propose hardware for the bias-preserving CX without simulating the resulting gate, based on SNAILs  \cite{Puri2020} or on a multi-loop SQUID with engineered nonlinearity \cite{Hua2025}. For dissipatively stabilized cats, Ref.~\cite{Cottet2024} proposed pumping an ATS at the control frequency to create the conditional-displacement coupling of the CX, with stabilization switched off during the gate and the resulting linear drive cancelled by a differential-flux pump. Ref.~\cite{Ye2026} recently proposed a unitary cat--cat CX based on a vacuum-conditional beam splitter.

To the best of our knowledge, no dedicated parametric coupler has been proposed and validated at the full-circuit level for activating the bias-preserving CX interaction between two Kerr-cat qubits. Here, we show that the ATS becomes a viable element for this gate when it is used as a coupler rather than as the nonlinear element of the cat modes. Placed between two resonators, each hosting a Kerr-cat qubit, a single ATS \cite{Lescanne2020} activates both two-mode interactions of the bias-preserving CX of Ref.~\cite{Puri2020}: a three-wave-mixing term $g_3$ and a conditional displacement $g_{\mathrm{CD}}$. The Kerr confinement of both qubits stays active throughout the gate, and the parasitic fourth-order nonlinearity of the ATS vanishes at the operating bias point, as for the linear quantum coupler \cite{linc}. Our contribution is threefold. First, we identify the ATS bias point and pump tones that jointly activate $g_3$ and $g_{\mathrm{CD}}$ through a single coupler, characterize the parasitic processes that constrain the circuit design, and show that the residual higher-order displacement terms can be compensated. Second, we show that a photon-number asymmetry between control and target ($\lvert\beta\rvert^2>\lvert\alpha\rvert^2$) suppresses non-adiabatic excitation of the control and enables gate times $T\approx1/K$ without derivative-based pulse shaping. Third, we simulate the full nonlinear coupler circuit with effective Kerr-cat terms and characterize the gate by logical process tomography, noise bias and bit-flip lifetimes. At $T=50$~ns the gate reaches $99.69\%$ average fidelity without loss and, under single-photon loss at $\kappa_{\rm 1ph}/2\pi=100$~kHz, a noise bias $\eta\approx9922$, about 570 times that of a naive CX.

The remainder of this paper is organized as follows. In Section~\ref{sec:theory}, we review the Kerr-cat qubit and the criteria for bias-preserving operations, then describe the circuit and its Hamiltonian under the parametric drives that generate CX interactions. In Section~\ref{sec:results}, we present numerical simulations of the interaction strengths and the full CX gate, including its tomographic noise-bias characterization. Finally, Section~\ref{sec:conc} concludes by stating the advantages of the ATS coupler and discussing prospects for scaling this architecture.
\section{Theory} \label{sec:theory}

\subsection{Kerr-Cat Qubits} \label{subsec:kerrcat}

\begin{figure}[h!] 
    \centering
    \includegraphics[width=\linewidth]{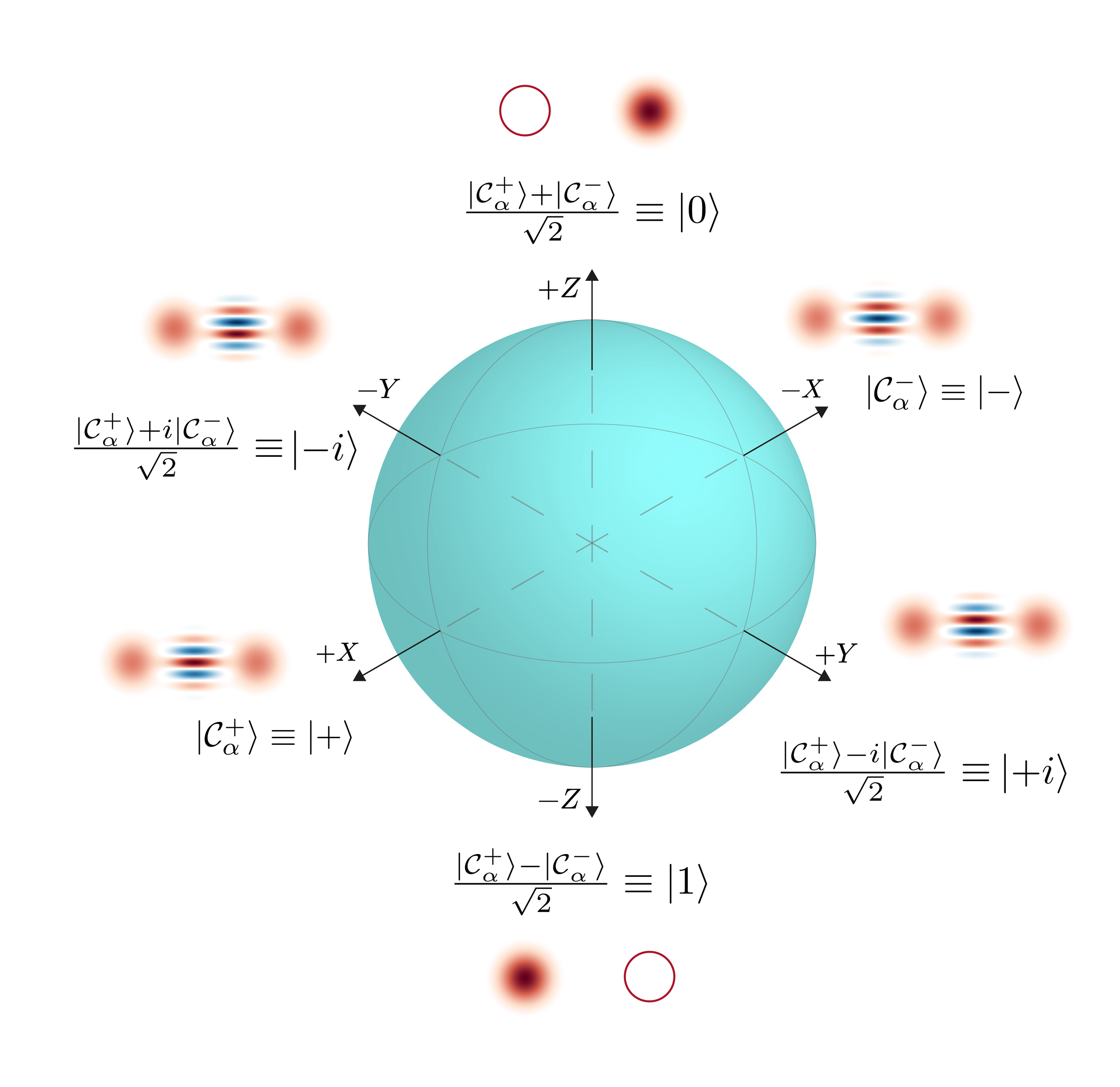}
    \caption{\justifying Bloch-sphere representation of the Kerr-cat logical subspace. The logical computational states lie at the $\pm Z$ poles, while the even and odd cat states lie at the $\pm X$ poles.}
    \label{fig:cat_qubit}
\end{figure}

The logical basis states are the nearly orthogonal coherent states $\ket{\pm\alpha}\simeq(\ket{C_\alpha^+}\pm\ket{C_\alpha^-})/\sqrt2$, where $\ket{C_\alpha^\pm}$ are the even and odd cat states (\autoref{fig:cat_qubit}). Their overlap falls exponentially with the mean photon number $|\alpha|^2$, which protects against bit flips from local noise such as single-photon loss, while phase flips grow only linearly in $|\alpha|^2$ \cite{Grimm2020, Puri2017, Cochrane1999}. The confinement is achieved by engineering a Hamiltonian that combines a Kerr nonlinearity with a parametric two-photon drive. In a frame rotating at half the two-photon drive frequency, and assuming zero detuning, the effective Hamiltonian is given by \cite{Puri2017, Grimm2020}
\begin{align}
\hat{H}_\text{cat}/\hbar =- K\hat{a}^{\dagger2} \hat{a}^2 + \epsilon_{2} \hat{a}^{\dagger2} + \epsilon_{2}^\ast \hat{a}^2,
\label{eq:1}
\end{align}
where $\epsilon_2$ is the two-photon drive strength, $K$ is the Kerr-nonlinearity, and $\hat{a}$ is the annihilation operator.

The amplitude of the cat state is tunable in situ via the two-photon drive strength $\epsilon_2$, which stabilizes the coherent states at displacements $\pm\alpha = \pm\sqrt{\epsilon_2/K}$. Several superconducting nonlinear elements capable of providing Kerr nonlinearity and a parametric two-photon drive have been demonstrated or proposed for Kerr-cat stabilization, most notably the SNAIL \cite{Grimm2020} and the Symmetrically Threaded SQUID (STS) \cite{Bhandari2024}.

\subsection{Bias-Preserving CX Gate} \label{subsec:biaspreservingcxgate}

Entangling gates used for error correction should not convert the dominant phase-flip errors into bit-flip errors; that is, they must preserve the noise bias. Our proposed ATS coupler implements such bias-preserving entangling gates between Kerr-cat qubits. In this subsection, we outline the conditions required for a gate to be bias-preserving following the framework of Ref. \cite{Puri2020} and describe how the ATS coupler satisfies these requirements.

\subsubsection{Not All Gates Preserve Bias\label{subsubsec:reqbias}} 
To demonstrate that bias preservation is not an intrinsic property of all logical gates, we analyze the performance of two different CX gate implementations under loss. 

As detailed below, the first implementation fails to preserve bias. The Hamiltonian for it is defined as:
\begin{equation}
    \label{eq:nativeCXPauli}
    \hat{H}_{CX}/\hbar = \Omega \left[\left(\frac{\hat{I}_1 + \hat{Z}_1}{2}\right) \otimes \hat{I}_2 + \left(\frac{\hat{I}_1 - \hat{Z}_1}{2}\right) \otimes \hat{X}_2 \right],
\end{equation}
where $\hat{X}_{i}$ is a logical Pauli operator of cat qubit $i$ of the form
\begin{align*}
    \hat{X}_i = &\ket{\alpha}_i\bra{-\alpha}_i + \ket{-\alpha}_i\bra{\alpha}_i,
\end{align*}
with similar definitions for $\hat{Z}_{i}$ and $\hat{I}_{i}$ on the Bloch sphere of the cat encoding. The unitary operator of this CX gate up to a global phase, controlled by qubit 1, is realized by evolving the Hamiltonian in \autoref{eq:nativeCXPauli} for a time $t = T = \frac{\pi}{2\Omega}$:
\begin{equation}
    \label{eq:nativeCXUnitary}
    \hat{U}(T) = \left[\left(\frac{\hat{I}_1 + \hat{Z}_1}{2}\right) \otimes \hat{I}_2 + \left(\frac{\hat{I}_1 - \hat{Z}_1}{2}\right) \otimes \hat{X}_2 \right].
\end{equation}
Consider a phase-flip error $\hat{Z}_2$ occurring on the target at time $t = \tau$. Since $\{\hat{X}_2,\hat{Z}_2\}=0$, the target phase-flip operator anticommutes with the conditional-$X$ term and does not commute with the full gate Hamiltonian. The resulting unitary operator at time $t = T$ becomes:
\begin{equation}
    \label{eq:nativeCXUnitaryPhase}
    \hat{U}^\prime(T) = \hat{I}_1 \otimes \hat{Z}_2 e^{i\Omega (T - \tau) (\hat{I}_1 - \hat{Z}_1) \otimes \hat{X}_2} \hat{U}(T).
\end{equation}
The above unitary matrix implements both a phase and bit-flip error in the target qubit, thus partially converting phase-flip errors into bit-flip errors on the target cat qubit and reducing the noise bias of the target cat qubit. 

In contrast to the above gate, the second CX gate implementation is designed to preserve the bit-flip protection of the control and target cat qubits and is written as \cite{Guillaud2019, Puri2020}
\begin{equation}
    \hat{U}(t) = |\beta\rangle\langle\beta| \otimes \hat{I} + |-\beta\rangle\langle-\beta| \otimes e^{i\frac{\pi}{T} t \hat{a}_t^\dagger \hat{a}_t}, \label{eq:ideal_bias_preserving_cx_gate}
\end{equation}
where $\hat{a}_t$ is the annihilation operator for the target qubit and a CX gate is realized at time $t = T$. This unitary corresponds to a conditional rotation of the target mode in phase space. The target single-photon-loss operators form a subspace closed under commutation with the conditional-rotation generator $\hat{a}_t^\dagger\hat{a}_t$. As shown below, this closure ensures that a propagated target-loss error contains only logical $Z$-type components at the gate output when $\langle \alpha |-\alpha \rangle \approx 0$.

Single-photon loss constitutes the dominant error mechanism in superconducting resonators, ultimately limiting the logical lifetime of the encoded cat qubits \cite{Ofek2016, Reglade2024}. Consequently, we focus our analysis on determining the bias of the CX gate under the single-photon loss channel. 

First, we consider a single-photon loss error on the control cat qubit at time $\tau \in [0,T]$
\begin{align*}
    \hat{U}_\text{c,err} &= \hat{U}(T-\tau) (\hat{a}_c \otimes \hat{I})\hat{U}(\tau)\numberthis\\
    &= \beta|\beta\rangle\langle\beta| \otimes \hat{I} - \beta |-\beta\rangle\langle-\beta| \otimes e^{i\pi\hat{a}_t^\dagger \hat{a}_t} \\
    \hat{U}_\text{c,err} &= (\hat{a}_c \otimes \hat{I}) \hat{U}(T) \propto \hat{Z}_1 \hat{U}_\text{CX} \numberthis,
\end{align*}
where $\hat{a}_c$ is the annihilation operator for the control qubit. As shown, single-photon loss translates into a phase flip error on the control qubit. 

Next, we consider single-photon loss on the target qubit at time $\tau$
\begin{align*}
    \hat{U}_\text{t,err} &= \hat{U}(T-\tau) (\hat{I} \otimes \hat{a}_t)\hat{U}(\tau)\numberthis\\
    &= (\hat{I} \otimes \hat{a}_t) (|\beta\rangle\langle\beta| \otimes \hat{I} \\&\ \ + e^{-i\pi\frac{T-\tau}{T}}|-\beta\rangle\langle-\beta| \otimes e^{i\pi\hat{a}_t^\dagger \hat{a}_t}) \\
    \hat{U}_\text{t,err} &\propto \hat{U}_\text{err}(\tau) \hat{U}_\text{CX} \numberthis,
\end{align*}
where $\hat{U}_\text{err}(\tau)$ in Pauli basis can be expressed as:
\begin{equation}
    \hat{U}_\text{err}(\tau) = \frac{1}{2}(\hat{I} + \hat{Z}_1) \hat{Z}_2 + \frac{1}{2}e^{-i\pi\frac{T-\tau}{T}} (\hat{I} - \hat{Z}_1)\hat{Z}_2 \label{eq:error_channel_ptm_target_qubit}.
\end{equation}
\autoref{eq:error_channel_ptm_target_qubit} demonstrates that single-photon loss on the target yields \textit{only} phase-flip errors (terms proportional to $\hat{Z}$). Unlike the naive implementation, which converts phase errors into bit errors, this gate preserves the noise bias. Accordingly, the primary objective of this work is to realize a hardware approximation of this ideal unitary.

\subsubsection{Bias-Preserving CX Gate Hamiltonian\label{subsubsec:cxbias}}

An approximation of the ideal, bias-preserving CX gate defined in \autoref{eq:ideal_bias_preserving_cx_gate} was proposed by Ref.~\cite{Puri2020} and is detailed in \autoref{eq:CXgateonCatqubits}. In this subsection, we identify the specific Hamiltonian terms required to engineer this effective evolution, with a particular focus on the entangling terms with strengths $g_3$ and $g_{\rm CD}$. We take the initial target and control amplitudes, $\alpha$ and $\beta$, to be real and positive, assume equal Kerr coefficients $K$, and choose $\phi(0)=0$ and $\phi(T)=\pi$.

\begin{align}
    &\hat{H}_{CX}/\hbar = -K (\hat{a}_c^{\dagger 2} - \beta^2) (\hat{a}_c^{2} - \beta^2) - \nonumber\\
    & K \left[\hat{a}_t^{\dagger 2} - \alpha^2 e^{-2i\phi(t)} \left(\frac{\beta - \hat{a}_c^\dagger}{2\beta}\right) - \alpha^2 \left(\frac{\beta + \hat{a}_c^\dagger}{2\beta}\right) \right] \nonumber\\
    & \times \left[\hat{a}_t^{2} - \alpha^2 e^{2i\phi(t)} \left(\frac{\beta - \hat{a}_c}{2\beta}\right) - \alpha^2 \left(\frac{\beta + \hat{a}_c}{2\beta}\right) \right]\nonumber\\
    &-\frac{\dot{\phi}(t)}{4\beta} \hat{a}_t^\dagger \hat{a}_t \left(2\beta - \hat{a}_c^\dagger - \hat{a}_c \right).
    \label{eq:CXgateonCatqubits}
\end{align}

To see how \autoref{eq:CXgateonCatqubits} works, replace $\hat a_c$ by its value in the two control states, $\hat a_c\approx+\beta$ for $\ket{0_\beta}\approx\ket{\beta}$ and $\hat a_c\approx-\beta$ for $\ket{1_\beta}\approx\ket{-\beta}$. For a control in $\ket{0_\beta}$, the target terms reduce to the static Kerr-cat Hamiltonian $-K(\hat a_t^{\dagger2}-\alpha^2)(\hat a_t^2-\alpha^2)$ and the last term vanishes, so the target is unchanged. For a control in $\ket{1_\beta}$, they become $  -K\big(\hat a_t^{\dagger2}-\alpha^2e^{-2i\phi(t)}\big) \big(\hat a_t^{2}-\alpha^2e^{2i\phi(t)}\big) -\dot\phi(t)\,\hat a_t^\dagger\hat a_t$. The first term is a Kerr-cat Hamiltonian whose stabilized states $\pm\alpha e^{i\phi(t)}$ rotate with $\phi(t)$. In the frame $e^{i\phi(t)\hat a_t^\dagger\hat a_t}$ that co-rotates with the cat, the second term cancels the term generated by the frame change, leaving the static Kerr-cat Hamiltonian. At $\phi(T)=\pi$ the rotation maps $\ket{\pm\alpha}\to\ket{\mp\alpha}$, a logical $\hat X$ on the target. The target therefore follows the rotation exactly, at any speed, and the gate error instead enters through the control, on which the last term of \autoref{eq:CXgateonCatqubits} exerts a force analyzed in Section~\ref{sec:gate-time-kerr}. \autoref{eq:CXgateonCatqubits} can be rearranged and regrouped as follows:

\begin{subequations}    \label{eq:CXgateonCatqubits_splitted}
\begin{align}
    \hat{H}&_{CX}/\hbar = -K (\hat{a}_c^{\dagger 2} - \beta^2) (\hat{a}_c^{2} - \beta^2)  \\
    & -K(\hat{a}_t^{\dagger 2} - \alpha^2\cos\phi\, e^{-i\phi}) (\hat{a}_t^{2} - \alpha^2\cos\phi \, e^{i\phi})  \\
    &+ \Delta_t \hat{a}_t^{\dagger}\hat{a}_t+ \Delta_c \hat{a}_c^{\dagger}\hat{a}_c \\
    & + g_1(\hat{a}_ce^{i\pi/2}+\hat{a}_c^{\dagger}e^{-i\pi/2}) \\
    & + g_3(\hat{a}_t^{\dagger2}\hat{a}_ce^{i(\phi+\pi/2)} + \hat{a}_t^{2}\hat{a}_c^{\dagger}e^{-i(\phi+\pi/2)}) 
    \label{eq:g3_in_full_cx}\\
    &  + g_{\rm CD}(\hat{a}_t^{\dagger} \hat{a}_t (\hat{a}_c +\hat{a}_c^{\dagger})).
    \label{eq:gcd_in_full_cx}
\end{align}
\end{subequations}

By decomposing the bias-preserving CX gate Hamiltonian as shown in \autoref{eq:CXgateonCatqubits_splitted}, we highlight the distinct interaction terms required to synthesize this gate:
\begin{itemize}
    \item[a.] \textbf{Control Stabilization:} A static Kerr nonlinearity combined with a two-photon drive (amplitude $K\beta^2$, zero phase) on the control resonator.
    \item[b.] \textbf{Target Stabilization:} A static Kerr nonlinearity combined with a time-dependent two-photon drive (amplitude $K\alpha^2\cos(\phi(t))$, phase $\phi(t)$) on the target resonator.
    \item[c.] \textbf{Dynamic Detunings:} A time-dependent frequency shift on the target $\Delta_t = - \frac{\dot{\phi}(t)}{2}$, and a required detuning on the control $\Delta_c = \frac{-K\alpha^4}{\beta^2}\sin^2(\phi(t))$.
    \item[d.] \textbf{Control Displacement Drive:} A single-photon drive on the control resonator, $g_1$, with amplitude $\frac{K\alpha^4}{2\beta}\sin(2\phi(t))$ and phase $\pi/2$.
    \item[e.] \textbf{Three-Wave Mixing ($g_3$):} A parametric interaction between the Kerr-cat modes with coupling strength $g_3=\frac{-K\alpha^2}{\beta}\sin(\phi(t))$ and phase $\phi(t)+\pi/2$.
    \item[f.] \textbf{Conditional Displacement ($g_{\rm CD}$):} A conditional displacement interaction ($ \propto \hat{a}_t^\dagger \hat{a}_t (\hat{a}_c + \hat{a}_c^\dagger)$) with strength $g_{\rm CD} = \frac{\dot{\phi}(t)}{4\beta}$.
\end{itemize}

Of these, the most challenging terms to implement are the two-mode interactions $g_3$ and $g_{\rm CD}$. In this paper, we propose using the asymmetrically threaded SQUID (ATS) as a tunable parametric coupler between the Kerr-cat qubit resonators. By driving the ATS at two pump frequencies, we can selectively activate the desired entangling interactions required for the bias-preserving CX Hamiltonian given by  \autoref{eq:CXgateonCatqubits}.

\subsection{ATS Coupler \label{subsec:couplerinteraction}} 

\begin{figure}[h!] 
    \centering
    \includegraphics[width=0.8\linewidth]{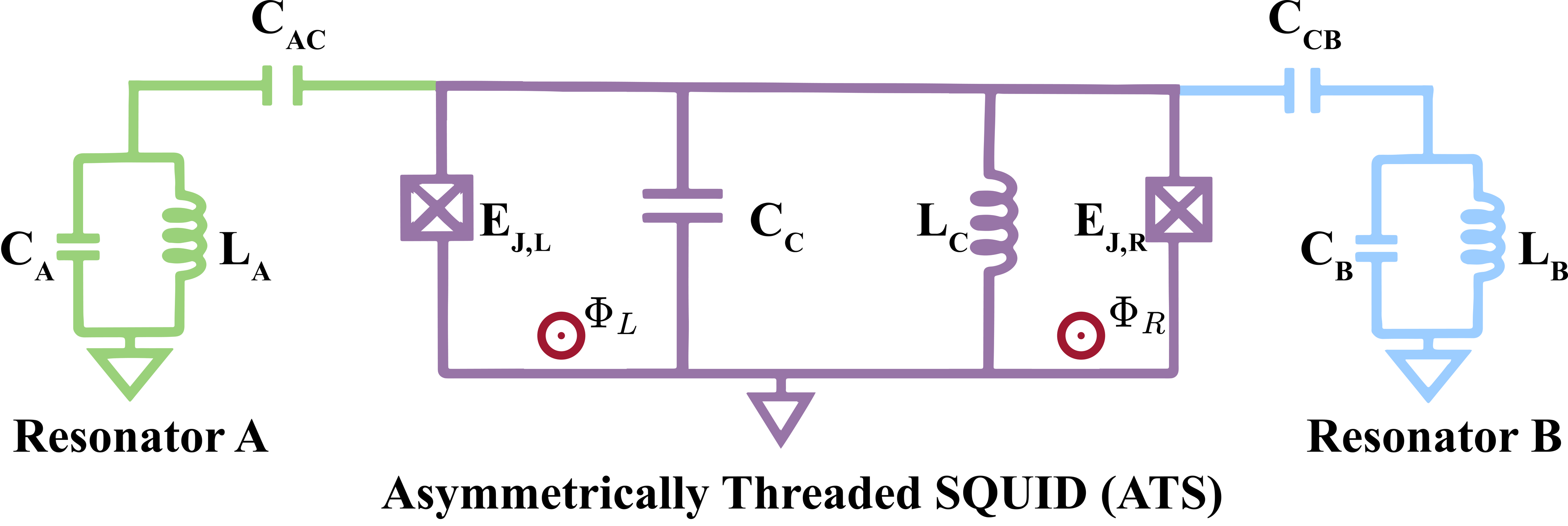}
    \caption{\justifying Schematic circuit diagram of the Resonator-ATS-Resonator architecture modeled in this study.}
    \label{fig:chip_design}
\end{figure}

The architecture proposed in this work comprises two superconducting planar resonators capacitively coupled to a common asymmetrically threaded SQUID (ATS), \autoref{fig:chip_design}. Each resonator has a fixed capacitive coupling to the ATS. The resonators are driven by individual drive lines and the ATS via an on-chip flux drive line. The nonlinear circuit elements used to stabilize Kerr-cat qubits within each resonator are not specified to allow flexibility in the application of our CX gate scheme. Their contributions are represented by effective Kerr nonlinearities and two-photon drives, which are added separately to the coupling-circuit Hamiltonian below. In this section, we describe how the ATS can be used as a three-wave mixing element to activate desired interactions and minimize unwanted nonlinearities.

\subsubsection{Introduction to the Resonator-ATS-Resonator System}

Following techniques adapted from Black-Box Quantization \cite{Nigg2012,jha2024smthesis}, we first diagonalize the quadratic Hamiltonian formed by the capacitances and linear inductances, and then express the ATS Josephson potential in the resulting normal-mode basis as discussed in Appendix \ref{app:rar-methodology}. This gives the resonator–ATS–resonator (RAR) Hamiltonian
\begin{align}
\hat{H}_{\rm RAR}/\hbar &= \sum_{o = \{a,b,c\}}{\omega_{o} \hat{o}^\dagger \hat{o}} \notag \\
&\quad - 2 E_J \cos(\hat{\varphi} + \varphi_{\Delta})\cos(\varphi_{\Sigma}) \notag \\
&\quad + 2 \Delta E_J \sin(\hat{\varphi} + \varphi_{\Delta})\sin(\varphi_{\Sigma}) \notag \\
&\quad + 2 E_{J,2} \cos(2\hat{\varphi} + 2\varphi_{\Delta})\cos(2\varphi_{\Sigma}),\label{eq:RAR_sys_hamiltonian}
\end{align}
where the operators $\hat a$, $\hat b$, and $\hat c$ annihilate excitations in the normal modes associated with resonator A (target qubit, $\hat{a}_t$), resonator B (control qubit, $\hat{a}_c$), and the ATS, respectively. Here $E_J$ is the mean junction energy, $\Delta E_J$ the junction asymmetry, and $E_{J,2}$ the second-harmonic Josephson energy. The ATS coupler is biased by two external flux biases, including the common flux $\varphi_{\Sigma}= \frac{\varphi_R + \varphi_L}{2}$ through both the superconducting loops and the differential flux $\varphi_\Delta = \frac{\varphi_L - \varphi_R}{2}$ between the two loops. Throughout this work, $E_J$, $\Delta E_J$, $E_{J,2}$, $E_C$, and $E_L$ denote the corresponding physical energies divided by $\hbar$, and therefore have units of angular frequency.

In the following sections, we use $\Delta E_J=E_{J,2}=0$ and $\varphi_\Sigma=\pi/2$, where the ATS Josephson potential vanishes, leaving a linear coupling circuit. This operating point eliminates ATS-induced static nonlinear interactions in the ideal model, while flux modulation activates the desired entangling interactions. We direct the reader to Appendix \ref{app:ej2_dej} for a detailed discussion of the implications of residual, non-zero $\Delta E_J$ and $E_{J,2}$. 

\subsection{Bias-Preserving Gates using ATS} \label{subsec:bias-preserving-Gates-using-ATS}

The ATS coupler can be driven to parametrically activate a bias-preserving CX gate between two Kerr-cat qubits, while suppressing unwanted nonlinearities. At a different bias point\footnote{The ATS, at this bias point, has also been called the LINC \cite{linc} or STS \cite{Bhandari2024}.}, the same coupler also provides a beam-splitter interaction, the ingredient of a bias-preserving ZZ gate~\cite{Puri2020}. We derive this interaction in Appendix~\ref{app:zzgate}.

As discussed in Section~\ref{subsubsec:cxbias}, the three-wave-mixing and conditional-displacement interactions, with strengths $g_3$ and $g_{\rm CD}$, are required to implement a bias-preserving CX gate. To realize these interactions, we bias the ATS at its standard asymmetric bias point of $(\varphi_{\Sigma}, \varphi_{\Delta}) = (\pi/2, \pi/2)$ and pump along its common flux $\varphi_{\Sigma}$ with a drive $\epsilon(t)$, yielding:

\begin{equation}
\hat{H}_{\rm RAR,CX}/\hbar = \sum_{o = \{a, c, b\}}\omega_o \hat{o}^\dagger \hat{o} - 2E_J \sin[\epsilon(t)]\sin\hat{\varphi}.
\label{eq:driver_ham}
\end{equation}

To activate our desired interactions, we may keep the coupler in its ground state, approximate $\sin(\epsilon(t)) \approx \epsilon(t)$, transform into the rotating frame of all three normal modes, and remove fast rotating terms.

In particular, to realize the $g_3$ interaction, the pump tone $\omega_p = 2\omega_a - \omega_b$ is used, activating:
\begin{align}
    \tilde{\hat{H}}_{\rm RAR,CX}/\hbar &\approx \underbrace{\frac{1}{2}E_J \epsilon_p(t) \varphi_a^2\varphi_b}_{g_{\rm 3}}\left(\hat{a}^{\dagger2}\hat{b} + \hat{a}^2\hat{b}^\dagger\right).
    \label{eq:ham_g3}
\end{align}

Similarly, to realize the $g_{\rm CD}$ interaction, the pump tone $\omega_p = \omega_b$ is used, activating:
\begin{align}
    \tilde{\hat{H}}_{\rm RAR,CX}/\hbar &\approx \underbrace{E_J \epsilon_p(t) \varphi_a^2\varphi_b}_{g_{\rm CD}}\ \hat{a}^\dagger \hat{a}\left(\hat{b} + \hat{b}^\dagger\right).
    \label{eq:ham_g_cd}
\end{align}
This pump also produces a resonant unconditional displacement of the control mode and additional higher-order contributions. Their compensation is discussed in Appendix \ref{app:nonlinear_terms}.  
\section{Simulations} \label{sec:results}

We first examine the dependence of the effective gate dynamics on gate time and cat amplitudes, and then investigate the circuit implementation.

\subsection{Gate Time, Kerr Nonlinearity, and Cat-Size Relation}
\label{sec:gate-time-kerr}

We first investigate how the gate time $T$, Kerr coefficient $K$, and cat amplitudes $\alpha$ (target) and $\beta$ (control) affect the lossless performance of the effective CX Hamiltonian in \autoref{eq:CXgateonCatqubits}. To interpret the numerical results, we develop a simplified analytical model that estimates how strongly the control mode is excited out of its cat subspace when the gate phase $\phi(t)$ increases linearly in time.

The dominant error of the lossless effective gate is non-adiabatic excitation of the control mode. The conditional displacement exerts a force on the control mode proportional to the target photon number, while the Kerr confinement opens a gap $\Delta_{\mathrm{gap},c}\simeq4K\beta^2$ that protects the cat subspace \cite{Puri2019}. Treating the control fluctuations as a harmonic mode driven by this force for a gate phase that increases linearly in time, $\phi(t)=\pi t/T,\; 0\leq t\leq T$, the excitation probability is governed by the dimensionless parameter
\begin{equation}
    \xi\equiv\frac{\alpha^2}{\beta^3KT}.
    \label{eq:xi}
\end{equation}
Assuming this leakage dominates the error, the infidelity is approximately
\begin{align}
    1-\mathcal F \;&\approx\;1-e^{-x}\label{eq:infid-combined}\\
    x&\simeq\frac{\pi^2}{64}\xi^2\sin^2(2K\beta^2T),
\end{align}
where the oscillatory factor prevents an exact single-parameter description of the dynamics. Replacing $\sin^2(2K\beta^2T)$ by unity gives an approximate upper envelope, $1-e^{-\pi^2\xi^2/64}$. A detailed derivation is given in Appendix \ref{app:excitation-model}.

\begin{figure}[h]
    \centering
    \includegraphics[width=\linewidth]{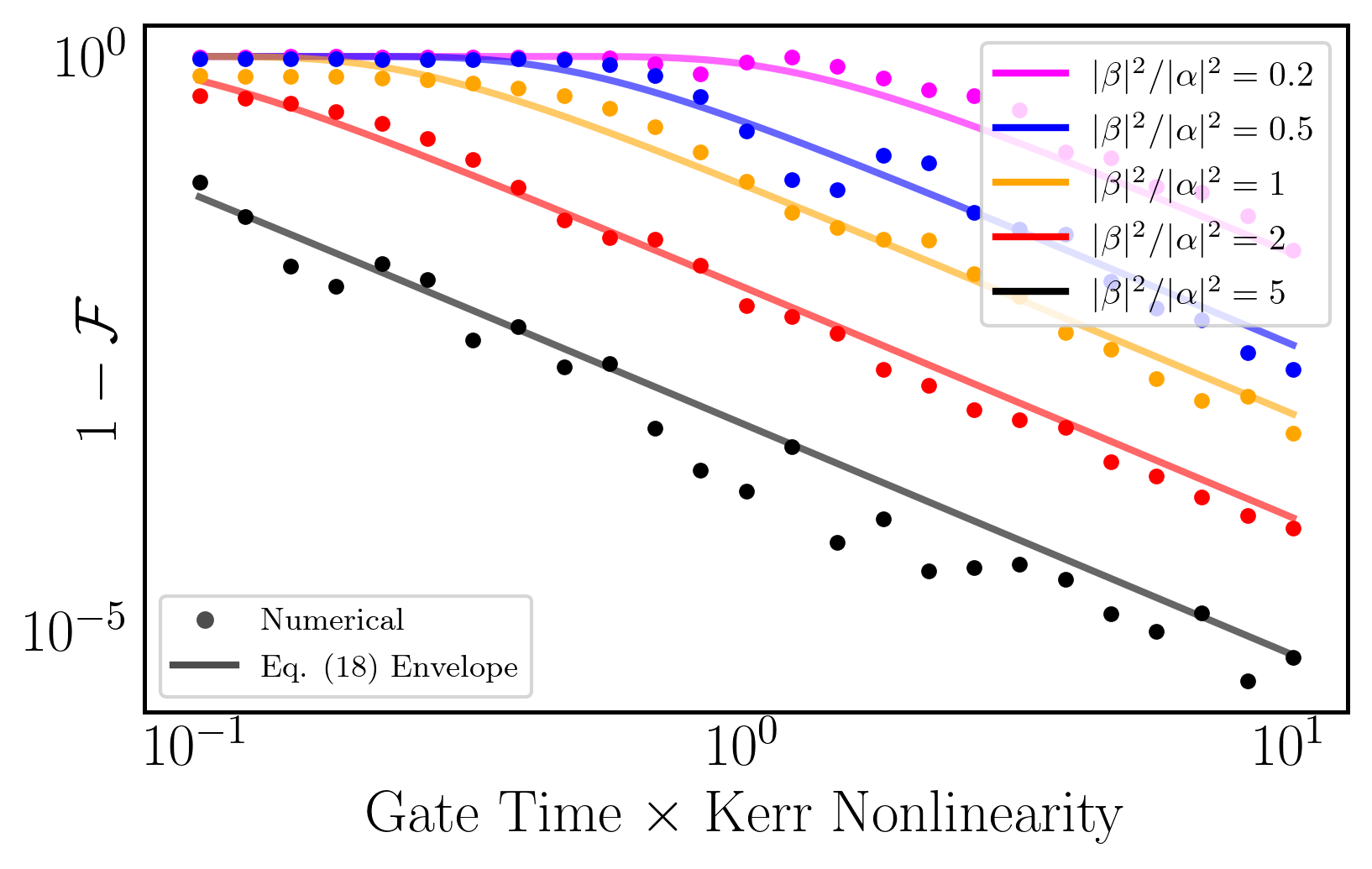}
    \caption{\justifying Infidelity of \autoref{eq:CXgateonCatqubits} vs. $KT$, for fixed $\alpha^2=2$, $\beta^2\in\{2,4,10\}$ and fixed $\beta^2=2$, $\alpha^2\in\{4,10\}$. Numerical results, markers, are compared with the envelope obtained from \autoref{eq:infid-combined}, solid lines.}
    \label{fig:xi-collapse}
\end{figure}

\autoref{fig:xi-collapse} compares the envelope of \autoref{eq:infid-combined} with direct simulation of \autoref{eq:CXgateonCatqubits}. Over the full parameter range, the simulated infidelity follows the envelope, with point-to-point oscillations that reflect the factor $\sin^2(2K\beta^2T)$, sampled more coarsely than its period. A joint power-law fit of the infidelity against $\xi$ in the non-saturated regime ($1-\mathcal F<0.3$) returns an exponent $2.17 \pm 0.03$ ($R^2 = 0.983$), close to the quadratic scaling of the envelope. Points can exceed the envelope slightly because the model neglects target photon-number fluctuations and target-control correlations. We therefore use \autoref{eq:infid-combined} as an estimate of the scaling rather than a strict bound.

\autoref{eq:xi} explains why a small target and a large  control cat favor fast gates: the conditional-displacement force on the control scales as $\alpha^2/\beta$, while the control gap grows as $\beta^2$. For symmetric cats ($\alpha=\beta$), $\xi=1/(\alpha KT)$; unequal amplitudes add a free parameter, so increasing $\beta$ at fixed $\alpha$, $K$ and $T$ lowers the excitation without the derivative-based pulse corrections of Ref.~\cite{Xu2022}. Because the lossless infidelity has no floor and keeps falling with $T$, gate fidelity does not set the target size, but the required bit-flip suppression, $\propto e^{-4\alpha^2}$, does. Ref.~\cite{Puri2020} simulated the effective gate at $T=10/K$ with $\alpha=\beta=2$. We show this is not a lower bound on the gate time.

Using the larger cat as the control allows faster gates. If an architecture requires the opposite CX direction, the simplest option is to run the same scheme with control and target interchanged, at the cost of a slower gate when the larger cat now serves as the target. Alternatively, the direction can be reversed with the standard identity
\begin{equation}
    \text{CX}_{2\to1} = (\hat H\otimes\hat H)\;\text{CX}_{1\to2}\;(\hat H\otimes\hat H),
    \label{eq:cx-reversal}
\end{equation}
but the Hadamard gates are not free. They interchange logical $Z$ and $X$ errors, and no native bias-preserving Hadamard exists for cat qubits \cite{Puri2020}, so $\hat H$ must be synthesized by the measurement-based basis changes and state injection already required for non-Clifford operations \cite{Chamberland2022}. A comparison of these options at the architecture level lies beyond the present work.

\subsection{Circuit Parameters}
\label{sec:parameter-choice}

\autoref{eq:xi} shows a larger $K$ permits a shorter gate, but the achievable $K$ is set by the underlying Kerr-cat platform, not by the coupler design. Reported values span more than an order of magnitude, from $K/2\pi\approx0.3$~MHz in Kerr-suppressed designs \cite{Frattini2024} and $\approx 1.2$~MHz in 2D devices \cite{Hajr2024} to $6.7$~MHz in the first Kerr-cat experiment, at reduced coherence \cite{Grimm2020}. We choose $K/2\pi=3$~MHz and $T=50$~ns ($KT\approx0.94$), with $\alpha^2=2$ and $\beta^2=10$; Eq.~(\ref{eq:xi}) then gives $\xi\approx0.067$ and a non-adiabatic envelope of $\approx7\times10^{-4}$. $\alpha^2$ and $\beta^2$ are chosen to exploit the cat-size asymmetry of Section~\ref{sec:gate-time-kerr} while keeping full-circuit simulation computationally tractable at large Hilbert spaces.

As explained in Section~\ref{subsubsec:cxbias}, $g_3$ and $g_{\rm CD}$ are explicitly related to system parameters as
\begin{align}
    g_3(t) &= -\frac{K \alpha^2 \sin(\phi(t))}{\beta}\label{eq:g_3_prop}\\
    g_{\rm CD} &= \frac{\dot{\phi}(t)}{4 \beta}, \label{eq:g_cd_prop}
\end{align}
which are related to circuit parameters with \autoref{eq:ham_g3} and \autoref{eq:ham_g_cd}. Inverting \autoref{eq:ham_g3} and \autoref{eq:ham_g_cd} for the RF drive amplitudes required to realize $g_3$ and $g_{\rm  CD}$ gives:
\begin{align}
    \epsilon_{g_3}(t) &= \frac{2 g_3(t)}{E_J \varphi_a^2 \varphi_b} \cos(\omega_{g_3}t - \pi t/T - \pi/2) \label{eq:g3_driver_strength} \\
    \epsilon_{g_{\rm CD}}(t) &= \frac{ g_{\rm CD}}{E_J \varphi_a^2 \varphi_b} \cos(\omega_{g_{\rm CD}}t) \label{eq:gcd_driver_strength}.
\end{align}
Substituting $K/2\pi=3$~MHz, $T=50$~ns, $\alpha^2=2$, $\beta^2=10$ into \autoref{eq:g_3_prop} and \autoref{eq:g_cd_prop}:
\begin{align}
    g_3(t)/2\pi &\approx -1.89\,\sin\!\left(\frac{\pi t}{50\text{ ns}}\right)\text{ MHz}, \\
    g_{\rm  CD}/2\pi & \approx 0.79\text{ MHz}.
\end{align}

The trade-off between $\hat n_c^{\rm floor}\propto1/(p_a\Delta_{ac}^2)$ and the drive cost $\epsilon_{g_3}\propto1/\varphi_a^2$, explained in Appendix~\ref{app:design-methodology} in detail, implies that neither pushing $\Delta_{ac}$ to its largest available value nor maximizing $p_a$ alone is optimal. We choose bare frequencies $\omega_a^{(0)}/2\pi=5.72$~GHz, $\omega_b^{(0)}/2\pi=4.14$~GHz, which \autoref{fig:physicality-window}(f) places in the low $\max\langle\hat n_c\rangle$ region and away from the degeneracy and resonance regions, at $g_{ac}/\Delta_{ac}=0.7$, $g_{cb}/\Delta_{bc}=-0.4$, where the appreciable mode hybridization is treated by exact diagonalization of the quadratic Hamiltonian. The pump carrier frequencies are determined from the dressed normal-mode frequencies as $\omega_{g_3}=2\omega_a-\omega_b$ and $\omega_{g_{\rm CD}}=\omega_b$.

$\varphi_a=0.128$ and $\varphi_b=0.078$ result in peak drive amplitudes $\epsilon_{g_3}^{\rm peak}=0.0982$, $\epsilon_{g_{\rm CD}}=0.0205$ using \autoref{eq:g3_driver_strength} and \autoref{eq:gcd_driver_strength}, with the combined modulation satisfying $\epsilon_{g_3}^{\rm peak} + \epsilon_{g_{\rm CD}} \approx 0.12 \ll1$, consistent with the small-angle approximation $\sin[\epsilon(t)]\approx\epsilon(t)$ underlying \autoref{eq:driver_ham}.

\autoref{tab:circuit_params} details the resulting circuit parameters.

\begin{table}[h]
    \centering
    \begin{tabular}{lrlr}
        \toprule
        \textbf{Parameter} & \textbf{Value} & \textbf{Parameter} & \textbf{Value} \\
        \midrule
        $E_J/2\pi$           & 30.0 GHz & $\varphi_\Sigma$   & $\pi/2 + \epsilon(t)$ \\
        $E_C/2\pi$           & 80.0 MHz & $\varphi_\Delta$   & $\pi/2$ \\
        $E_L/2\pi$           & 41.34 GHz & $g_\text{ac}/2\pi$ & 0.4 GHz \\
        $\omega_a^{(0)}/2\pi$ & 5.72 GHz  & $g_\text{cb}/2\pi$ & 0.4 GHz \\
        $\omega_b^{(0)}/2\pi$ & 4.14 GHz  &                    & \\
        \bottomrule
    \end{tabular}
    \caption{\justifying Circuit parameters for Resonator-ATS-Resonator system. $E_J, E_C$, and $E_L$ values denote the characteristic energies of the ATS, while $\omega_{a,b}^{(0)}$ represent the bare frequencies of the target and control resonators, respectively. The parameters $g_\text{ac}$ and $g_\text{cb}$ quantify the capacitive coupling strength between resonators and ATS. DC bias of ATS is set to asymmetric flux bias point $\varphi_\Sigma = \varphi_\Delta = \pi/2$, and an additional RF bias $\epsilon(t)$ is applied to $\varphi_\Sigma$.}
    \label{tab:circuit_params}
\end{table}

\subsection{Interaction Simulations} \label{subsec:inter_sims}

Adopting the parameter set defined in Section~\ref{sec:parameter-choice}, we simulate the full-circuit Hamiltonian in \autoref{eq:driver_ham}, driven by the required control fields. Simulations were performed using the \texttt{QuTiP} library \cite{Johansson2012} and GPU-accelerated solvers via \texttt{jaxquantum} \cite{jha2024jaxquantum}.

\subsubsection{$g_3$ Interaction} \label{subsubsec:g3_sims}
The $g_3$ interaction mediates a three-wave mixing process wherein two photons in the target mode (A) are exchanged for a single photon in the control mode (B). To validate this coupling, we compare the evolution generated by the ideal effective Hamiltonian, $\hat{H}_{{\rm eff}}/\hbar = g_3 (\hat{a}^{\dagger 2} \hat{b} + \hat{a}^2 \hat{b}^\dagger)$, against the full physical Hamiltonian:
\begin{equation}
    \hat{H}_{{\rm full}}/\hbar = \hat{H}_{{\rm static}}/\hbar - 2 E_J \sin[\epsilon_{g_3}(t)] \sin\hat{\varphi}.
\end{equation}
For this verification, we treat $\epsilon_{g_3}(t)$ as a constant-envelope tone (neglecting the gate-specific time-dependence) to isolate the resonant interaction physics. The simulations include calibrated compensation of parasitic drive terms, as described in Appendix \ref{app:nonlinear_terms}. \autoref{fig:g3_sim} demonstrates close agreement between the theoretical prediction and the full circuit dynamics.
\begin{figure}[h]
    \centering
    \includegraphics[width=0.8\linewidth]{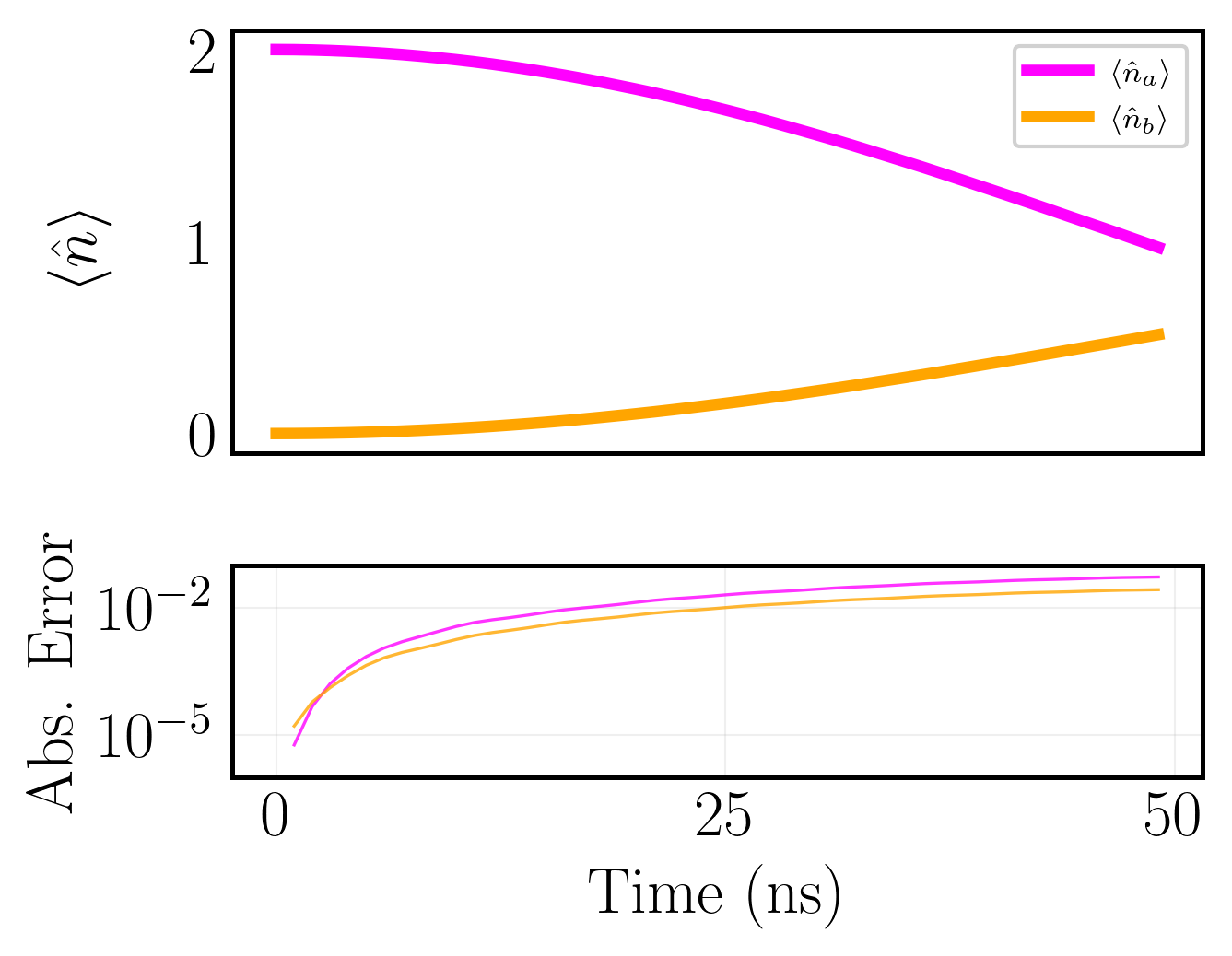}
    \caption{\justifying Simulation of the $g_3$ interaction realized by driving the ATS at the difference frequency $\omega_p = 2\omega_a - \omega_b$. The system is initialized with Qubit A (Target) in the two-photon state $\ket{2}$ and Qubit B (Control) in the vacuum state $\ket{0}$. The plot displays the absolute error between the ideal theoretical evolution and the full circuit simulation. Physically, the interaction drives a transition $\ket{0}_b\ket{2}_a \to \ket{1}_b\ket{0}_a$. Over 50 ns, the mean occupation of mode A decreases by about 1 photon, while that of B increases by about 0.5, confirming the expected 2:1 exchange ratio. The upper panel shows the mode occupations, and the lower panel shows their absolute differences from the effective model.}
    \label{fig:g3_sim}
\end{figure}

\subsubsection{$g_{\rm CD}$ Interaction} \label{subsubsec:gcd_sims}
The $g_{\rm CD}$ term implements a conditional displacement, acting as a displacement force on the control mode (B) that is proportional to the photon number of the target mode (A). We validate this interaction by comparing the ideal effective Hamiltonian, $\hat{H}_{\text{eff}}/\hbar = g_{\rm CD} \hat{a}^\dagger \hat{a} (\hat{b} + \hat{b}^\dagger)$, against the full driven system:
\begin{equation}
    \hat{H}_{{\rm full}}/\hbar = \hat{H}_{{\rm static}}/\hbar - 2 E_J \sin[\epsilon_{g_{\rm CD}}(t)] \sin\hat{\varphi}.
\end{equation}
Similar to the $g_3$ analysis, the full Hamiltonian is simulated in the rotating frame, with drive parameters calibrated. For quantitative analysis, we use the dimensionless quadrature operators $\hat{x} = (\hat{b} + \hat{b}^\dagger)/\sqrt{2}$ and $\hat{p} = i(\hat{b}^\dagger - \hat{b})/\sqrt{2}$, which satisfy the commutation relation $[\hat{x}, \hat{p}] = i$ and vacuum variance $\langle 0|\hat{x}^2|0\rangle = 1/2$.

\begin{figure}[h]
    \centering
    \begin{subfigure}[b]{0.5\linewidth}
        \includegraphics[width=0.9\linewidth]{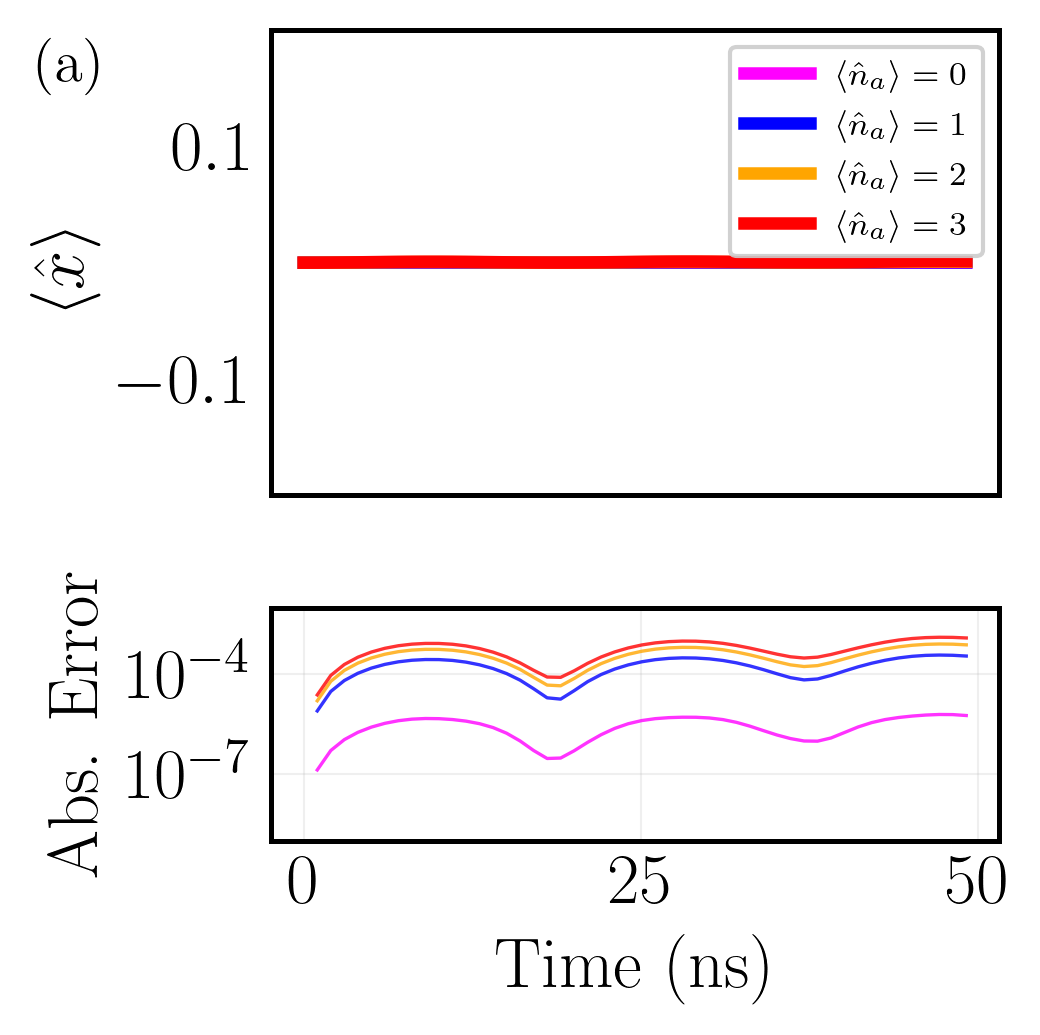}
    \end{subfigure}%
    \begin{subfigure}[b]{0.5\linewidth}
        \includegraphics[width=0.9\linewidth]{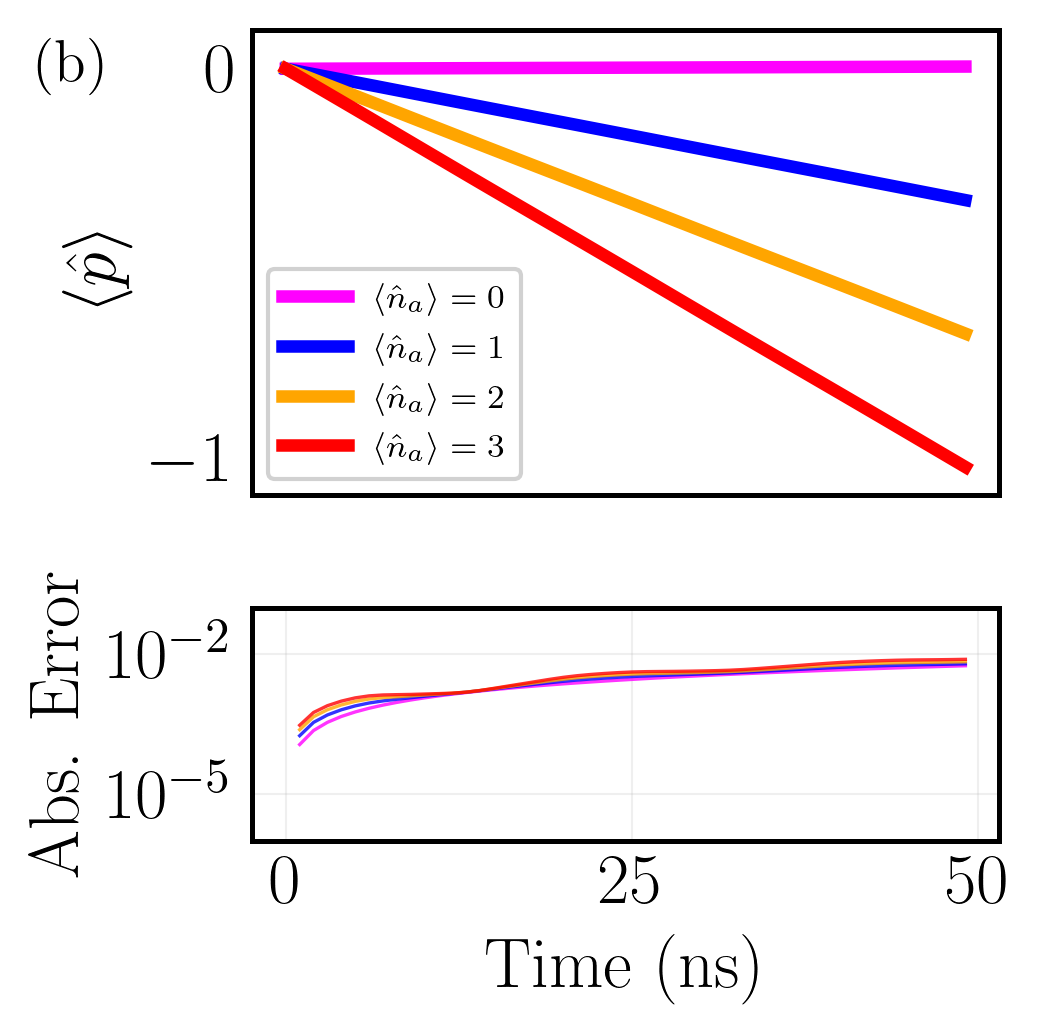}
    \end{subfigure}
    \caption{\justifying Simulation of the Conditional Displacement ($g_{\rm CD}$) interaction. The ATS is driven at the control frequency $\omega_b$ with amplitude $\epsilon_{g_{\rm CD}} \propto 1/\beta$. The plots show the time-evolution of the position $\langle \hat{x}_b \rangle$ and momentum $\langle \hat{p}_b \rangle$ of the control qubit, conditioned on the target qubit being in photon number states $n_a \in \{0, 1, 2, 3\}$. The lower panels show the absolute differences between the full-circuit and effective-model quadrature expectations. The momentum shift $\Delta p$ scales linearly with the target photon number ($ \Delta p(n_a=2) \approx 2\Delta p(n_a=1)$, etc.), verifying the conditional-displacement coupling $\hat{n}_a \hat{x}_b$.}
    \label{fig:gcd_sim}
\end{figure}

\autoref{fig:gcd_sim} presents the simulation results. We initialize the target mode A in various Fock states $\ket{n}_a$ and observe the trajectory of the control mode B in phase space. The simulated momentum displacement follows the predicted linear dependence on the target photon number.

\subsection{Bias-Preserving CX Gate with System Hamiltonian}
\label{subsec:cxsims}
\subsubsection{CX Gate on Cat States} 
\label{subsubsec:cx_sims}
To validate the conditional logic of the gate, we simulate the time-evolution of the full RAR Hamiltonian under the shaped pulse envelopes derived in Section \ref{sec:parameter-choice}. We examine two representative logical inputs, $\ket{1_\beta1_\alpha}$ and $\ket{0_\beta1_\alpha}$, using control-target ordering: $\ket{i_\beta j_\alpha} =\ket{i_\beta}_b\otimes\ket{j_\alpha}_a$. Here, $\alpha$ and $\beta$ denote the target and control cat amplitudes, respectively.

\begin{figure}[h]
    \centering
    \begin{subfigure}[b]{0.5\linewidth}
        \includegraphics[width=0.9\linewidth]{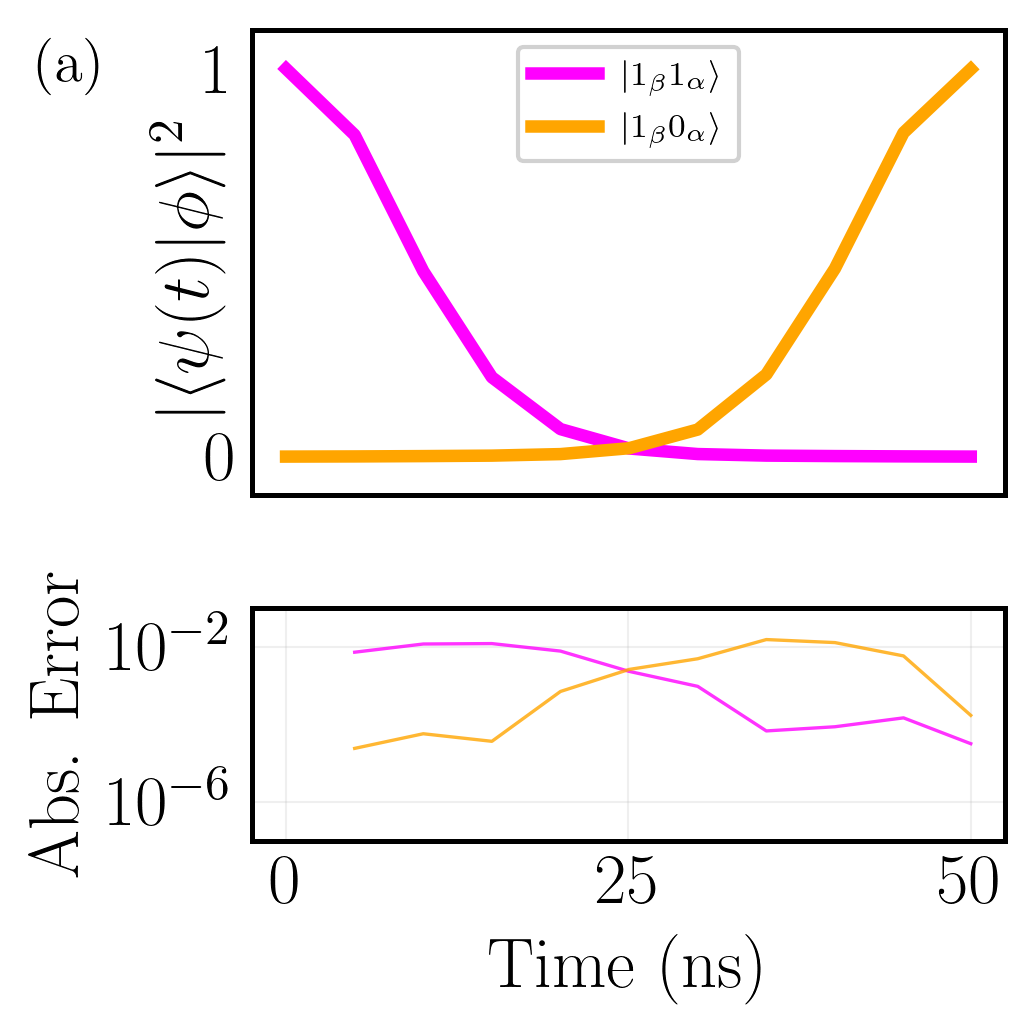}
    \end{subfigure}%
    \begin{subfigure}[b]{0.5\linewidth}
        \includegraphics[width=0.9\linewidth]{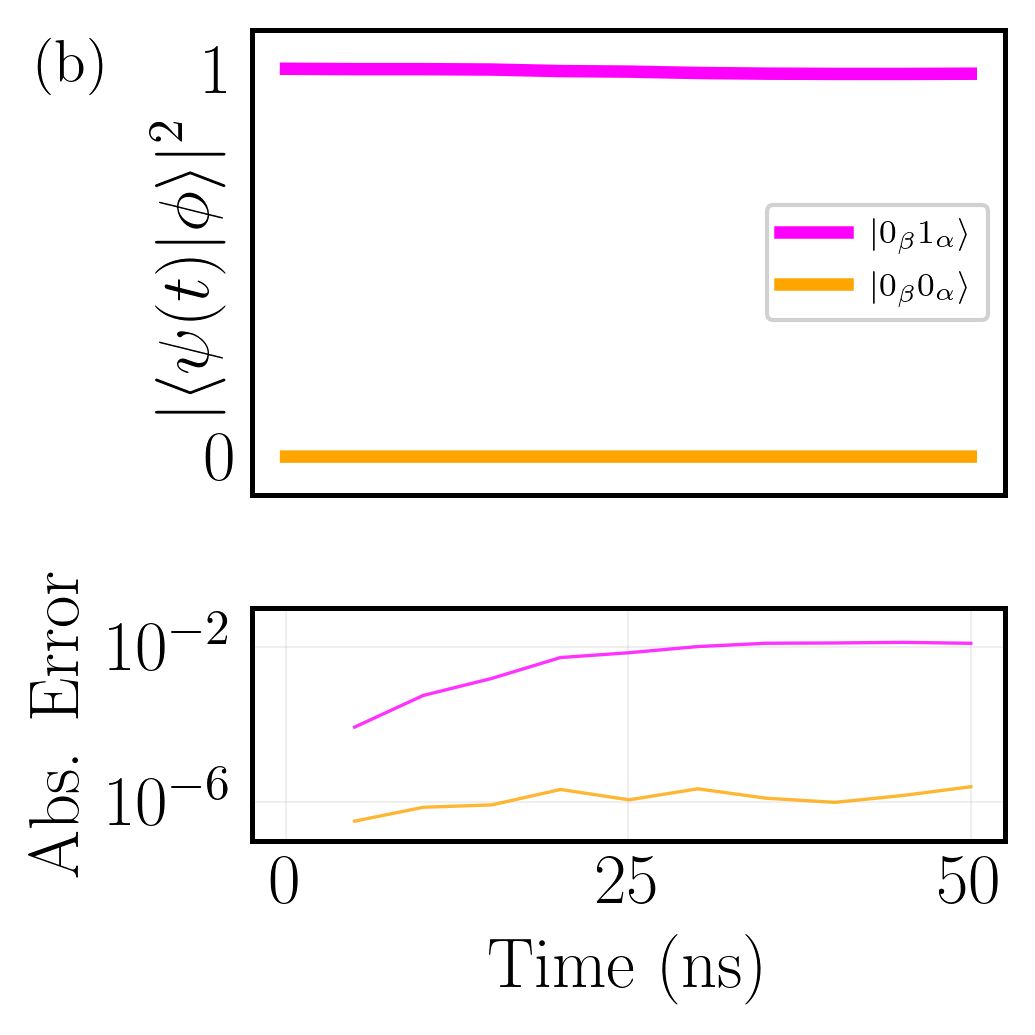}
    \end{subfigure}
    \caption{\justifying Time-domain simulation of the bias-preserving CX gate under the full driven RAR Hamiltonian. The two ATS pump tones are applied together with the local controls required by \autoref{eq:CXgateonCatqubits_splitted}, using the waveforms specified in Section~\ref{sec:parameter-choice} for a gate duration $T=50$~ns. (a) Evolution from $\ket{\psi(0)}=\ket{1_\beta1_\alpha}$, showing transfer of the joint-state overlap from $\ket{1_\beta1_\alpha}$ to $\ket{1_\beta0_\alpha}$. (b) Evolution from $\ket{\psi(0)}=\ket{0_\beta1_\alpha}$, showing that the overlap with the initial logical state remains close to unity.}
    \label{fig:full_sim}
\end{figure}

\autoref{fig:full_sim} illustrates the conditional action of the CX gate for these two inputs. When the control is in $\ket{1_\beta}$, the evolution approximately implements $\ket{1_\beta1_\alpha}\to\ket{1_\beta0_\alpha}$. When the control is in $\ket{0_\beta}$, the state remains close to $\ket{0_\beta1_\alpha}$.

\subsubsection{Fidelity and Bias Preservation Estimates} \label{subsubsec:fidelity_estimates}
Throughout the PTM and $\chi$-matrix analysis, we use control-target tensor-product ordering, $\hat{P}_\mu\otimes\hat{P}_\nu$, with the first operator acting on the control and the second on the target.

We quantify the gate performance by reconstructing the Pauli Transfer Matrix (PTM) and the Process Matrix ($\chi$) from the simulated dynamics (see Appendix \ref{app:ptm} for methodology). Specifically, we isolate the error channel $\mathcal{E}_{\text{err}}$ by computing the product $R_\text{err} =  R_{\text{ATS}} R_{\text{ideal}}^{-1}$, where $R_{\text{ATS}}$ is the simulated PTM and $R_{\text{ideal}}$ represents the ideal bias-preserving CX operation.

\begin{figure*}[t]
    \centering
    \begin{subfigure}[b]{0.32\textwidth}
        \centering
        \includegraphics[width=0.95\linewidth]{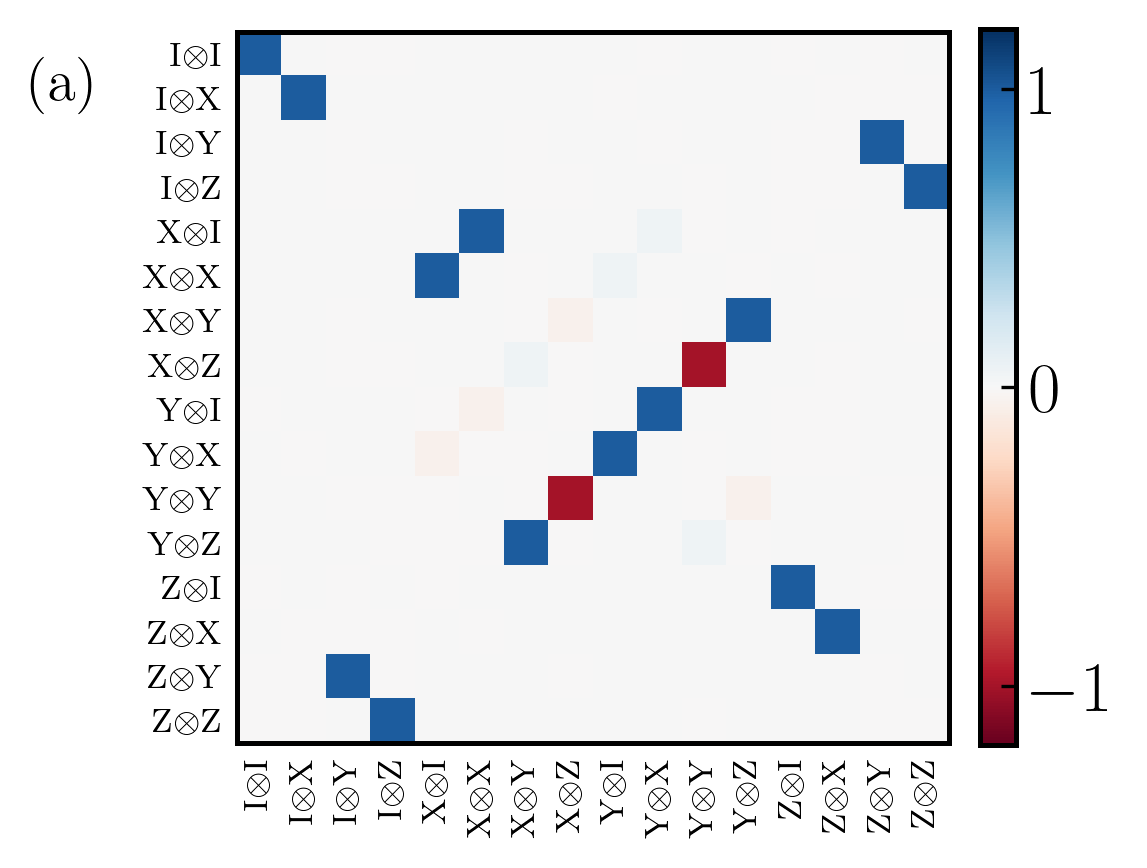}
    \end{subfigure}%
    \begin{subfigure}[b]{0.32\textwidth}
        \centering
        \includegraphics[width=\linewidth]{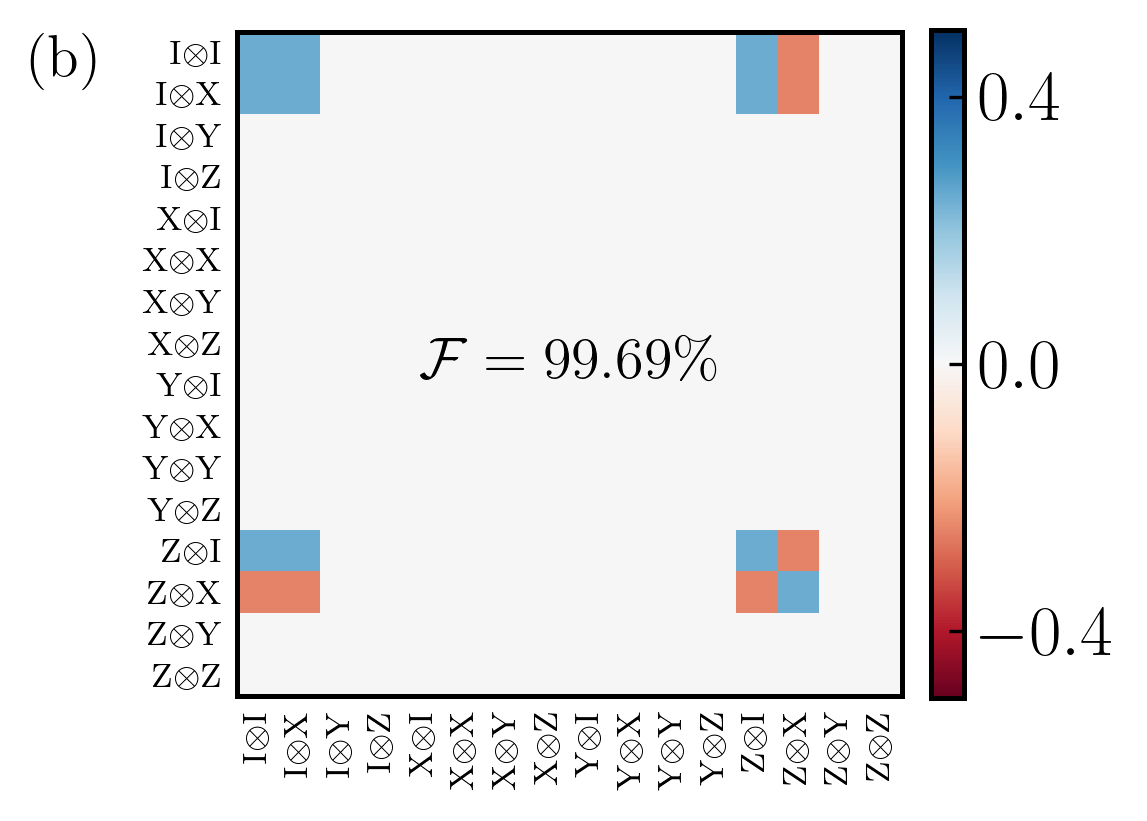}
    \end{subfigure}%
    \begin{subfigure}[b]{0.34\textwidth}
        \centering
        \includegraphics[width=\linewidth]{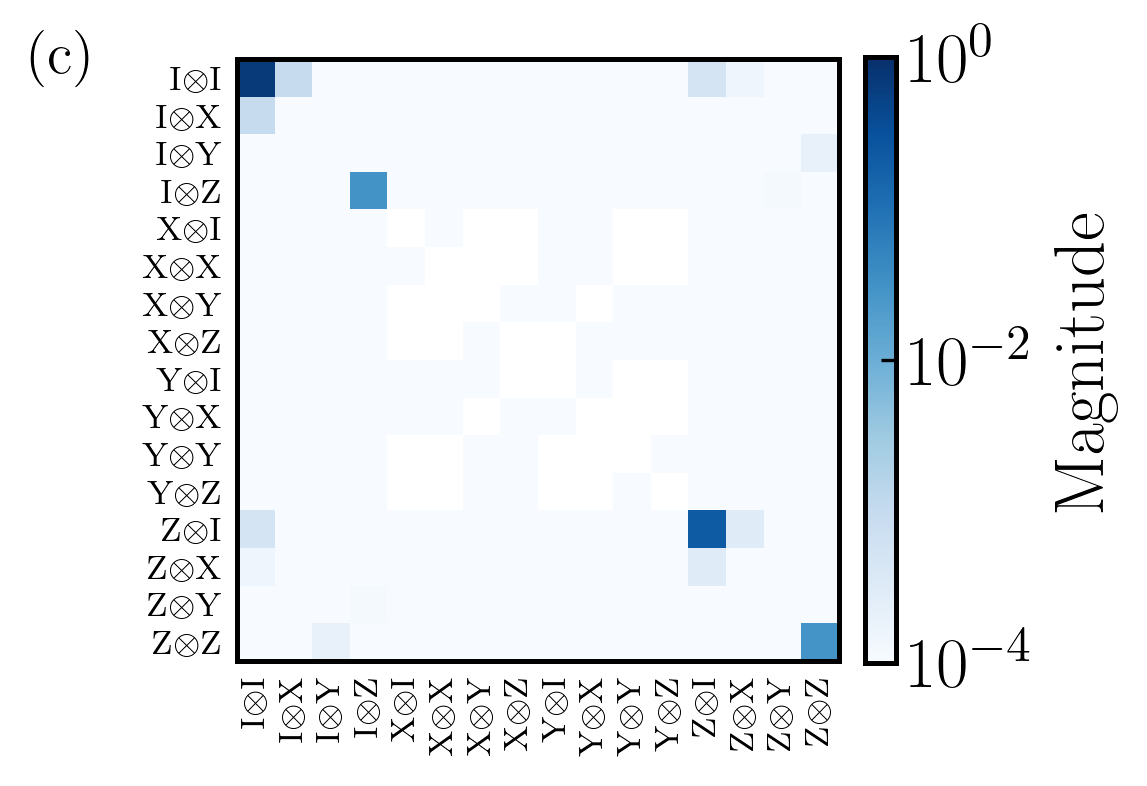}
    \end{subfigure}
    
    \vspace{0.5cm}

    \begin{subfigure}[b]{0.32\textwidth}
        \centering
        \includegraphics[width=0.95\linewidth]{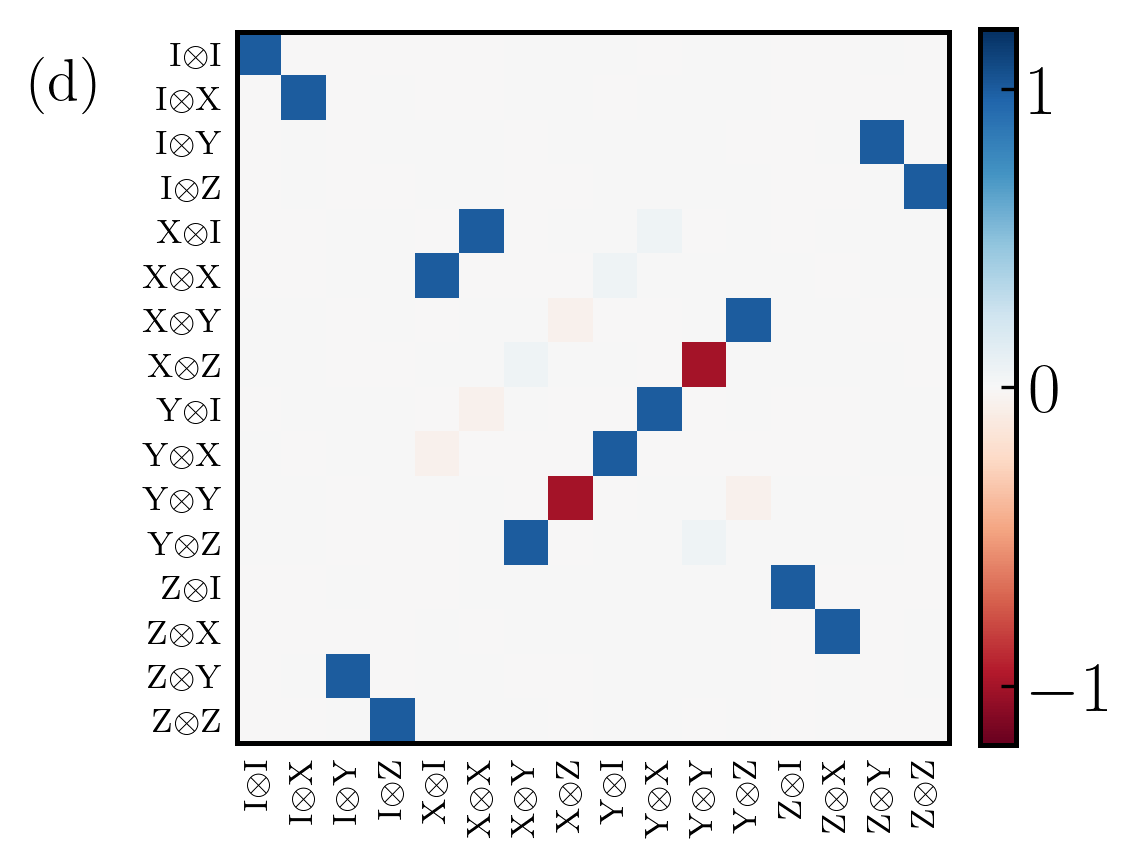}
    \end{subfigure}%
    \begin{subfigure}[b]{0.32\textwidth}
        \centering
        \includegraphics[width=\linewidth]{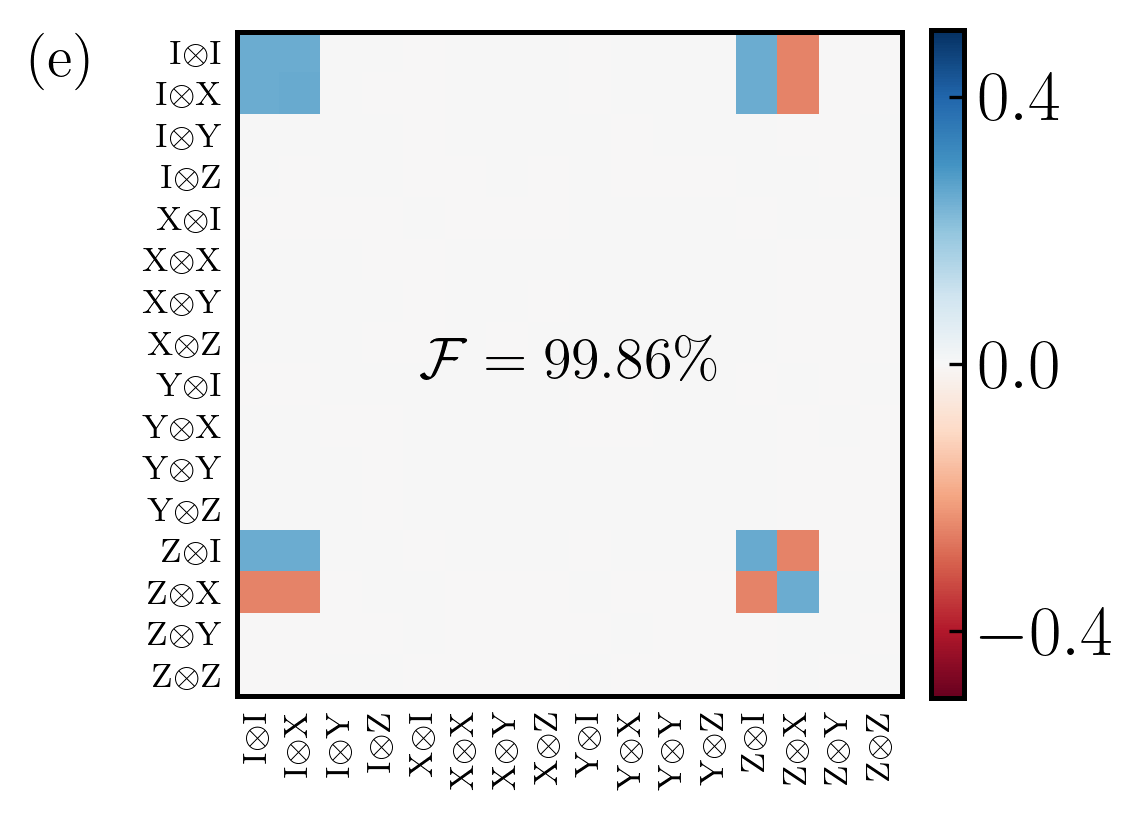}
    \end{subfigure}%
    \begin{subfigure}[b]{0.34\textwidth}
        \centering
        \includegraphics[width=\linewidth]{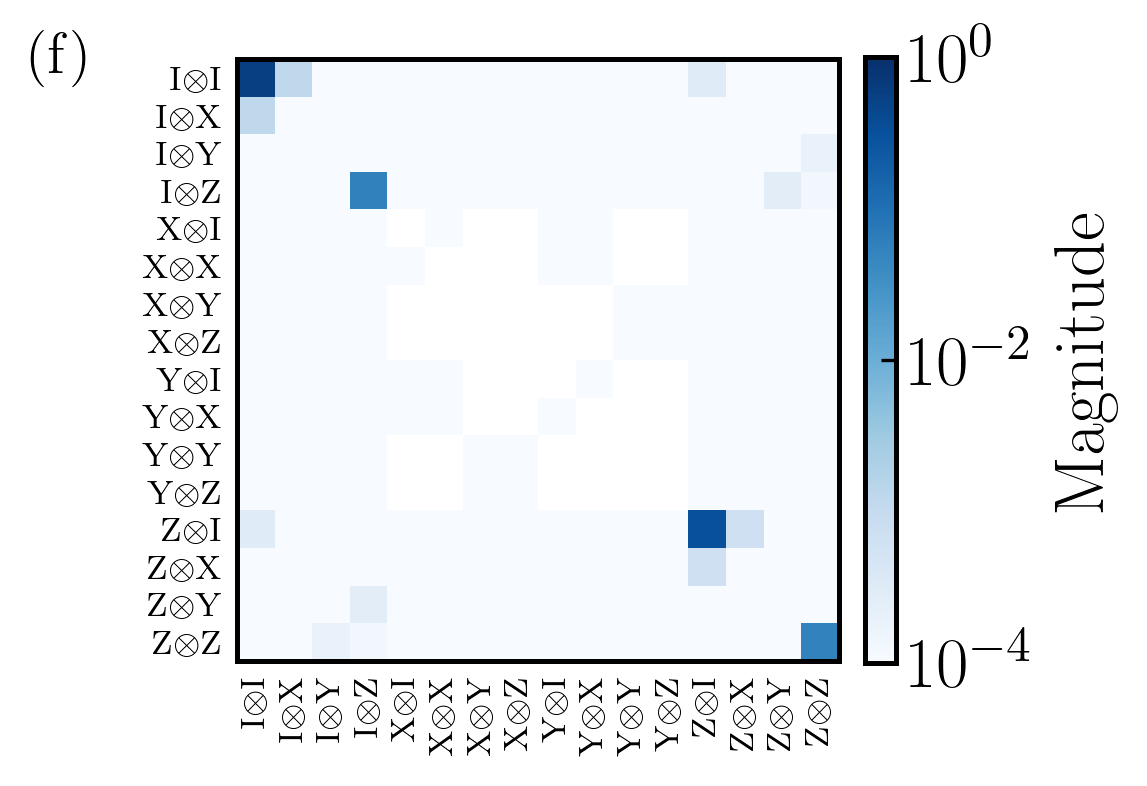}
    \end{subfigure}
    
    \caption{\justifying Tomographic verification of the ATS Bias-Preserving CX Gate. Top row (a-c) corresponds to $T=50$ ns, while the bottom row (d-f) corresponds to $T=100$ ns. (a, d) The simulated Pauli Transfer Matrices (PTM) in the absence of photon loss. (b, e) The corresponding Process Matrices ($\chi$) in the absence of photon loss. (c, f) The error process matrices $\chi_{\text{err}}$ calculated in the presence of single-photon loss ($\kappa_{\rm 1ph}/2\pi = 100$ kHz). The dominance of Z-type errors in (c) and (f) visually confirms that the gate preserves the noise bias. We achieve $\mathcal{F}_{\text{avg}} = 99.69\%$ for the 50 ns operation without loss.}
    \label{fig:ptm_chi}    
\end{figure*}

The computed matrices are displayed in \autoref{fig:ptm_chi}. To extract a scalar metric, we convert the PTM to the Process Matrix representation ($\chi$) via the standard superoperator basis transformation \cite{Greenbaum2015, Hashim2025}. The Average Gate Fidelity $\mathcal{F}_{\text{avg}}$ is then derived from the entanglement fidelity $\mathcal{F}_e$ (the first diagonal element $\chi_{00}$) using the relation \cite{Nielsen2002, Wood2018}
\begin{equation}
    \mathcal{F}_{\text{avg}} = \frac{D \mathcal{F}_e + {\rm tr} (\chi_{err})}{D + 1},
\end{equation}
where $D=4$ is the Hilbert space dimension for two qubits.

\begin{table*}[t]
\renewcommand{\arraystretch}{1.5}
\centering
\begin{tabular}{|c|c|c|c|c|c|}
\hline
\textbf{Metric} & \textbf{Symbol} & \begin{tabular}[c]{@{}c@{}}\textbf{Naive Non-BP}\\ \textbf{CX Gate}\end{tabular} & \begin{tabular}[c]{@{}c@{}}\textbf{Ideal BP}\\ \textbf{CX Gate}\end{tabular} & \begin{tabular}[c]{@{}c@{}}\textbf{ATS BP} \\ \textbf{CX Gate}\end{tabular} & \textbf{Idling}\\
\hline
Dephasing & $\lambda_\text{deph}$ & $0.261$ & $0.280$ & $0.279$ & $0.278$\\
\hline
Non-dephasing & $\lambda_\text{nondeph}$ & $0.015$ & $2.11\times 10^{-5}$ & $2.81\times 10^{-5}$ & $2.17\times 10^{-5}$\\
\hline
\textbf{Bias} & $\eta = \lambda_\text{deph}/\lambda_\text{nondeph}$ & \textbf{17.44} & \textbf{13191} & \textbf{9922} & \textbf{12854}\\
\hline
\end{tabular}
\caption{\justifying Comparison of bias-preservation metrics under single-photon loss ($\kappa_\text{1ph}/2\pi = 100$ kHz). Metrics are derived from the PTM of the error channel. The Naive Non-BP (Non-Bias-Preserving) CX uses a logical Hamiltonian in \autoref{eq:nativeCXPauli}, showing a substantial reduction of the noise bias ($\eta \approx 17$). The Ideal BP CX represents the theoretical limit using \autoref{eq:CXgateonCatqubits_splitted}. The proposed ATS-coupled gate achieves $\eta \approx 9922$, maintaining the bias required for repetition codes. Idling is simulated for the equivalent gate duration ($T=50\,$ns).}
\label{tab:ptm_bias_metrics}
\end{table*}

\begin{figure*}[t]
    \centering
    \begin{subfigure}[b]{0.32\textwidth}
        \includegraphics[width=1\linewidth]{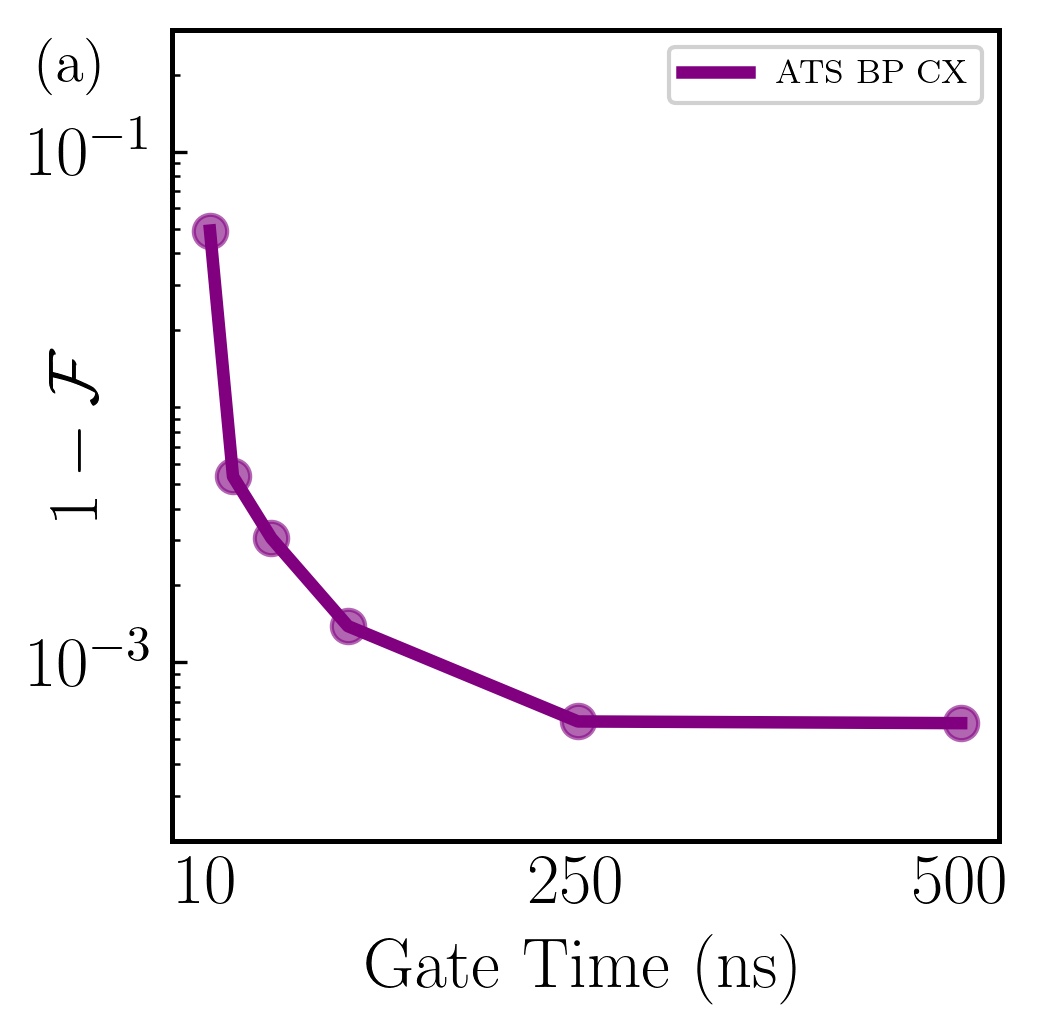}
    \end{subfigure}%
    \begin{subfigure}[b]{0.32\textwidth}
        \includegraphics[width=1\linewidth]{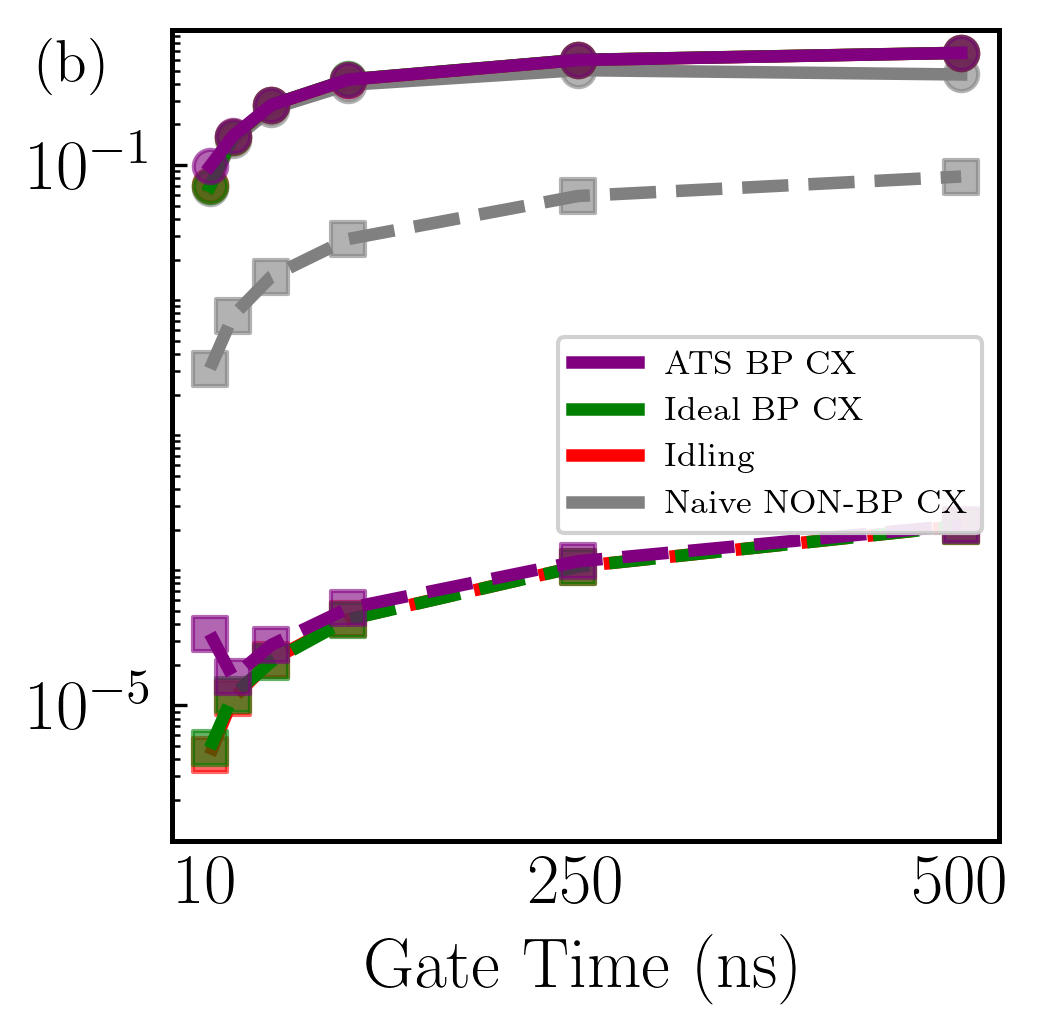}
    \end{subfigure}
        \begin{subfigure}[b]{0.32\textwidth}
        \includegraphics[width=1\linewidth]{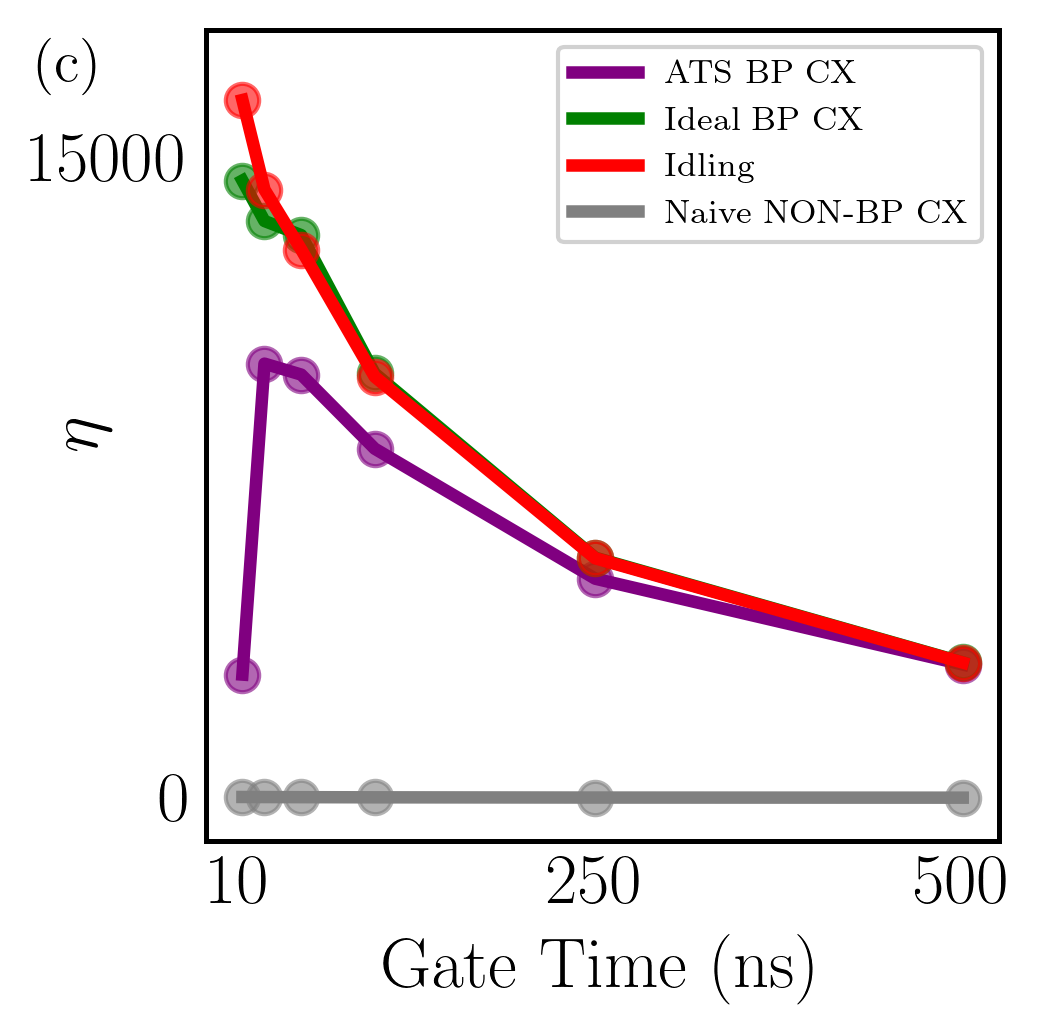}
    \end{subfigure}
    \caption{\justifying (a) shows lossless gate infidelity as a function of gate duration $T$ under no loss. Comparing the simulated PTM to the ideal target operation, we observe that increasing the gate time improves fidelity by suppressing non-adiabatic leakage, highlighting the trade-off between speed and leakage in the ATS-mediated coupling. (b) and (c) are simulated under single-photon loss, $\kappa_{\rm 1ph}/2\pi=100$ kHz. (b) shows error probabilities obtained from noise channel for 4 different cases. Solid lines correspond to dephasing errors, while dashed lines correspond to non-dephasing errors. (c) shows bias, $\eta = \lambda_{\rm dephasing} / \lambda_{\rm nondephasing}$, calculated by deriving PTMs for each gate time. At longer gate durations, the dephasing error weight approaches saturation while the non-dephasing error weight continues to grow, reducing the bias. Nevertheless, $\eta$ remains much larger than unity over the simulated range.}
    \label{fig:gate_time_vs_bias_err_infid}
\end{figure*}

For our target gate time of $T=50$ ns, we achieve an average gate fidelity of $\mathcal{F}_{\text{avg}} = 99.69\%$ in the lossless simulation, and it rises to $99.86\%$ at $T=100$ ns as shown in \autoref{fig:ptm_chi}(b,e). The infidelity versus gate time is plotted in \autoref{fig:gate_time_vs_bias_err_infid}(a), and it is observed that lossless fidelity improves with increase in gate time as expected in Section~\ref{sec:gate-time-kerr}. The dominant contribution to the lossless gate error is leakage out of the code space.

Decomposing the error process matrix shows the process infidelity splits into three distinct contributions: leakage out of the code space ($1-\mathrm{tr}(\chi_\text{err})\approx0.19\%$); a coherent control-qubit phase error ($\approx0.07\%$), identified from a near-rank-one $\chi_{II,ZI}$ block~\cite{Korotkov2013} corresponding to a residual rotation of $\theta\approx3.1^\circ$ and in principle correctable via calibration or a virtual-$Z$ correction; and a remaining contribution ($\approx0.07\%$) concentrated in the $IZ$ and $ZZ$ channels. Unlike the coherent phase term, this remainder is not a single rank-one rotation and may stem from residual excitation outside the code space, which is hard to correct by calibration.

Most importantly, the error process matrices (\autoref{fig:ptm_chi}(c,f)) confirm the bias-preserving nature of the gate under single-photon loss, $\kappa_{\rm 1ph}/2\pi$, of 100 kHz. The error weight is concentrated almost exclusively on the phase-flip (Z-type) terms, while bit-flip (X-type) errors are strongly suppressed. Under loss, the gate fidelity ($77.53\%$) is close to the idling fidelity over the same duration ($77.72\%$), so it is limited by photon loss, not by the gate. We chose this high rate ($\kappa_{\rm 1ph}=K/30$) to stress-test bias preservation. At $\kappa_{\rm 1ph}=K/4000$ ($\kappa_{\rm 1ph}/2\pi=0.75$~kHz), as in Ref.~\cite{Puri2020}, the gate reaches $99.48\%$ fidelity with $\eta\approx1150$. The bias is lower there because loss-induced errors, which are almost all phase flips, shrink with $\kappa_{\rm 1ph}$, while the small loss-independent gate error
does not.

\subsubsection{Bias Preservation under Single-Photon Loss} \label{subsubsec:cx_sims_w_loss}

To assess the fault-tolerant compatibility of our design, we analyze the performance of the ATS bias-preserving CX gate in the presence of single-photon loss, the dominant decoherence mechanism in superconducting cavities. We characterize the total error process in the presence of photon loss by factoring out the ideal CX operation, $R_{\rm err}^{\rm loss} =R_{\rm CX}^{\rm loss}(R_{\rm CX}^{\rm ideal})^{-1}$, simulating a dissipation rate of $\kappa_\text{1ph}/2\pi = 100$ kHz.

As shown in Section~\ref{subsec:biaspreservingcxgate}, control loss gives $Z_1$ and target loss gives $Z_2$ or $Z_1Z_2$. Neglecting off-diagonal process-matrix elements, we approximate the resulting error process by a Pauli channel
\begin{align*}
    \mathcal{E}(\rho) &\approx p_{I} \rho^\prime + p_{Z_1} Z_1 \rho^\prime Z_1  \numberthis\\
    &+ p_{Z_2} Z_2 \rho^\prime Z_2  + p_{Z_1Z_2} Z_1Z_2 \rho^\prime Z_1Z_2,
\end{align*}
where $\rho^\prime$ denotes the state following the ideal unitary evolution. Because $\bar n_b\kappa_{1\rm ph}T$ is not small at this stress-test rate, we replace the first-order probabilities of Ref.~\cite{Guillaud2019} by the probability of an odd number of loss events, $q_{a,b}=\frac12(1-e^{-2\bar n_{a,b}\kappa_{\rm 1ph}T})$, and combine the two modes as independent processes, giving
\begin{align}
    p_{Z_1} &= q_b (1-q_a) = \frac{1}{4}(1-e^{-2\bar{n}_b \kappa_{\rm 1ph}T})(1+e^{-2\bar{n}_a \kappa_{\rm 1ph}T}), \\
    p_{Z_2} &= p_{Z_1Z_2} = \frac{q_a}{2}= \frac{1}{4}(1-e^{-2\bar{n}_a \kappa_{\rm 1ph}T}).
\end{align}
For our operating parameters ($\bar{n}_a=2$, $\bar{n}_b=10$) and gate time ($T=50$ ns), the control qubit operates in a high-loss regime, yielding a phase-flip probability $p_{Z_1} \approx 0.2195$. However, because the gate is bias-preserving, this high error rate remains $Z$-type and does not convert into bit-flip errors. The target qubit, operating with lower photon number, contributes a smaller probability of $p_{Z_2} \approx 0.0295$. These theoretical predictions align with the simulated error process matrices, where $p_{Z_1} = 0.2200$ and $p_{Z_2} = 0.0299$.

Table \ref{tab:ptm_bias_metrics} summarizes the performance of the proposed architecture against standard benchmarks. The naive CX implementation disrupts the noise structure, collapsing the protection to $\eta \approx 17$. In contrast, the ATS design yields a bias of $\eta_{\rm ATS} \approx 9922$. The reduction from the ideal BP CX gate, $\eta_{\rm ideal}\approx13191$, to the proposed design's $\eta_{\rm ATS}\approx 9922$ can be traced to specific error channels in $\chi_\text{err}$. Across all four simulated cases, every Pauli channel involving an $X$ or $Y$ operator on the control qubit remains at the numerical noise floor ($\lesssim 10^{-10}$), confirming that control-qubit bit-flip weights remain below the numerical resolution regardless of gate implementation, consistent with the large control cat size ($\bar n_b=10$). All residual non-dephasing weight is instead confined to channels acting on the target qubit ($IX$, $IY$, $ZX$, $ZY$), reflecting the weaker bit-flip suppression of the smaller target cat ($\bar n_a=2$). Comparing the ATS BP CX to the ideal BP CX gate, the excess non-dephasing weight has two parts of similar size. About half is in the $IX$ and $ZX$ channels, from a small coherent X rotation of the target ($\approx 0.2^\circ$ per gate) that is present even without loss. The other half is a loss-induced increase of the $IY$ weight (by $\approx 30\%$), while $ZY$ changes only marginally. The two contributions have different origins. The $IX$/$ZX$ part is present already in the lossless simulation and saturates the Cauchy--Schwarz bound $|\chi_{II,IX}| \le \sqrt{\chi_{II,II}\,\chi_{IX,IX}}$, identifying it as a coherent rotation of the target qubit that we attribute to residual, non-adiabatic corrections to the ATS drive not captured by the idealized bias-preserving Hamiltonian. The $IY$ part appears only under loss, consistent with single-photon loss acting on transiently excited target population during the gate. Both effects act on the target qubit, whereas the control-qubit bit-flip sector remains at the numerical noise floor. Despite this reduction, the ATS implementation increases the noise bias by approximately $570$ times relative to the naive non-BP CX benchmark. The limiting factors from higher-order device nonlinearities and fabrication imperfections are analyzed in Appendices~\ref{app:nonlinear_terms} and~\ref{app:ej2_dej}.

Without loss, the fidelity rises from $95.09\%$ at $T=10$~ns, where non-adiabatic leakage dominates, to $99.69\%$ at 50~ns and $99.94\%$ at 500~ns, \autoref{fig:gate_time_vs_bias_err_infid}(a). Under loss, $\eta$ decreases with gate time, \autoref{fig:gate_time_vs_bias_err_infid}(c), for example from 13191 at 50~ns to 3178 at 500~ns for the effective-model gate, because the control phase-flip probability saturates while the much smaller non-dephasing weight keeps growing roughly linearly. Over the whole simulated range $\eta\gg1$.

To further check bias-preserving capabilities of the proposed CX gate, we extract the characteristic bit-flip ($T_{\text{bit}}$) and phase-flip ($T_{\text{phase}}$) lifetimes of the logical qubits. We simulate the time-evolution of the relevant logical observables under the single-photon loss ($\kappa_{\rm 1ph}/2\pi = 100$ kHz), more information on the derivation of these times is given in Appendix \ref{app:fliptimes}.

\autoref{fig:bit_phase_flip_times} illustrates the decay of the phase coherence, monitored via the parity operator $\langle \Pi \rangle = \langle e^{i\pi \hat{a}^\dagger \hat{a}}\rangle$ \cite{Royer1977}, and the bit polarization, monitored via $\langle \sigma_z \rangle$. For phase coherence, we observe a rapid decay consistent across all gate implementations. This behavior is theoretically expected: single-photon loss events scramble the photon number parity, leading to a phase-flip rate of $\Gamma_{\text{phase}} = 2\bar{n}\kappa_{\rm 1ph}$. For our target qubit ($\bar{n}=2$), this corresponds to a theoretical limit of $T_{\text{phase}} = (2\bar{n}\kappa_{\rm 1ph})^{-1} = 0.398~\mu\text{s}$, which matches our simulation results, $T_{\text{phase}} \approx 0.404~\mu\text{s}$.

In contrast, the bit-flip lifetimes ($\langle \sigma_z \rangle$) exhibit orders-of-magnitude separation between gate implementations, reflecting the exponential suppression of bit flips. Based on Appendix \ref{app:fliptimes}, expected bit-flip rate is $\Gamma_{\rm bit} = \kappa_{\rm 1ph}|\alpha|^2e^{-4|\alpha|^2}$, giving a decay time $T_{\rm bit} = 1/(2\Gamma_{\rm bit}) = 1.186$ ms for the idling case. The extracted lifetimes are summarized in \autoref{tab:bit_flip_times}. The naive non-bias-preserving CX gate fails to protect the manifold, yielding a bit-flip time of only $T_{\text{bit}} \approx 0.791~\mu\text{s}$, which is on the same order as the phase-flip time. Conversely, the proposed ATS bias-preserving CX gate achieves $T_{\text{bit}} \approx 312~\mu\text{s}$. While this is lower than the ideal theoretical limit due to residual non-adiabatic transitions and circuit nonlinearities, it represents a $\sim 394\times$ improvement over the naive case. We compare the fitted lifetimes with estimates obtained by propagating the relevant input states through the PTM of a complete two-gate cycle. Further details on the fitting methodology and PTM-based extraction are provided in Appendix \ref{app:fliptimes}.

\begin{table}[h]
\renewcommand{\arraystretch}{1.2}
\centering
\begin{tabular}{|c|c|c|c|c|}
\hline
\begin{tabular}[c]{@{}c@{}}Flip Times \\ ($\mu$s)\end{tabular} & \begin{tabular}[c]{@{}c@{}}Naive Non-BP \\ CX Gate\end{tabular} & \begin{tabular}[c]{@{}c@{}}Ideal BP \\ CX Gate\end{tabular} & \begin{tabular}[c]{@{}c@{}}ATS BP \\ CX Gate\end{tabular} & Idling\\
    \hline
       $T_\text{bit}$  & 0.791 &  1056.67 & 311.95 & 1185.97 \\
    \hline
        $T_\text{bit, PTM}$ & 0.809 &  1216.21 & 312.41 & 1165.40 \\
    \hline
       $T_\text{phase}$  & 0.393 & 0.397 & 0.404 & 0.397 \\
    \hline
        $T_\text{phase, PTM}$ & 0.403 & 0.398 & 0.403 & 0.398 \\
    \hline
        $T_\text{bit}/T_\text{phase}$ & 2.01 & 2661.6 & 772.2 & 2987.3 \\
    \hline
    \end{tabular}
    \caption{\justifying Bit-flip times $T_\text{bit}$ and phase-flip times $T_\text{phase}$ are extracted from repeated $\text{CX}^2$ simulations under single-photon loss at rate $\kappa_\text{1ph}/2\pi = 100$ kHz. Numerical bit-flip times $T_\text{bit, PTM}$ and phase flip times $T_\text{phase, PTM}$ are obtained using the Pauli Transfer Matrix (PTM). The results show agreement with the time-domain simulations, and relatively high discrepancy on bit-flip for ideal bias-preserving CX gate and idling is due to ultra-low rates being sensitive to numerical errors. The ATS bias-preserving CX gate achieves substantially enhanced $T_\text{bit}$ compared to the non-bias-preserving gate, showing bias preservation.}
    \label{tab:bit_flip_times}
\end{table}

The lifetime ratio $T_{\rm bit}/T_{\rm phase}$ and the single-gate $\eta$ characterize different aspects of the noise. They differ in their time dependence and treatment of correlated errors, and need not coincide numerically.

\begin{figure*}[t]
  \centering
  \hspace*{1cm}\begin{subfigure}[t]{0.4\linewidth}
    \centering
    \raisebox{10mm}{\includegraphics[width=0.8\linewidth]{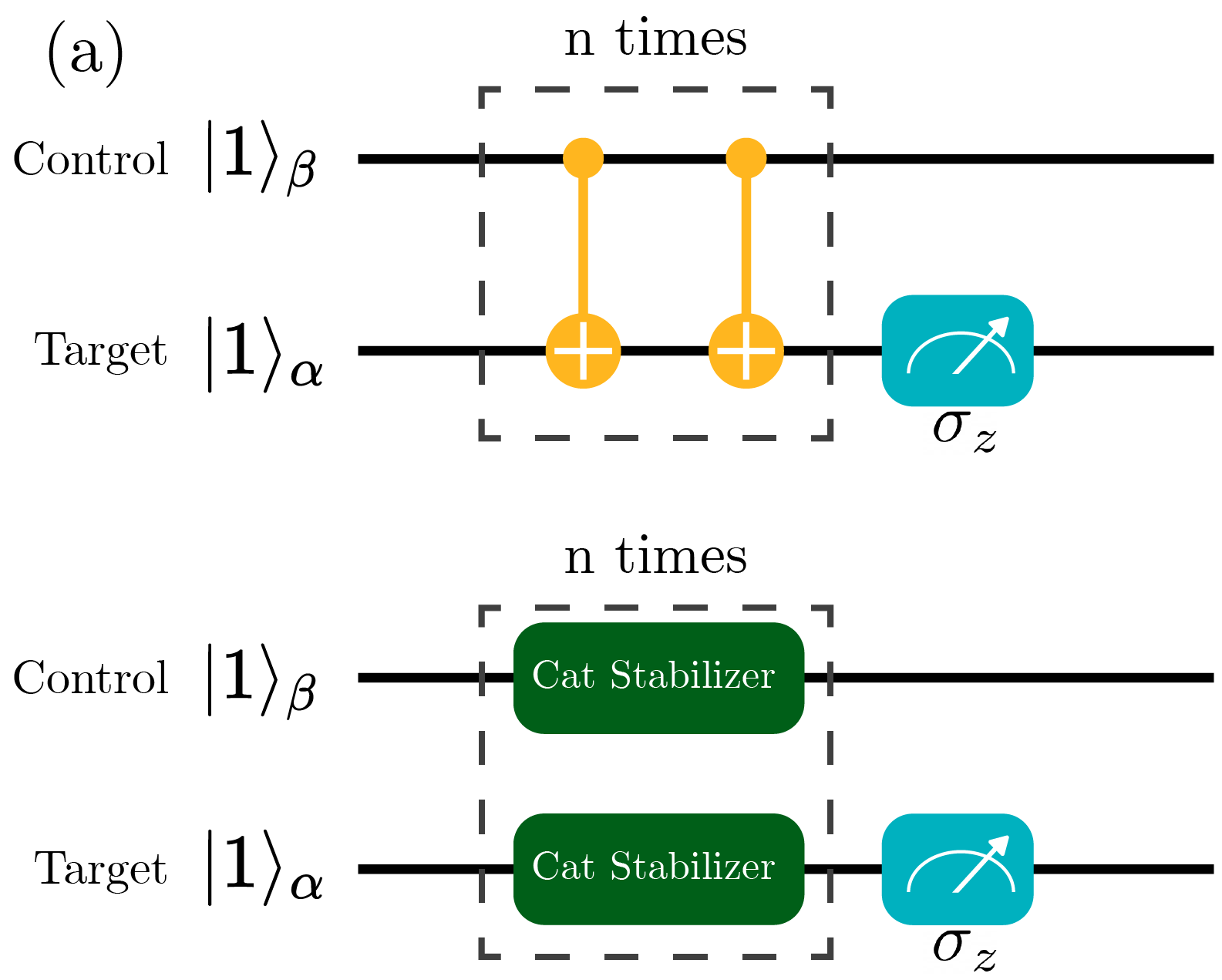}}
    \label{fig:z_diag}
  \end{subfigure}%
  \begin{subfigure}[t]{0.6\linewidth}
    \centering
    \includegraphics[width=0.7\linewidth]{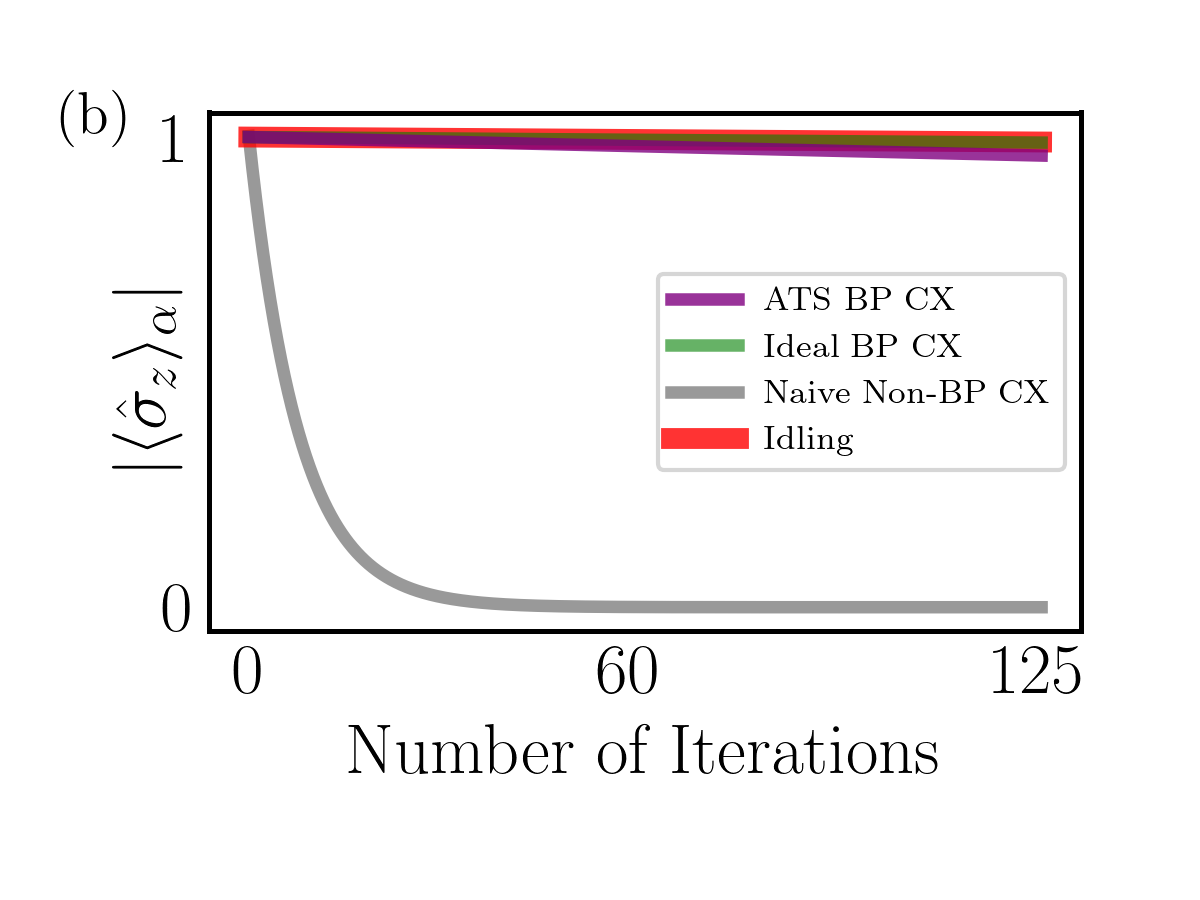}
    \label{fig:cx_z}
  \end{subfigure}
  \hspace*{1cm}\begin{subfigure}[t]{0.4\linewidth}
    \centering
    \raisebox{10mm}{\includegraphics[width=0.8\linewidth]{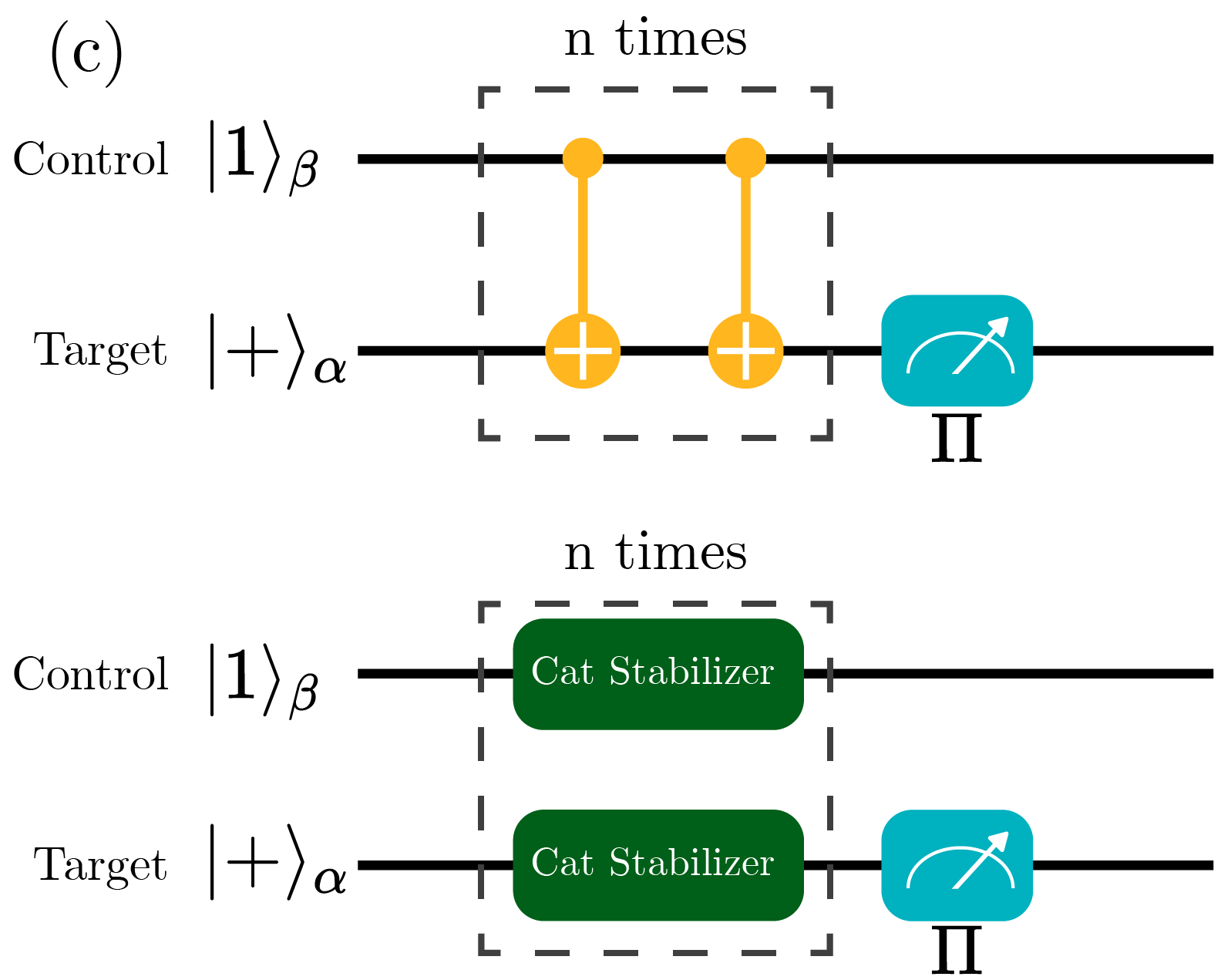}}
    \label{fig:x_diag}
  \end{subfigure}%
  \begin{subfigure}[t]{0.6\linewidth}
    \centering
    \includegraphics[width=0.7\linewidth]{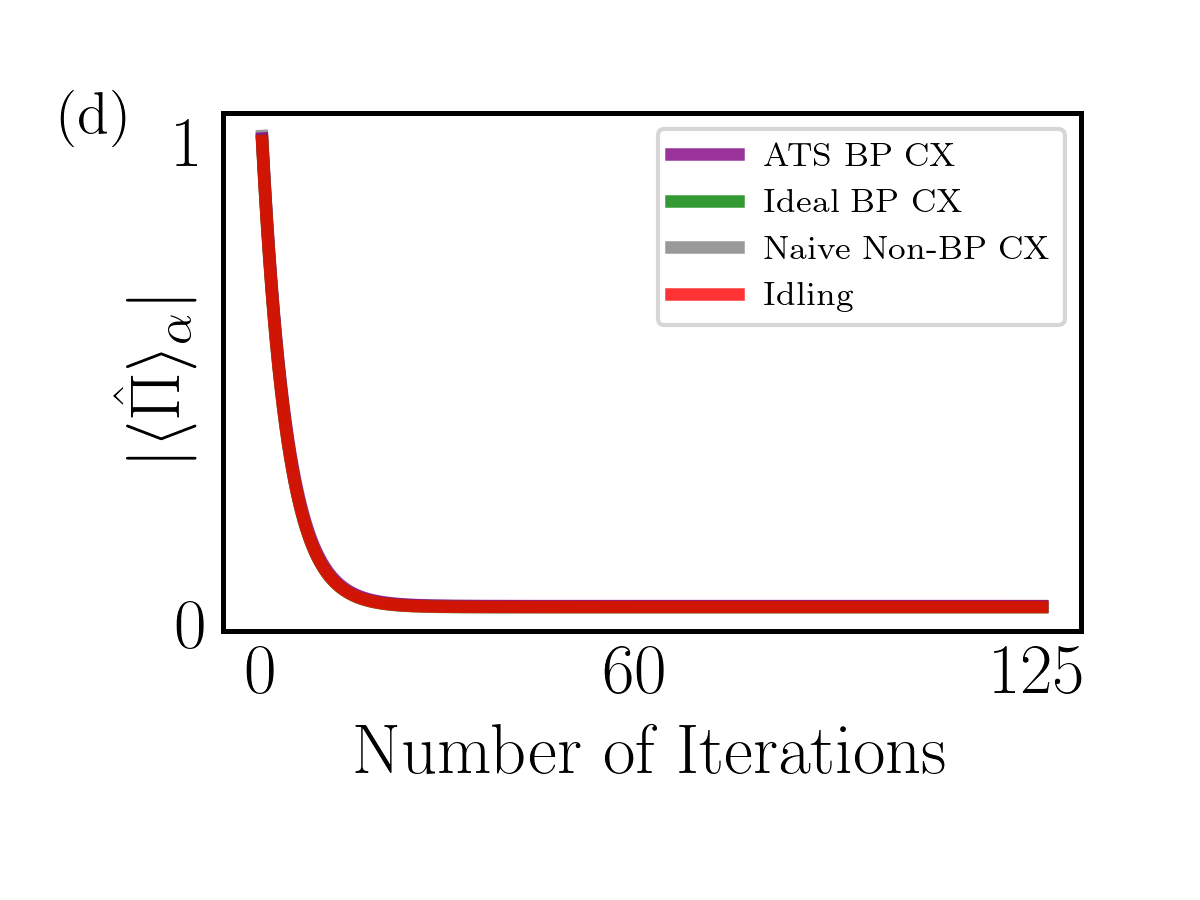}
    \label{fig:cx_x}
  \end{subfigure}
  \caption{\justifying To estimate bit-flip time, $\ket{1_\beta1_\alpha}$ is evolved according to the diagram (a) for CX gates and idling. The CX gate is applied twice for identity operation spanning 100 ns, and idling is realized by 100 ns cat stabilizers. (b) shows the evolution of $\langle \sigma_z \rangle_\alpha$. The idling and ideal bias-preserving CX curves are similar; the ATS gate has a much longer bit-flip time than the naive CX gate. To calculate phase-flip time, same procedure is applied with $\ket{1_\beta+_\alpha}$ as in diagram (c). (d) shows that all cases have similar phase-flip times as expected.}
  \label{fig:bit_phase_flip_times}
\end{figure*}

\section{Conclusions}  \label{sec:conc}
We proposed an ATS coupler that activates both two-mode interactions of the bias-preserving CX between two Kerr-cat qubits: the three-wave-mixing term $g_3$ and the conditional displacement $g_{\rm CD}$, using the bias point $\varphi_\Sigma=\varphi_\Delta=\pi/2$ and two pump tones. A larger control cat ($\beta^2>\alpha^2$) suppresses non-adiabatic leakage and allows gate times $T\approx1/K$ without derivative-based pulse corrections. In simulations of the coupler circuit with effective Kerr-cat terms, the gate reaches $99.69\%$ fidelity in 50~ns without loss and keeps a noise bias $\eta\approx9922$ at $\kappa_{\rm 1 ph}/2\pi=100$~kHz, about 570 times that of a naive CX. Beyond cat codes, the three-wave mixing of the ATS may also be useful for other stabilized bosonic encodings, such as Gottesman–Kitaev–Preskill (GKP) codes.

A primary challenge is the strong parametric drive on the ATS. It produces residual linear drives, which we compensate (Appendix~\ref{app:nonlinear_terms}),  and it can excite the coupler (Appendix~\ref{app:design-methodology}). Future work could reduce coupler excitation with derivative-based pulse corrections such as DRAG \cite{Motzoi2009}, or with numerical pulse optimization such as GRAPE \cite{Khaneja2005} or CRAB \cite{Caneva2011}. Another option is engineered coupler dissipation that returns excited coupler states to the ground state without disturbing the logical subspace, but its benefit must be weighed against the logical errors the dissipative channel itself introduces.

Extending the scheme beyond a single coupler raises further questions that we have not addressed here. In a register of many cat qubits, each ATS is activated by its own pump tones at $2\omega_a-\omega_b$ and $\omega_b$, so the frequency plan must keep these tones and their low-order mixing products clear of the transitions of neighboring couplers and resonators. Pump selectivity and flux-line crosstalk then become design constraints. A related question is whether a single ATS can be shared among more than two cat qubits, which would reduce the hardware overhead but requires that the unwanted pairwise interactions remain detuned. Finally, the photon-number asymmetry that enables fast gates, $|\beta|^2 > |\alpha|^2$, makes the gate directional: reversing the roles of control and target either slows the gate or requires the basis changes, whose cost must be weighed at the architecture level.

Taken together, these results show that a single parametric coupler can supply required two-mode ingredients of the bias-preserving CX gate between Kerr-cat qubits, at gate times comparable to the inverse Kerr nonlinearity and with a noise bias that stays orders of magnitude above that of a naive CX across the loss rates we studied. Because the ATS is already a standard element in superconducting bosonic hardware, the scheme requires no new nonlinear components, only a choice of bias point and pump tones. The natural next steps are an experimental demonstration of the $g_3$ and $g_{\rm CD}$ interactions on a two-resonator device and a circuit-level evaluation of the gate inside a repetition-code syndrome cycle, which together would establish the ATS coupler as a building block for concatenated cat-qubit architectures.
\section*{Acknowledgements} \label{sec:acknowledgements}
S.R.J. and K.Y. conceptualized the project and formulated the related theory. K.Y. developed and ran all presented numerical simulations, building on a modeling framework created by S.R.J. R.P. performed feasibility analyses of the chosen circuit parameters through the design and simulation of a physical chip layout. K.Y. wrote the manuscript with contributions from A.L. and M.M.K., and feedback from S.R.J. S.R.J. and M.W.A. supervised the project. All authors contributed to discussions of the results and the manuscript.

The authors would like to thank QWorld for providing the initial platform and support for this work, which originated as a project during the QIntern’24 program and was subsequently continued independently. We also sincerely thank Wiktor Gondek, Adair Campos-Uscanga, and Max Hays for valuable discussions throughout the development of this work. 

\bibliographystyle{apsrev4-2}
\bibliography{ref}

\appendix*
\begin{appendices}
\section{Derivation of the Non-Adiabatic Error Estimate}
\label{app:excitation-model}
\numberwithin{equation}{section}

Here we derive the excitation estimate entering \autoref{eq:infid-combined}. We take the gate phase to increase linearly in time,
\begin{equation}
    \phi(t)=\pi t/T,\qquad 0\leq t\leq T,
\end{equation}
so that the conditional-displacement coefficient of \autoref{eq:g_cd_prop} is constant during the gate, $g_{\rm CD}=\pi/(4\beta T)$. Approximating the target photon-number operator by its mean, $\hat n_t\simeq\alpha^2$, this coefficient exerts a constant force on the control fluctuations with strength
\begin{equation}
    J=g_{\rm CD}\alpha^2
    =\frac{\pi\alpha^2}{4\beta T}.
\end{equation}

Near either stabilization point $\pm\beta$, we approximate the control fluctuations by a harmonic mode with gap magnitude $\Delta_{\mathrm{gap},c}\simeq4K\beta^2$ \cite{Puri2019}. An initially unexcited harmonic mode subjected to this square force pulse evolves into a coherent state. Its probability of excitation relative to the initial local vacuum is
\begin{align}
    P_{\mathrm{exc}}^{(\mathrm{harm})}(T)
    &=1-e^{-x},\\
    x&=4\left(\frac{J}{\Delta_{\mathrm{gap},c}}\right)^2
    \sin^2\!\left(\frac{\Delta_{\mathrm{gap},c}T}{2}\right).
    \label{eq:Pleak}
\end{align}
Using the dimensionless parameter $\xi$ of \autoref{eq:xi} gives $J/\Delta_{\mathrm{gap},c}\simeq(\pi/16)\xi$ and
\begin{equation}
    x\simeq\frac{\pi^2}{64}\xi^2
    \sin^2(2K\beta^2T).
    \label{eq:harmonic-excitation}
\end{equation}
The excitation therefore depends on both $\xi$ and the accumulated harmonic phase $2K\beta^2T$. Bounding the oscillatory factor by unity gives
\begin{equation}
    P_{\mathrm{exc}}^{(\mathrm{harm})}
    \leq 1-\exp\!\left[-\frac{\pi^2}{64}\xi^2\right]
    \leq \frac{\pi^2}{64}\xi^2.
    \label{eq:excitation-envelope}
\end{equation}
Thus, $\xi$ controls an upper envelope of the harmonic excitation, which scales quadratically with $\xi$ for small $\xi$.

To connect this excitation estimate to gate fidelity, we assume that non-adiabatic leakage provides the dominant additional error. When the component remaining in the logical subspace closely follows the ideal CX evolution, the gate infidelity is approximately the leakage probability averaged over logical input states. Approximating this average leakage by the harmonic excitation probability gives \autoref{eq:infid-combined} in the small-error regime.

\section{Resonator-ATS-Resonator: Derivation and Methodology}
\label{app:rar-methodology}
\subsection{Bare ATS Hamiltonian}
\label{app:ats-hamiltonian}

We collect here, in the notation used throughout this manuscript, the elements of the ATS potential needed to define the two bias points employed below. The device itself was introduced in Ref.~\cite{Lescanne2020}; the flux parametrization and Fock-basis treatment we adopt follow Ref.~\cite{jha2024smthesis}.

The ATS consists of two Josephson junctions, of energies $E_{J,L}$ and $E_{J,R}$, interrupting a SQUID loop that is split at its midpoint by a linear superinductance. The two resulting sub-loops carry independent external flux biases $\varphi_{{\rm ext},L}$ and $\varphi_{{\rm ext},R}$, which we trade for common-mode and differential-mode combinations
\begin{align*}
    \varphi_\Sigma &= \tfrac{1}{2}(\varphi_{{\rm ext},L}+\varphi_{{\rm ext},R})\\
    \varphi_\Delta &= \tfrac{1}{2}(\varphi_{{\rm ext},L}-\varphi_{{\rm ext},R}), \numberthis
    \label{eq:app-flux-def}
\end{align*}
each independently set by external bias lines. Writing $\hat\varphi$ for the phase across the shared superinductance and parametrizing the junction asymmetry as $E_{J,L}=E_J+\Delta E_J$, $E_{J,R}=E_J-\Delta E_J$, the Josephson energy $-E_{J,L}\cos(\hat\varphi+\varphi_{{\rm ext},L})-E_{J,R}\cos(\hat\varphi-\varphi_{{\rm ext},R})$ can be rewritten purely in terms of $\varphi_\Sigma,\varphi_\Delta$: substituting $\varphi_{{\rm ext},L}=\varphi_\Sigma+\varphi_\Delta$, $\varphi_{{\rm ext},R}=\varphi_\Sigma-\varphi_\Delta$ and expanding both cosines with the angle-sum identity, the terms proportional to $E_J$ collect into $\cos(\hat\varphi+\varphi_\Delta)$ and those proportional to $\Delta E_J$ into $\sin(\hat\varphi+\varphi_\Delta)$, giving $-2E_J\cos(\varphi_\Sigma)\cos(\hat\varphi+\varphi_\Delta)+2\Delta E_J\sin(\varphi_\Sigma)\sin(\hat\varphi+\varphi_\Delta)$. We include equal second-harmonic contributions from the two junctions, with coefficient $E_{J,2}$ and the sign convention specified in \autoref{eq:app-ats-potential}. Together with the superinductive energy $\tfrac{1}{2}E_L\hat\varphi^2$ and a second-harmonic Josephson correction $E_{J,2}$ motivated by observed percent-level higher-harmonic contributions in tunnel junctions \cite{Willsch2024}, the full ATS Hamiltonian follows:
\begin{align}
    \hat H_{\rm ATS}/\hbar &= 4E_C\hat n^2 + \tfrac{1}{2}E_L\hat\varphi^2 + \hat\varepsilon_{\rm pot,ATS}(\hat\varphi),
    \label{eq:app-ats-full} \\
    \hat\varepsilon_{\rm pot,ATS}(\hat\varphi) &=
        - 2E_J\cos(\varphi_\Sigma)\cos(\hat\varphi+\varphi_\Delta)
        \nonumber\\
        &\quad+ 2\Delta E_J\sin(\varphi_\Sigma)\sin(\hat\varphi+\varphi_\Delta)
        \nonumber\\
        &\quad+ 2E_{J,2}\cos(2\varphi_\Sigma)\cos(2\hat\varphi+2\varphi_\Delta).
    \label{eq:app-ats-potential}
\end{align}
Here, $\hat n$ is the dimensionless charge operator conjugate to $\hat\varphi$, satisfying $[\hat\varphi,\hat n]=i$; it is distinct from the oscillator photon-number operator $\hat c^\dagger\hat c$. As fixed by the convention stated after \autoref{eq:RAR_sys_hamiltonian}, every energy scale above is quoted, and used, as an angular frequency.

Within the linear-superinductance model, $\hat\varphi$ is noncompact. The quadratic Hamiltonian $4E_C\hat n^2+\tfrac12E_L\hat\varphi^2$ therefore defines a harmonic oscillator, whose Fock basis provides a convenient computational basis for the ATS mode \cite{jha2024smthesis}. In terms of its ladder operators $\hat c,\hat c^\dagger$
\begin{align}
    \hat H_{\rm ATS}/\hbar &= \underbrace{\sqrt{8E_CE_L}}_{\displaystyle\omega_c^{\rm bare}}\,\hat c^\dagger\hat c
        + \hat\varepsilon_{\rm pot,ATS}(\hat\varphi),
    \label{eq:app-ats-fock} \\
    \hat\varphi &= \Phi_{\rm ZPF}(\hat c+\hat c^\dagger),
    \qquad
    \Phi_{\rm ZPF} \equiv \left(\frac{2E_C}{E_L}\right)^{1/4},
    \label{eq:app-phi-zpf}
\end{align}
having dropped an overall zero-point constant. The quantities $\omega_c^{\rm bare}$ and $\Phi_{\rm ZPF}$ define the uncoupled ATS reference scales used below. The resonator detunings are defined relative to $\omega_c^{\rm bare}$, while the phase participations are referenced to $\Phi_{\rm ZPF}$. The dressed normal-mode frequencies and participations are obtained from the coupled circuit.

Setting $\Delta E_J=E_{J,2}=0$, two choices of bias point $(\varphi_\Sigma,\varphi_\Delta)$ recur throughout the manuscript:

\textbf{Asymmetric point}, $\varphi_\Sigma=\pi/2+\epsilon(t)$, $\varphi_\Delta=\pi/2$: here $\cos(\hat\varphi+\varphi_\Delta)=-\sin\hat\varphi$, and to first order in the small pump $\epsilon(t)$, $\cos(\varphi_\Sigma)=\cos(\pi/2+\epsilon(t))\approx-\epsilon(t)$. Substituting both into \autoref{eq:app-ats-potential} gives
\begin{equation}
    \hat H_{\rm ATS}/\hbar = \omega_c^{\rm bare}\hat c^\dagger\hat c
        - 2E_J\epsilon(t)\sin(\hat\varphi),
    \label{eq:app-asym-bias}
\end{equation}
which is exactly the bias point underlying the CX Hamiltonian, \autoref{eq:driver_ham}.

\textbf{Symmetric point}, $\varphi_\Sigma=\pi/2+\epsilon(t)$, $\varphi_\Delta=0$: here instead $\cos(\hat\varphi+\varphi_\Delta)=\cos\hat\varphi$, and $\cos(\varphi_\Sigma)=\cos(\pi/2+\epsilon(t))\approx-\epsilon(t)$, so
\begin{equation}
    \hat H_{\rm ATS}/\hbar = \omega_c^{\rm bare}\hat c^\dagger\hat c
        + 2E_J\epsilon(t)\cos(\hat\varphi),
    \label{eq:app-sym-bias}
\end{equation}
which is the bias point underlying the ZZ Hamiltonian.

\subsection{Capacitive Coupling and Normal Modes}
\label{app:coupling-normal-modes}

Following Ref.~\cite{jha2024smthesis}, consider two capacitively-shunted inductive elements connected by a coupling capacitance $C_c$. The capacitance matrix and its inverse give a kinetic (charge) coupling term
\begin{equation}
    \hat H_{\rm coupling} = J_{1,2}\,\hat n^{(1)}\hat n^{(2)}
        = -\hbar g_{1,2}(\hat a_1-\hat a_1^\dagger)(\hat a_2-\hat a_2^\dagger),
    \label{eq:app-coupling-def}
\end{equation}
with $J_{1,2}=4e^2[C^{-1}]_{1,2}$ and $g_{1,2}=n^{(1)}_{\rm zpf}n^{(2)}_{\rm zpf}J_{1,2}/\hbar$~\cite{jha2024smthesis}. For an ATS-resonator pair, the explicit form is:
\begin{equation}
    g_{1,2} = \left(\frac{E_L^{(1)}}{32E_C^{(1)}}\right)^{1/4}\left(\frac{E_L^{(2)}}{32E_C^{(2)}}\right)^{1/4}J_{1,2}/\hbar.
    \label{eq:app-g-formula}
\end{equation}
Applying \autoref{eq:app-coupling-def} to the resonator--ATS--resonator (RAR) chain, with resonator A and resonator B each capacitively coupled to the central ATS mode, retaining the resonator--ATS charge couplings and neglecting the residual direct resonator--resonator charge coupling gives:
\begin{equation}
    \hat H_{\rm int}/\hbar = -g_{ac}(\hat a-\hat a^\dagger)(\hat c-\hat c^\dagger)
        - g_{cb}(\hat c-\hat c^\dagger)(\hat b-\hat b^\dagger).
    \label{eq:app-rar-coupling}
\end{equation}

\autoref{eq:app-rar-coupling} is a coupling between charge (momentum-like) operators, not a beam-splitter. Writing $\hat p_i=-i(\hat d_i-\hat d_i^\dagger)/\sqrt2$ so that $(\hat d_i-\hat d_i^\dagger)=i\sqrt2\,\hat p_i$, \autoref{eq:app-rar-coupling} becomes $\hat H_{\rm int}/\hbar=2 g_{ac}\hat p_a\hat p_c+2 g_{cb}\hat p_c\hat p_b$, which is purely kinetic. The full quadratic Hamiltonian of the linear (pre-ATS-nonlinearity) system is therefore:
\begin{align*}
    \hat H/\hbar = \tfrac{1}{2}\mathbf x^T &V\mathbf x + \tfrac{1}{2}\mathbf p^T M\mathbf p,
    \qquad
    V = {\rm diag}(\omega_a^{(0)},\omega_c^{\rm bare},\omega_b^{(0)}),
    \\
    M &= \begin{pmatrix}
        \omega_a^{(0)} & 2g_{ac} & 0\\
        2g_{ac} & \omega_c^{\rm bare} & 2g_{cb}\\
        0 & 2g_{cb} & \omega_b^{(0)}
    \end{pmatrix}. \numberthis
    \label{eq:app-quadratic-ham}
\end{align*}
In the bare-coordinate representation, the capacitive couplings appear only in the kinetic matrix $M$, while $V$ is diagonal. The ATS phase is proportional to the bare coordinate $\hat x_c$; its expansion in normal modes follows from the coordinate transformation below.

For a single resonator--ATS pair, transformation to the interaction picture of the uncoupled oscillators gives
\begin{align*}
    \hat H_{2,{\rm LC}}/\hbar &= g_{ac}\hat a^\dagger\hat c\, e^{i\Delta_{ac}t}
        + g_{ac}\hat a\hat c^\dagger e^{-i\Delta_{ac}t}\\
        &- g_{ac}\hat a\hat c\, e^{-i\Sigma_{ac}t}
        - g_{ac}\hat a^\dagger\hat c^\dagger e^{i\Sigma_{ac}t}, \numberthis
    \label{eq:app-rotating-frame}
\end{align*}
with $\Delta_{ac}=\omega_a-\omega_c$, $\Sigma_{ac}=\omega_a+\omega_c$. If $|\Sigma_{ac}|\gg|\Delta_{ac}|$, the last two (counter-rotating) terms are dropped, giving the beam-splitter form $\hat H_{2,{\rm LC}}\approx\hbar\omega_a\hat a^\dagger\hat a+\hbar\omega_c\hat c^\dagger\hat c+\hbar g_{ac}(\hat a^\dagger\hat c+\hat a\hat c^\dagger)$, diagonalized in closed form with dressed frequencies $\tilde\omega_{1,2}=\tfrac12(\omega_a+\omega_c)\mp\tfrac12\sqrt{\Delta_{ac}^2+4g_{ac}^2}$.

Keeping instead both the co- and counter-rotating pieces of \autoref{eq:app-rotating-frame} and applying first-order (Rayleigh--Schr\"odinger) perturbation theory to the vacuum of $c$, the co-rotating term contributes an admixture $g_{ac}/\Delta_{ac}$ and the counter-rotating term an admixture $g_{ac}/\Sigma_{ac}$ (the latter surviving because $\hat a^\dagger\hat c^\dagger$ connects the vacuum to a two-excitation state at energy $\Sigma_{ac}$, with no resonant denominator to suppress it). Summing both
\begin{equation}
    \frac{\varphi_a}{\varphi_c} \simeq g_{ac}\left(\frac{1}{\Delta_{ac}}+\frac{1}{\Sigma_{ac}}\right),
    \qquad p_a\equiv\left(\frac{\varphi_a}{\varphi_c}\right)^2,
    \label{eq:phi-perturbative}
\end{equation}
and likewise for $\varphi_b/\varphi_c$ with $(g_{cb},\Delta_{bc},\Sigma_{cb})$.

Although the counter-rotating terms oscillate rapidly for the chosen parameters, we retain them to calculate the dressed frequencies and phase participations without the rotating-wave approximation (RWA). The counter-rotating terms change the participations by $\Delta/\Sigma \approx 5-10\%$, so we diagonalize \autoref{eq:app-quadratic-ham} directly, following the same Bogoliubov--Valatin construction Ref.~\cite{jha2024smthesis} sets up for two modes extended here to the three-mode chain $a$--$c$--$b$.

Diagonalizing the symmetric matrix $M^{1/2}VM^{1/2}=O\Lambda O^T$ gives the normal-mode frequencies and the ATS-phase participations
\begin{equation}
    \omega_k = \sqrt{\lambda_k},
    \qquad
    \varphi_k = \Phi_{\rm ZPF}\big[M^{1/2}O\Lambda^{-1/4}\big]_{ck},
    \label{eq:phi-decomp}
\end{equation}
where the row index $c$ selects the ATS coordinate and $k\in\{a,b,c\}$ runs over the three normal modes. The transformation $\mathbf x=T\mathbf X$, $T=M^{1/2}O\Lambda^{-1/4}$, is canonical by construction ($\langle\hat X_k^2\rangle=1/2$ for each normal mode).

Because \autoref{eq:phi-decomp} is a canonical transformation restricted to the row selecting the ATS coordinate
\begin{equation}
    \varphi_a^2+\varphi_b^2+\varphi_c^2 = \Phi_{\rm ZPF}^2\left[1-\mathcal O(g^2/\omega^2)\right],
    \label{eq:app-sumrule-derived}
\end{equation}
recovering exactly $\Phi_{\rm ZPF}^2$ at $g_{ac}=g_{cb}=0$ and decreasing smoothly as coupling grows.

The perturbative expressions provide intuition for the distribution and total magnitude of the ATS phase fluctuations. All numerical phase participations reported in this manuscript are obtained from the exact quadratic normal-mode transformation in \autoref{eq:phi-decomp}.

\begin{figure*}[t]
    \centering
    \includegraphics[width=1.05\textwidth]{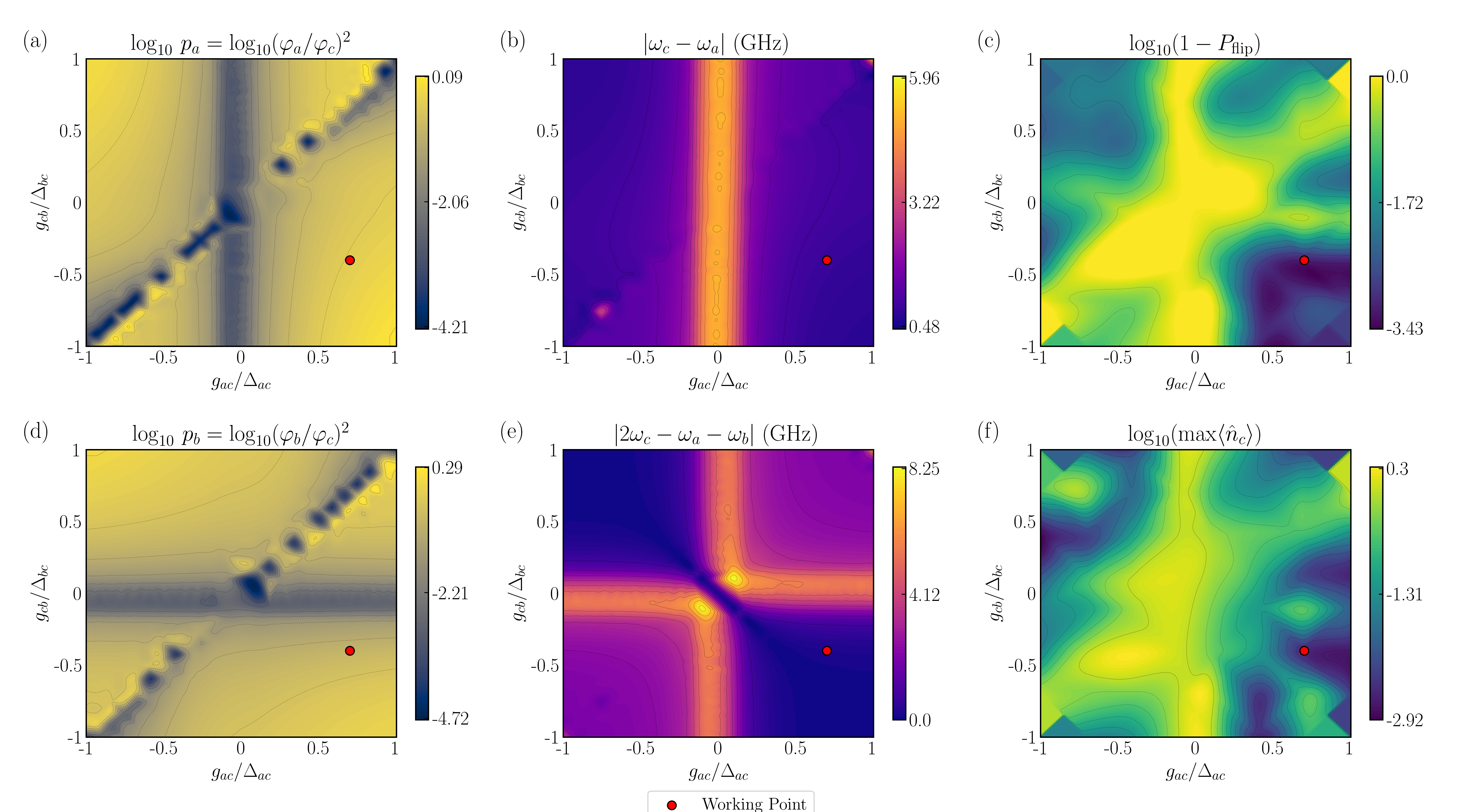}
    \caption{\justifying Design landscape as a function of the signed hybridization ratios $g_{ac}/\Delta_{ac}$ and $g_{cb}/\Delta_{bc}$, each spanning $[-1,1]$. (a,d) Target and control participation ratios $p_a,p_b$ (log scale), from the exact diagonalization. The dark trench running along $g_{ac}/\Delta_{ac}\approx g_{cb}/\Delta_{bc}$ marks the $\omega_a\approx\omega_b$ near-degeneracy, where hybridization redistributes participation away from both target and control modes simultaneously and which is excluded from candidate operating points on that basis alone. (b) Dressed detuning $|\omega_c-\omega_a|=|\Delta_{ac}|$, the quantity directly setting the coupler-leakage floor of \autoref{eq:coupler-floor}. (e) Spectral clearance $|2\omega_c-\omega_a-\omega_b|$ from the parasitic resonance. Smaller values identify operating points requiring closer examination for resonant coupler excitation. (c,f) Direct time-domain simulation of the full circuit Hamiltonian. Infidelity $1-P_{\rm flip}$, calculated as the flip rate of initial $|11\rangle$ state, and $\max\langle\hat n_c\rangle$ is the peak coupler population. The favorable region in (c,f), away from the degeneracy trench of (a,d) and the resonance cross of (e), confirms that the fast proxies in (a,b,d,e) qualitatively predict the simulated performance.}
    \label{fig:physicality-window}
\end{figure*}

\subsection{Design Methodology: Coupler Leakage, Drive Law, and Frequency Selection}
\label{app:design-methodology}

The gate-time relation of Section~\ref{sec:gate-time-kerr} fixes the target interaction strengths $g_3$ and $g_{\rm CD}$ that the ATS coupler must deliver. Converting those target strengths into a physical circuit, which requires a choice of resonator frequencies $\omega_a,\omega_b$, capacitive couplings $g_{ac},g_{cb}$, and RF drive amplitudes $\epsilon_{g_3},\epsilon_{g_{\rm CD}}$, is governed by a small number of closed-form relations, derived once from the normal-mode structure of \autoref{eq:driver_ham} and reused throughout the design.

We fix $E_C$, $E_L$, and $E_J$ before optimizing the resonator frequencies. The first two determine the bare ATS frequency and zero-point phase fluctuation, while $E_J$ sets the strength of the parametrically activated interactions. The frequency-selection procedure below is therefore conditional on this choice of ATS parameters.

For fixed target interaction strengths, the required pump amplitudes scale as $1/(|\varphi_a|^2|\varphi_b|)$. Small target-mode participation therefore imposes a particularly strong drive penalty. We use this dependence to identify candidate operating points with sufficiently large phase participations in both resonator modes.

The dominant, unavoidable channel competing with the intended $g_3$ process is the nearby process $\hat a^\dagger\hat b\hat c^\dagger$, in which one of the two target photons is replaced by a coupler photon, detuned by $\Delta_{ac}=\omega_a-\omega_c$ from the $g_3$ pump. Treating the resonator amplitudes semiclassically, this parasitic interaction gives an off-resonant coupler-excitation scale:
\begin{equation}
    \hat n_c^{\rm floor} \;\propto\; \frac{K^2\alpha^6}{p_a\,\Delta_{ac}^2},
    \qquad p_a\equiv\varphi_a^2/\varphi_c^2,
    \label{eq:coupler-floor}
\end{equation}
The explicit dependence on $\beta$ cancels after substituting the required $g_3$ strength. This expression is a scaling estimate for this channel, and the actual excitation depends on the drive waveform and the remaining circuit interactions. The target participation $p_a$ and detuning $\Delta_{ac}$ in this case are the only remaining levers. A separate, avoidable resonance is worth flagging as a hard design constraint. The channel $\hat a\hat c^{\dagger2}$, activated on the $g_{\rm  CD}$ pump, becomes resonant at
\begin{equation}
    2\omega_c \approx \omega_a+\omega_b,
    \label{eq:resonance-constraint}
\end{equation}
and can dominate the total coupler excitation if the chosen $(\omega_a,\omega_b)$ falls close to resonance. Unlike \autoref{eq:coupler-floor}, this one is simply avoided by frequency choice.

The drive amplitudes and parasitic excitation depend jointly on the dressed detunings and phase participations. Their leading dispersive scalings do not by themselves determine an optimal detuning. We therefore search numerically for an operating point where the detuning and participation are large enough to keep the drive amplitudes small while maintaining a low coupler-excitation floor. \autoref{fig:physicality-window}(a,d,f) shows this trade-off clearly, and we choose $(\omega_a,\omega_b)$ by locating the low $\max\langle\hat n_c\rangle$ region of panel (f) and low $1-P_{\rm flip}$ in (c) that clears both exclusion zones, rather than by pushing $\Delta_{ac}$ toward either extreme.

We restrict the numerical search to $|g/\Delta|\leq1$ on both links as a chosen design window. Within this window, we obtain the normal-mode frequencies and phase participations by exact diagonalization of the positive-definite quadratic Hamiltonian.

\section{Quantum Process Tomography and Error Quantification} \label{app:ptm}
\numberwithin{equation}{section}
\subsection{Pauli Transfer Matrix and Noise Channel Construction}
To characterize an unknown quantum process, quantum process tomography (QPT) is used \cite{Manenti2023, Greenbaum2015}. The Pauli Transfer Matrix (PTM) provides a quantitative representation for analyzing quantum channels and is particularly effective for characterizing the noise bias of bosonic operations \cite{Puri2020}. Here, we employ the PTM formalism to isolate the error channel of the implemented CX gate and quantify the ratio of dephasing to non-dephasing errors.

Consider a quantum channel $\mathcal{E}$ mapping an input density matrix $\rho$ to an output $\rho^\prime = \mathcal{E}(\rho)$. Any $N$-qubit density matrix can be decomposed in the Pauli basis $\mathcal{P}_N = \{I, X, Y, Z\}^{\otimes N}$ as $\rho = \sum_k c_k P_k$, where $\vec{c}$ is a $4^N \times 1$ coefficient vector and $P_k$ are the Pauli operators spanning the $N$-qubit operator space. Since the channel $\mathcal{E}$ is a linear map, there exists a $4^N \times 4^N$ real matrix $R$, defined as the Pauli Transfer Matrix, such that the evolution of the coefficients is given by:
\begin{equation}
    \vec{c}^\prime = R \vec{c}.
\end{equation}
To derive the elements of $R$, we use the Hilbert-Schmidt orthogonality relation, $\text{Tr}(P_l P_j) = 2^N \delta_{l,j}$. The coefficients of the state are extracted via $c_l = \text{Tr}(P_l \rho)/2^N$. The linear relation implies:
\begin{equation}
    c_i^\prime = \sum_j R_{i,j} c_j.
\end{equation}
To determine $R_{ij}$, we apply the linear map $\mathcal{E}$ to the basis operator $P_j$. Although $P_j$ is generally not a density matrix, its evolution is defined by linear extension from physical input states. Experimentally, the action on each basis operator can be inferred from suitable linear combinations of outputs obtained from physical input states. In this case, the input coefficient vector has a single non-zero entry $c_l = \delta_{l,j}$. Substituting this into the channel definition:
\begin{align}
    c_i^\prime &= \frac{1}{2^N} \text{Tr}(P_i \rho^\prime) = \frac{1}{2^N} \text{Tr}(P_i \mathcal{E}(P_j)).
\end{align}
Comparing this result with the matrix multiplication $c_i^\prime = \sum_l R_{i,l} \delta_{l,j} = R_{i,j}$, we arrive at the formal definition of the PTM coefficients:
\begin{equation}
    R_{i,j} = \frac{1}{2^N}\text{Tr}(P_i\mathcal{E}(P_j)). \label{eq:ptm_derived}
\end{equation}
Experimentally, this corresponds to preparing eigenstates of the Pauli operators, evolving them under the channel, and measuring the expectation values of the Pauli basis elements.

Ideally, the channel can be reconstructed from the PTM by inverting this decomposition:
\begin{equation}
\mathcal{E}(\rho) = \sum_{k,l} \frac{1}{2^N} R_{l,k} \text{Tr}(P_k \rho) P_l. \label{eq:channel_derived}
\end{equation}
For the CX gate analysis, we define the ideal and noisy channels as
\begin{align}
\mathcal{E}_\text{ideal}(\rho) &= U_\text{CX} \rho U_\text{CX}^\dagger, \\
\mathcal{E}_\text{noisy}(\rho) &= \mathcal{E}_\text{err}( \mathcal{E}_\text{ideal}(\rho)),
\end{align}
where $U_{\text{CX}}$ is the unitary for the ideal CX gate, and $\mathcal{E}_\text{err}$ represents the error process occurring after the ideal operation. The effective error channel is therefore isolated as $\mathcal{E}_\text{err} = \mathcal{E}_\text{noisy} \circ \mathcal{E}_\text{ideal}^{-1}$. In the PTM representation, channel composition corresponds to matrix multiplication. Thus, if $R_\text{ideal}$ and $R_\text{noisy}$ denote the PTMs of the ideal and implemented gates respectively, the PTM of the error channel, $R_\text{err}$, is given by:
\begin{equation}
    R_\text{err} = R_\text{noisy} R_\text{ideal}^{-1}.
    \label{eq:ptm_error}
\end{equation}

The error PTM fully specifies the reconstructed logical error map. Its entries describe how Pauli components of an input operator contribute to the output. To characterize the distribution of errors among Pauli components, we convert $R_{\rm err}$ to the process matrix $\chi_{\rm err}$.

\subsection{Transformation to the $\chi$-Matrix Representation}
To physically interpret the error channel derived in \autoref{eq:ptm_error}, we use the process matrix (or $\chi$-matrix) representation. In this basis, the action of the error channel $\mathcal{E}_\text{err}$ on a density matrix $\rho$ is expanded in the Pauli basis as
\begin{equation}
\mathcal{E}_\text{err}(\rho) = \sum_{m,n} \chi_{mn} P_m \rho P_n^\dagger,
\end{equation} where indices $m,n$ run from $0$ to $4^N-1$. The $\chi$-matrix is positive semi-definite by definition. The diagonal elements $\chi_{kk}$ are the weights of the Pauli-twirled channel, $\mathcal{E}_{\rm tw}(\rho)=\sum_k \chi_{kk}P_k\rho P_k$, and can be read as the probability of the Pauli error $P_k$ after twirling. The off-diagonal elements, which arise from coherent errors as well as from incoherent non-Pauli processes such as photon loss, are discarded by this projection. The diagonal weights therefore provide a convenient summary of the error budget and of the noise bias, which we use below.

The transformation from the PTM, $R$, to the $\chi$-matrix constitutes a change of basis in the vector space of superoperators. We derive this mapping by expanding the PTM definition:
\begin{align}
R_{ij} &= \frac{1}{2^N} \text{Tr}\left( P_i \mathcal{E}(P_j) \right) \nonumber \\
&= \frac{1}{2^N} \sum_{m,n} \chi_{mn} \text{Tr}\left( P_i P_m P_j P_n \right).
\end{align}
To invert this relation and extract $\chi$ from a calculated PTM, we use the orthogonality of the Pauli basis. The transformation can be expressed as a tensor contraction:
\begin{equation}
\chi_{mn} = \frac{1}{8^N} \sum_{i,j} R_{ij} \text{Tr}(P_m P_i P_n P_j).
\label{eq:ptm_to_chi}
\end{equation}
Numerically, this operation is implemented via Einstein summation. We construct a rank-4 tensor $\mathcal{T}_{minj} = \text{Tr}(P_m P_i P_n P_j)$ and contract it with the PTM $R_{ij}$ to recover the process matrix.

From the computed $\chi$-matrix, we extract the probabilities for the relevant error generators. Specifically, for the bias-preserving CX gate, we are interested in the ratio of phase-flip errors to bit-flip errors. These are quantified by summing the diagonal populations of the relevant Pauli subsets
\begin{align}
\lambda_\text{nondeph} &= \sum_{P_k \in \mathcal{S}_X} \chi_{kk}, \\
\lambda_\text{deph} &= \sum_{P_k \in \mathcal{S}_Z} \chi_{kk},
\end{align}
where $\mathcal{S}_X$ and $\mathcal{S}_Z$ denote the sets of Pauli operators containing at least one $X/Y$ component or consisting of $ZI, IZ,$ and $ZZ$, respectively. Consequently, bias is defined as $\eta = \lambda_\text{deph}/\lambda_\text{nondeph}$.

\section{Bit and Phase Flip Time Estimations of the System}\label{app:fliptimes}
\numberwithin{equation}{section}
The dynamics of the system, including the resonator-ATS coupling and single-photon loss channels, are governed by the Lindblad master equation
\begin{equation}
    \dot\rho
    =-\frac{i}{\hbar}[\hat H_{\rm gate},\rho]
    +\kappa_a\mathcal D[\hat a]\rho
    +\kappa_b\mathcal D[\hat b]\rho.
\end{equation}
where $H_\text{gate}$ is the Hamiltonian realizing the idle cat qubits (\autoref{eq:1}) or the driven CX gate (\autoref{eq:CXgateonCatqubits}). The single-photon loss rates for the two resonators are denoted by $\kappa_{a}$ and $\kappa_{b}$, with the standard dissipator defined as $\mathcal{D}[\hat{L}]\rho = \hat{L}\rho \hat{L}^\dagger - \frac{1}{2}\{\hat{L}^\dagger \hat{L}, \rho\}$. In the theoretical analysis, we neglect intrinsic losses of the ATS coupler during the pump-activated intervals, assuming the coupler is far detuned from the resonator modes and the suppression of higher-order leakage terms.

Bit-flip errors in a stabilized cat qubit correspond to transitions between the logical computational states, approximated by the coherent states $|0_L\rangle \approx |\alpha\rangle$ and $|1_L\rangle \approx |-\alpha\rangle$. These states are stabilized against the Hamiltonian evolution but remain susceptible to dissipative channels. The dominant mechanism for bit-flips is single-photon loss, described by the jump operator $\hat{a}$ occurring at rate $\kappa_{\rm 1ph}$ \cite{Mirrahimi2014}.

To derive the single-photon-loss contribution to the logical bit-flip rate, we project the loss operator onto the confined cat manifold. We use the orthonormal logical basis
\begin{equation}
    \ket{0_L}
    =\frac{\ket{\mathcal C_\alpha^+}
           +\ket{\mathcal C_\alpha^-}}{\sqrt{2}},
    \qquad
    \ket{1_L}
    =\frac{\ket{\mathcal C_\alpha^+}
           -\ket{\mathcal C_\alpha^-}}{\sqrt{2}},
\end{equation}
where the normalized even and odd cat states are:
\begin{equation}
    \ket{\mathcal C_\alpha^\pm}
    =
    \frac{\ket{\alpha}\pm\ket{-\alpha}}
    {\sqrt{2(1\pm s)}},
    \qquad s=e^{-2|\alpha|^2}.
\end{equation}
For large $|\alpha|$, these logical states approach $\ket{\alpha}$ and $\ket{-\alpha}$, respectively.

The annihilation operator exchanges the even and odd cat states according to:
\begin{align}
    \hat a\ket{\mathcal C_\alpha^+}
    &=\alpha\sqrt{\frac{1-s}{1+s}}
      \ket{\mathcal C_\alpha^-},\\
    \hat a\ket{\mathcal C_\alpha^-}
    &=\alpha\sqrt{\frac{1+s}{1-s}}
      \ket{\mathcal C_\alpha^+}.
\end{align}
Consequently, its projection onto the logical subspace is \cite{Puri2020}
\begin{equation}
    \hat a_L
    =P_{\mathcal C}\hat aP_{\mathcal C}
    =\frac{\alpha}{\sqrt{1-s^2}}
      \left(\hat Z_L-is\hat Y_L\right),
\end{equation}
where $P_{\mathcal C}$ projects onto the cat manifold. The dominant $\hat Z_L$ component produces phase errors, while the exponentially small $\hat Y_L$ component connects the logical computational states. The corresponding transition matrix element is:
\begin{equation}
    \mathcal M_{0\to1}
    =\bra{1_L}\hat a_L\ket{0_L}
    =\frac{\alpha e^{-2|\alpha|^2}}
           {\sqrt{1-e^{-4|\alpha|^2}}}.
    \label{eq:M01}
\end{equation}

Within the projected loss model $\kappa_{\rm 1ph}\mathcal D[\hat a_L]$, the symmetric logical bit-flip rate is therefore
\begin{align}
    \Gamma_x
    &=\kappa_{\rm 1ph}
      \left|\mathcal M_{0\to1}\right|^2
      \nonumber\\
    &=\kappa_{\rm 1ph}
      \frac{|\alpha|^2e^{-4|\alpha|^2}}
           {1-e^{-4|\alpha|^2}}
      \nonumber\\
    &\approx\kappa_{\rm 1ph}
      |\alpha|^2e^{-4|\alpha|^2},
\end{align}
where the final approximation assumes $e^{-4|\alpha|^2}\ll1$.

Using $|\alpha|^2=|\epsilon_2|/K$ for the resonant Kerr-cat Hamiltonian with $K>0$, the leading-order rate becomes:
\begin{equation}
    \Gamma_x
    \approx
    \kappa_{\rm 1ph}\frac{|\epsilon_2|}{K}
    \exp\!\left(-\frac{4|\epsilon_2|}{K}\right).
\end{equation}
Thus, within the confined-manifold approximation, the bit-flip time $T_{\rm bit}=1/(2\Gamma_x)$ grows exponentially with cat size. This estimate describes the single-photon-loss contribution and does not include additional errors from gate-induced leakage or other noise mechanisms.

The Kerr nonlinearity and two-photon drive provide Hamiltonian confinement of the cat manifold. Within this manifold, single-photon loss acts predominantly as a logical phase flip. 

Phase-flip errors correspond to parity changes within the code manifold. The loss of a single photon applies the operator $\hat{a}$ to the state, mapping an even cat $|\mathcal{C}_\alpha^+\rangle$ to an odd cat $|\mathcal{C}_\alpha^-\rangle$, approximately $\alpha |\mathcal{C}_\alpha^-\rangle$. This parity change corresponds to a logical Pauli-$Z$ error \cite{Puri2017}. Since each photon-loss event independently flips the parity, the decay of the parity expectation value $\langle\hat\Pi\rangle$ follows from the Lindblad equation for the dissipator $\mathcal{D}[\hat a]$. For idle Kerr-cat confinement, the Hamiltonian commutes with parity. Approximating the photon number within the cat manifold by $\bar n\simeq|\alpha|^2$ gives
\begin{equation}
    \frac{d\langle\hat\Pi\rangle}{dt} \approx -2\kappa_{\rm 1ph}|\alpha|^2 \langle\hat\Pi\rangle,
\end{equation}
so that the parity coherence decays at rate $2\kappa_{\rm 1ph}\bar n$, where $\bar n = \langle \hat a^\dagger \hat a\rangle$. The phase-flip rate is thus:
\begin{equation}
\Gamma_{z} \approx 2\kappa_{\rm 1ph} |\alpha|^2 = \frac{2\kappa_{\rm 1ph} |\epsilon_2|}{K}.
\end{equation}

We define both lifetimes as $1/e$ decay times: for symmetric bit flips at rate $\Gamma_x$, $\langle\hat\sigma_z\rangle$ decays as $e^{-2\Gamma_x t}$, so $T_{\rm bit} = 1/(2\Gamma_x)$; the parity decays at $\Gamma_z$, so $T_{\rm phase} = 1/\Gamma_z$.

With these lifetime conventions
\begin{equation}
    \frac{T_{\rm bit}}{T_{\rm phase}}
    =\frac{\Gamma_z}{2\Gamma_x}
    \approx e^{4|\alpha|^2},
\end{equation}
which grows exponentially with the mean photon number \cite{Guillaud2019}. For the target mode ($\bar n_a = 2$), this predicts $\Gamma_z / (2\Gamma_x)  \approx 2981$, in close agreement with the asymptotic ratio $T_\text{bit}/T_\text{phase} \approx 2987$ extracted from the idling simulation in \autoref{tab:bit_flip_times}. We note this theoretical $\Gamma_z / (2\Gamma_x)$ is a rate ratio and should be compared against $T_\text{bit}/T_\text{phase}$ rather than the finite-gate-time $\eta$ of \autoref{tab:ptm_bias_metrics}.

The estimated bit- and phase-flip times agree with the simulated idling and ideal-CX values in \autoref{tab:bit_flip_times}.

\subsection{Numerical Simulation and Fitting Methodology}

To validate the analytical error models, we perform time-domain simulations of the logical observables under repeated gate operations. This section details the construction of these simulations and the statistical methods used to extract the characteristic lifetimes $T_{\text{bit}}$ and $T_{\text{phase}}$.

The bit-flip and phase-flip lifetimes are characterized by initializing the control-target system $|\psi\rangle = |\text{control}, \text{target}\rangle$ in specific logical states:
\begin{itemize}
    \item \textbf{Bit-flip measurement:} The system is initialized in the $|1_{\beta}1_{\alpha}\rangle$ state.
    \item \textbf{Phase-flip measurement:} The system is initialized in the $|1_{\beta}+_{\alpha}\rangle$ state.
\end{itemize}

The evolution is simulated for $2M$ iterations of the gate. To isolate the identity operation and monitor decay over time, the resulting density matrix $\rho$ is saved after every two gate applications, leveraging the property $\text{CX}^2 = \mathbb{I}$.

Ensuring the validity of numerical results requires monitoring the population at the boundaries of the truncated Hilbert space. We trace out the control/target modes to inspect the reduced density matrices of each resonator. The population of the two highest Fock states, $|N-1\rangle$ and $|N-2\rangle$, is calculated at every time step to ensure that leakage into unconfined states does not invalidate the simulation.

We track the decay of the logical manifold through two primary observables:
\begin{itemize}
    \item \textbf{Bit-flip decay:} We monitor the expectation value of the logical $\hat{\sigma}_z$ operator (the polarization in the cat basis).
    \item \textbf{Phase-flip decay:} We monitor the Wigner parity $\langle\hat{\Pi}\rangle = \langle e^{i\pi\hat{a}^\dagger\hat{a}} \rangle$. This observable is robust against frame rotations and provides a direct measure of phase coherence.
\end{itemize}

Lifetimes are obtained by fitting the saved expectation values over the accumulated gate time $t = \text{iterations} \times T$. Due to the differences in error scales between bit and phase flips, we use two different fitting models:
\begin{itemize}
    \item \textbf{Phase-flip (Exponential Fit):} The parity decay follows standard decoherence physics and is fitted to:
    \begin{equation}
        f_{\text{exp}}(t) = A e^{-t/T} + C.
    \end{equation}
    We discard the initial transient points to focus on the steady-state decay. The phase-flip time is then extracted as $T_{\text{phase}} = T$.
    \item \textbf{Bit-flip (Linear Fit):} Because $T_{\text{bit}}$ is significantly longer than the total simulation window, a linear rate equation provides a more robust estimation of the slope:
    \begin{equation}
        f_{\text{linear}}(t) = -\Gamma t + \text{Intercept}.
    \end{equation}
    The fit is performed on the absolute value of the $\hat{\sigma}_z$ data, ignoring the initial transient. The logical bit-flip time is defined as $T_{\text{bit}} = 1 / \Gamma$.
\end{itemize}

\subsection{Verification via Pauli Transfer Matrix Analysis}

To further validate the extracted lifetimes, we compare the results of the time-domain simulations with the decay rates predicted by the Pauli Transfer Matrix (PTM) of the noisy gate operation as derived in Appendix \ref{app:ptm}.

A common approach to estimate lifetimes is to use the diagonal entries of the error PTM, $R_{\rm err} = R_{\rm noisy}R_{\rm ideal}^{-1}$, for the target logical operators: $R_{3,3}$ for $\hat I\hat Z$ (bit flips) and $R_{1,1}$ for $\hat I\hat X$ (phase flips). With the chosen lifetime conventions,
\begin{align}
    T_{\text{bit}} &= \frac{-T_{\text{gate}}}{\ln R_{3,3}}, \\
    T_{\text{phase}} &= \frac{-T_{\text{gate}}}{\ln R_{1,1}}.
\end{align}
These expressions treat the error of each gate as an independent stochastic process. This is adequate for the phase-flip time and for the bit-flip time of the ideal gate ($1182\,\rm \mu$s from the diagonal entry versus $1216\,\rm \mu$s from the repeated-gate propagation below), but not for the ATS gate ($889\,\rm \mu$s versus $312\,\rm \mu$s). The difference originates in a small coherent rotation of the target qubit ($\approx 0.2^\circ$ per gate). Because $\hat I\hat X$ is invariant under conjugation by CX, this rotation is not undone between gates. Its amplitude adds over successive gates until loss-induced dephasing randomizes it, producing a faster effective bit-flip rate than the single-gate diagonal entry predicts. We therefore extract $T_{\rm bit}$ by propagating a full input state through repeated applications of the complete PTM, as described below, which retains these coherent contributions.

For the bit-flip lifetime we initialize the PTM representation of $|\psi\rangle=|11\rangle$,
\begin{equation}
    \vec{\rho}_{11} = \tfrac12\left(\vec{P}_{II} - \vec{P}_{IZ} - \vec{P}_{ZI} + \vec{P}_{ZZ}\right),
\end{equation}
in the normalized Pauli basis, and apply the noisy gate PTM twice per cycle, $\vec\rho_{n+1} = R_{\rm noisy}^{2}\,\vec\rho_n$, so that the ideal sequence is the identity ($\mathrm{CX}^2=\mathbb{I}$) and any change of the target polarization is due to errors. After each cycle we evaluate the code-space-normalized target polarization $|\langle \hat I\hat Z\rangle_n|/\langle \hat I\hat I\rangle_n$, which removes the effect of leakage, and fit it as a function of the elapsed time $t_n = 2nT_{\rm gate}$ with the decay model $|\langle\hat Z\rangle|(t) \simeq e^{-t/T_{\rm bit}}$. Unlike the single-gate diagonal entry, this propagation retains the coherent components of the error channel, which add over successive gates, and its result agrees with the time-domain simulations.

\autoref{tab:bit_flip_times} summarizes the consistency between our time-domain simulations and the PTM projection method. We observe that while the diagonal PTM extraction significantly misses the bit-flip time for the ATS-based gate, the projection method aligns closely with the numerical simulation results.

\section{Effect of $E_{J,2}$ and $\Delta E_J$}
\label{app:ej2_dej}

The analysis presented in the main text assumes an ideal ATS with vanishing junction asymmetry ($\Delta E_J = 0$) and second-harmonic Josephson energy ($E_{J,2} = 0$). Real devices can exhibit junction asymmetry and higher-harmonic contributions to the Josephson potential \cite{Willsch2024}. In this appendix, we analyze the physical consequences of non-zero $\Delta E_J$ and $E_{J,2}$ on the bias-preserving CX gate and present numerical benchmarks for a representative case of $\Delta E_J = E_{J,2} = 0.01\,E_J$.

\subsection{Origin of Residual Static Nonlinearities}

At the operating point $(\varphi_\Sigma, \varphi_\Delta) = (\pi/2, \pi/2)$, the three contributions to the ATS potential in \autoref{eq:RAR_sys_hamiltonian} evaluate to:
\begin{align}
    2E_J \sin(\hat{\varphi})\cos(\pi/2) &= 0, \label{eq:ej_vanish}\\
    +2\Delta E_J \cos(\hat{\varphi})\sin(\pi/2) &= 2\Delta E_J \cos(\hat{\varphi}), \label{eq:dej_static}\\
    -2E_{J,2}\cos(2\hat{\varphi})\cos(\pi) &= 2E_{J,2}\cos(2\hat{\varphi}). \label{eq:ej2_static}
\end{align}
The primary $E_J$ contribution vanishes at this bias point. A purely linear idle coupler is recovered only when $\Delta E_J=E_{J,2}=0$. However, the $\Delta E_J$ and $E_{J,2}$ terms do not vanish, and they introduce a residual static nonlinear potential that persists even in the absence of an RF drive.

\subsection{Induced Cross-Kerr Interaction}

The most consequential effect of these residual terms is the emergence of a cross-Kerr interaction between the resonator modes. Expanding the $E_{J,2}$ and $\Delta E_J$ contribution to fourth order in the phase operator $\hat{\varphi} = \sum_{o} \varphi_o(\hat{o} + \hat{o}^\dagger)$, these terms generate
\begin{align}
    \delta\hat H_{\rm static}/\hbar
    &= C-(\Delta E_J+4E_{J,2})\hat\varphi^2
    \nonumber\\
    &\quad+
    \frac{\Delta E_J+16E_{J,2}}{12}\hat\varphi^4
    +\mathcal O(\hat\varphi^6).
\end{align}
\begin{align*}
    \delta\hat H_{\rm static}^{(4)}/\hbar
    &\supset
    \chi_{ab}\hat n_a\hat n_b,
    \\
    \chi_{ab}
    &=2(\Delta E_J+16E_{J,2})
      \varphi_a^2\varphi_b^2, \numberthis
\end{align*}
which constitutes a direct cross-Kerr coupling $\chi_{ab}\,\hat{a}^\dagger\hat{a}\,\hat{b}^\dagger\hat{b}$ between the target and control resonators.

Our circuit parameters (\autoref{tab:circuit_params}) and $\Delta E_J = E_{J,2} = 0.01\,E_J$ yield a cross-Kerr shift of $\chi_{ab}/2\pi \approx 1.02$ MHz. This cross-Kerr reflects the fact that the ideal ATS was specifically engineered to eliminate static nonlinearity at the operating point, while junction imperfections reintroduce precisely the interaction that the bias point was designed to suppress. However, using the coherent states such that $\langle \hat{n}_a\rangle = |\alpha|^2$ for mode a and similarly for mode b, we can express the cross-Kerr term as
\begin{align*}
    \chi_{a,b} \hat{a}^\dagger\hat{a}\,\hat{b}^\dagger\hat{b}& = \chi_{a,b}|\alpha|^2\hat{n}_b + \chi_{a,b}|\beta|^2\hat{n}_a - \chi_{a,b} |\alpha|^2 |\beta|^2 \\&+ \chi_{a,b} (\hat{n}_a - |\alpha|^2)(\hat{n}_b - |\beta|^2), \numberthis
\end{align*}
which allows us to absorb the mean-field shifts into the mode frequencies, leaving only a coupling between photon-number fluctuations. Since the Kerr confinement suppresses transitions out of the cat manifold, we expect this residual term to cause only a small additional error.

\subsection{Consequences for Drive Frequencies}

The residual static potential also renormalizes the dressed mode frequencies and, consequently, the pump frequencies required to activate the $g_3$ and $g_\mathrm{CD}$ interactions. These corrections are incorporated self-consistently by including $\Delta E_J$ and $E_{J,2}$ in the system Hamiltonian prior to normal-mode diagonalization. The dressed transition frequencies are then extracted from the resulting eigenspectrum and used directly to define the rotating frame and the pump tones, following the same procedure described in Section \ref{sec:results} for the ideal case.

\subsection{Drive Corrections from $\Delta E_J$ and $E_{J,2}$}

Under the common flux modulation $\varphi_\Sigma = \pi/2 + \epsilon(t)$, the time-dependent parts of the $\Delta E_J$ and $E_{J,2}$ terms take the form
\begin{align}
    H_\mathrm{drive}^{\Delta E_J}/\hbar &= 2\Delta E_J \cos(\hat{\varphi})\big[\cos(\epsilon) - 1\big], \\
    H_\mathrm{drive}^{E_{J,2}}/\hbar &= 2E_{J,2}\cos(2\hat{\varphi})\big[\cos(2\epsilon) - 1\big],
\end{align}
where the subtraction of $\cos(0) = 1$ removes the static contribution already accounted for in the system Hamiltonian. The total driven Hamiltonian therefore becomes:
\begin{align*}
    \hat{H}_\mathrm{drive}/\hbar &= -2E_J \sin[\epsilon(t)]\sin\hat{\varphi} \\
    &+ H_\mathrm{drive}^{\Delta E_J}/\hbar + H_\mathrm{drive}^{E_{J,2}}/\hbar. \numberthis
    \label{eq:full_drive_dej}
\end{align*}
For drive amplitudes satisfying $\epsilon_{\rm peak} \ll 1$, the correction terms scale as $\cos(\epsilon)-1 \approx -\epsilon^2/2$, yielding contributions of order $\Delta E_J \cdot \epsilon_{\rm peak}^2 \sim 10^{-4}E_J$, three orders of magnitude below the primary drive term $E_J\sin(\epsilon) \approx E_J\epsilon_{\rm peak}$. Despite their small magnitude, we retain these drive corrections in all simulations presented below for completeness.

\subsection{Numerical Results}

To quantify the impact on gate performance, we repeat the CX gate simulation of Section \ref{subsec:cxsims} with varying $\Delta E_J = E_{J,2}$ included in the system construction, and the corresponding drive corrections $H_\mathrm{drive}^{\Delta E_J}$ and $H_\mathrm{drive}^{E_{J,2}}$ included alongside the primary $E_J$ drive. All other circuit parameters remain as specified in \autoref{tab:circuit_params}. The dressed mode frequencies and pump tones are recalculated from the updated eigenspectrum, and the rotating frame is defined using these corrected values.

\begin{figure}[h]
    \centering
    \includegraphics[width=\linewidth]{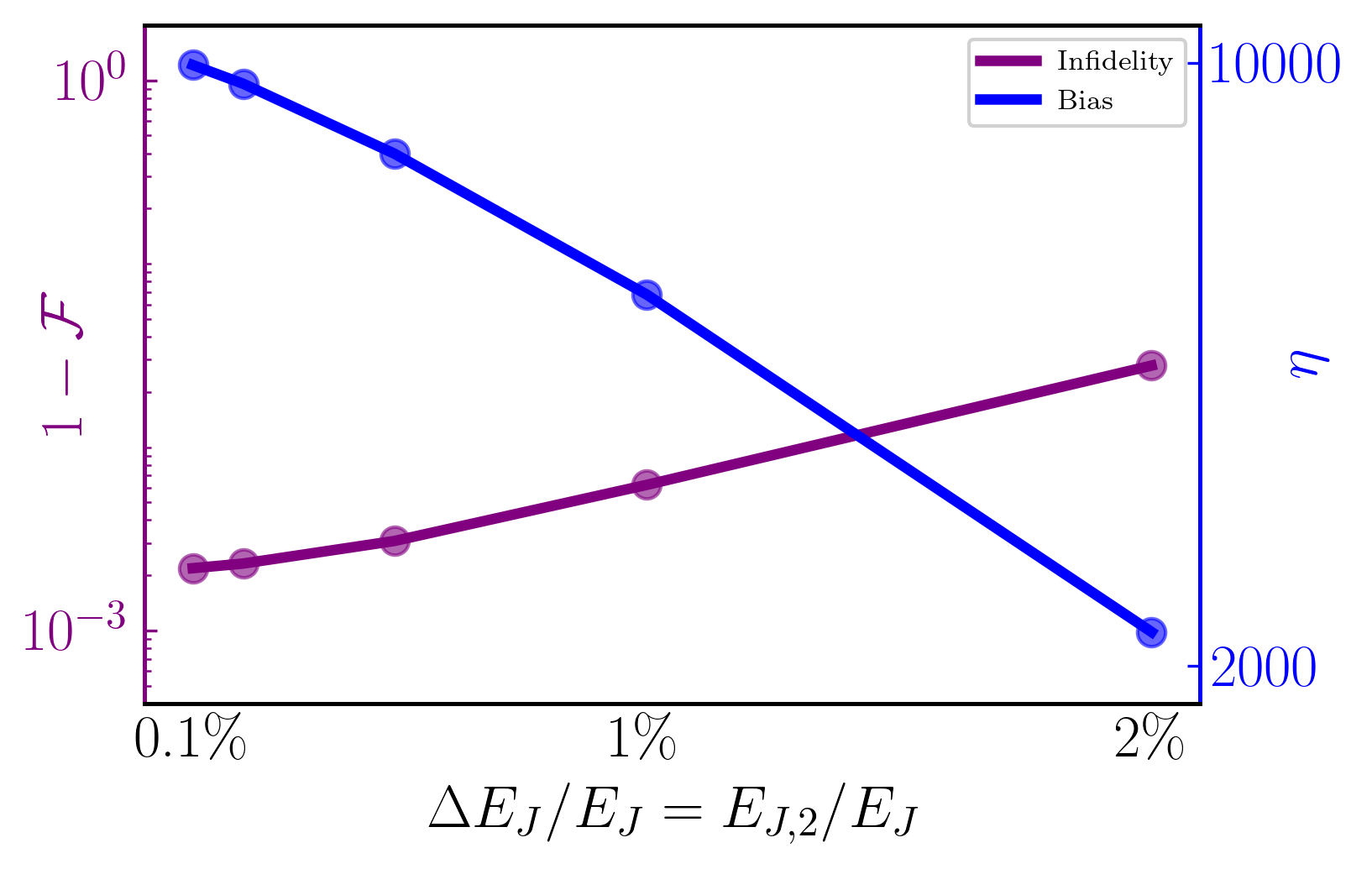}
    \caption{\justifying Infidelity and bias metrics obtained by sweeping $\Delta E_J/E_J = E_{J,2} / E_J$ from $0.1\%$ to $2\%$. Fidelity is calculated under no loss, and bias is calculated under $\kappa_{\rm 1ph}/2\pi = 100$ kHz.}
    \label{fig:dej_fid_bias}
\end{figure}

The results are shown in \autoref{fig:dej_fid_bias}. The average gate fidelity under no loss decreases from $\mathcal{F}_\mathrm{avg} = 99.69\%$ to $99.38\%$ at $\Delta E_J/E_J = E_{J,2}/E_J = 1\%$, reflecting the residual cross-Kerr interaction and associated higher-order nonlinearities that are not fully absorbed by the rotating frame. Under single-photon loss ($\kappa_{1\mathrm{ph}}/2\pi = 100$ kHz), the noise bias decreases from $\eta \approx 9922$ to $\eta \approx 6924$ at $1\%$. While this reduction is notable, the gate remains in the biased regime ($\eta \gg 1$), preserving the noise asymmetry required for concatenation with repetition codes. At fixed fractional imperfections, reducing $E_J$ reduces the residual static nonlinearities but increases the pump amplitudes required to realize the target interactions. The resulting trade-off must therefore account for both static errors and the stronger nonlinear dependence on flux modulation.

\section{Compensation of Unwanted Drive Terms} \label{app:nonlinear_terms}

Although the ATS enables selective activation of interactions through the pump frequencies, additional terms in the full nonlinear potential must be included when calibrating the gate. In the interaction picture of the three normal modes, the driven interaction is:
\begin{align}
    \hat{H}_\text{drive}/\hbar &= -2E_J \sin[\epsilon(t)] \sin\bigl(\varphi_a\left(\hat{a}e^{-i\omega_a t} + \hat{a}^\dagger e^{i\omega_a t}\right)  \nonumber\\
    &+ \varphi_b\left(\hat{b}e^{-i\omega_b t} + \hat{b}^\dagger e^{i\omega_b t}\right)  \nonumber\\
    &+ \varphi_c\left(\hat{c}e^{-i\omega_c t} + \hat{c}^\dagger e^{i\omega_c t}\right)\bigr).
\end{align}
By defining the total time-dependent phase operator as $\hat{\varphi}(t) = \sum_{j \in \{a,b,c\}} \varphi_j (\hat{j}e^{-i\omega_j t} + \text{h.c.})$, we expand the sine potential into a Taylor series:
\begin{equation}
    \hat{H}_\text{int}/\hbar = -2E_J \sin[\epsilon(t)] \sum_{m=0}^{\infty} \frac{(-1)^m}{(2m+1)!} \hat{\varphi}(t)^{2m+1}.
\end{equation}

Normal ordering the expansion produces mixing terms whose coefficients contain products of zero-point phase amplitudes. Their dynamical importance also depends on mode occupations and detunings. We use the expansion through fifth order to construct compensation terms and assess their performance using simulations of the full sine potential.

For a single-tone sinusoidal driver with frequency $\omega_p$ and strength $\epsilon_p$, resonant interactions are activated when $\omega_p \approx \omega_{\vec{k}}$ where $\vec{k}$ is any mixing term. Our protocol uses specific drives ($\omega_{g_{\rm 3}} = 2\omega_a - \omega_b$ and $\omega_{g_{\rm CD}} = \omega_b$) to engineer the entangling gates. However, even detuned low-order terms produce unwanted displacement drives on the modes. Specifically, $\varphi_a(\hat{a} + \hat{a}^\dagger)$ and $\varphi_c(\hat{c} + \hat{c}^\dagger)$ have very strong magnitude causing non-ideal effects despite large detuning. Similarly, the first-order term $\varphi_b(\hat{b} + \hat{b}^\dagger)$ poses a challenge; it has zero detuning, and its magnitude exceeds the desired $g_{\rm  CD}$ interaction by a factor of $1/\varphi_a^2$. High-fidelity gate operation thus necessitates an active cancellation drive to nullify this residual linear drive.

\begin{figure}[h]
    \centering
    \includegraphics[width=0.8\linewidth]{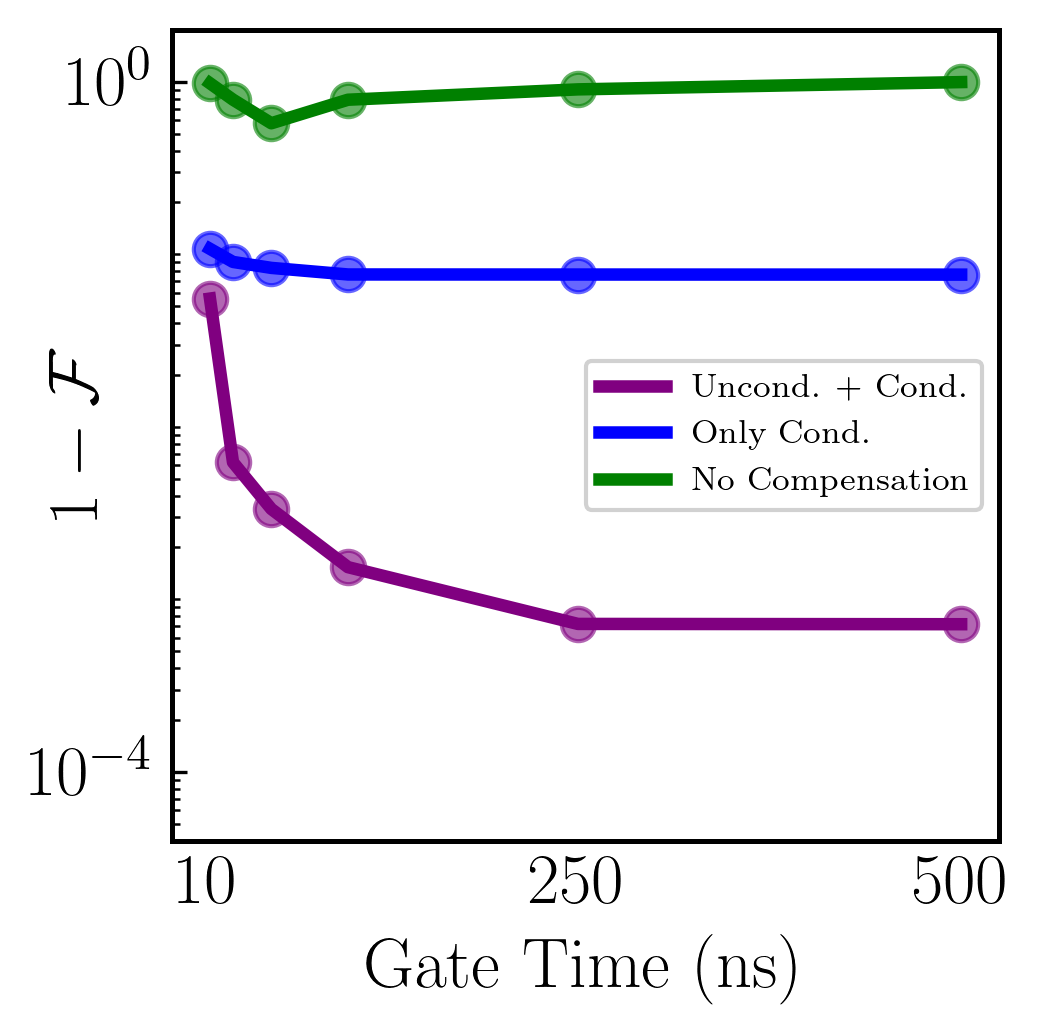}
    \caption{\justifying Lossless gate infidelity as a function of gate duration. The curves compare no compensation (green), compensation of occupation-dependent corrections only (blue), and the combination of this compensation with cancellation of the unconditional drive (purple).}
    \label{fig:comp_vs_fidelity}
\end{figure}

To evaluate the net effect across the cat manifold, we perform a fifth-order analytical expansion supplemented by numerical simulations of the full potential. Confining the dynamics to the cat subspace allows for a semiclassical approximation of the number operators ($\hat{n}_a \approx |\alpha|^2$ and $\hat{n}_b \approx |\beta|^2$) enabling calibration of the mean drive-induced shifts and displacements. Consequently, we compensate not only the primary linear term $\varphi_b(\hat{b} + \hat{b}^\dagger)$, but other resonant and near-resonant terms of the form $(\hat{o} + \hat{o}^\dagger)$, $\hat{n}_i(\hat{o} + \hat{o}^\dagger)$, and $\hat{n}_i \hat{n}_j (\hat{o} + \hat{o}^\dagger)$, where $\hat{o} \in \{\hat{a}, \hat{b}, \hat{c}\}$ and $i,j \in \{a,b\}$. This procedure compensates the mean contributions of unwanted terms while retaining the intended $g_{\rm CD}$ interaction. The impact of this compensation strategy on suppressing gate infidelity under no loss is illustrated in \autoref{fig:comp_vs_fidelity}.

The cancellation drive need not be applied separately to each mode. It can be generated at the coupler itself~\cite{Cottet2024}. Adding a differential RF flux, $\varphi_\Delta = \pi/2 + \epsilon_\Delta(t)$, to the common-flux pump $\varphi_\Sigma = \pi/2 + \epsilon_\Sigma(t)$, the ATS potential of \autoref{eq:driver_ham} (with $\Delta E_J = E_{J,2} = 0$) becomes
\begin{equation}
    U(\hat\varphi) = \frac{E_L}{2}\hat\varphi^2 - 2E_J \sin[\epsilon_\Sigma(t)]\,\sin\!\big[\hat\varphi + \epsilon_\Delta(t)\big].
\end{equation}
The differential flux shifts the minimum of the potential. Redefining the phase variable, $\hat\varphi \to \hat\varphi - \epsilon_\Delta(t)$, moves this shift into a linear term,
\begin{equation}
    \hat{H}_{\rm RAR,CX}/\hbar = -2E_J \sin[\epsilon_\Sigma(t)]\,\sin\hat{\varphi} - E_L\, \epsilon_\Delta(t)\, \hat{\varphi},
\end{equation}
up to a constant and a term proportional to $\dot\epsilon_\Delta \hat n$, which is also linear in the mode operators and only renormalizes the amplitude and phase of the required drive. Since $\hat\varphi = \sum_o \varphi_o(\hat o + \hat o^\dagger)$, the choice
\begin{equation}
    \epsilon_\Delta(t) = -\frac{2E_J}{E_L}\,\sin[\epsilon_\Sigma(t)]
\end{equation}
cancels the first-order displacement of all three modes simultaneously and at both pump frequencies, while leaving the third-order terms that generate $g_3$ and $g_{\rm CD}$ unchanged. Within the mean-field approximation $\hat n_i \approx \bar n_i$, the third- and fifth-order terms of the form $\hat n_i(\hat o + \hat o^\dagger)$ and $\hat n_i\hat n_j(\hat o + \hat o^\dagger)$ carry the same $\varphi_o$ structure to leading order, so they rescale this condition to
\begin{equation}
    \epsilon_\Delta(t) = -\frac{2E_J}{E_L}\,\sin[\epsilon_\Sigma(t)]
    \Big[1 - \tfrac12\textstyle\sum_{i}\varphi_i^2\,(2\bar n_i + 1) + \mathcal{O}(\varphi^4)\Big],
\end{equation}
which reduces to the condition of Ref.~\cite{Cottet2024} for a single cat mode. The small mode-dependent remainder is removed by fine-tuning the amplitude of $\epsilon_\Delta$ or by a weak local drive.

\section{ZZ Gate}
\label{app:zzgate}

As discussed in Ref. \cite{Puri2020}, a bias-preserving ZZ gate between Kerr-cat qubits only requires a beam-splitter interaction between the two resonators. Here we derive this interaction to leading order. Parametric beam-splitter interactions in the symmetric, LINC configuration have been also studied previously \cite{linc}. To activate this interaction, the ATS can be parked at its symmetric bias point $(\varphi_{\Sigma}, \varphi_{\Delta}) = (\pi/2, 0)$ and pumped along its common flux $\varphi_{\Sigma}$ with a drive $\epsilon(t)$. At this bias point, the resonator-ATS-resonator (RAR) system Hamiltonian can be simplified from \autoref{eq:RAR_sys_hamiltonian} to:
\begin{align}
    \hat{H}_{\rm RAR,ZZ}/\hbar &= \sum_{o = \{a,b,c\}}{\omega_{o} \hat{o}^\dagger \hat{o}} + 2E_J \sin[\epsilon(t)]\cos\hat{\varphi}.
\end{align}
It is useful to write the bare ATS phase operator $\hat{\varphi}$ in the normal mode basis
\begin{align}
\hat{\varphi} &= \hat{\varphi}_a + \hat{\varphi}_c  + \hat{\varphi}_b \notag \\
&= \varphi_a(\hat{a} + \hat{a}^\dagger) + \varphi_c(\hat{c} + \hat{c}^\dagger) + \varphi_b(\hat{b} + \hat{b}^\dagger),\label{eq:bare_ats_phase}
\end{align}
where $\varphi_o$ is the zero-point phase amplitude of mode $o$ across the ATS inductive branch. After substituting in \autoref{eq:bare_ats_phase} and performing a small angle expansion and approximating $\sin(\epsilon(t)) \approx \epsilon(t)$, the above expression becomes:
\begin{align*}
\hat{H}_{\rm RAR,ZZ}/\hbar &\approx \sum_{o = \{a, b,c\}}\omega_o \hat{o}^\dagger\hat{o} \notag\\
&\ \ + 2E_J \epsilon(t)[\cos\hat{\varphi}_c\cos{(\hat{\varphi}_a + \hat{\varphi}_b)} \notag\\
&\ \ - \sin\hat{\varphi}_c\sin{(\hat{\varphi}_a + \hat{\varphi}_b)} ].  \numberthis
\end{align*}
To activate a beam-splitter interaction, we set the common flux drive $\epsilon(t) = \epsilon_p \cos(\omega_p t)$ to a pump frequency $\omega_p = \omega_a - \omega_b$. The only terms that will survive a RWA on the above Hamiltonian will be non-energy conserving terms with energy equal to $\omega_p$, allowing us to remove the $\sin\hat{\varphi}_c$ term:
\begin{align}
\hat{H}_{\rm RAR,ZZ}/\hbar &\approx \sum_{o = \{a,b,c\}}\omega_o \hat{o}^\dagger\hat{o} \nonumber\\
&+ 2E_J \epsilon(t)\cos\hat{\varphi}_c\cos{(\hat{\varphi}_a + \hat{\varphi}_b)}. 
\end{align}
Next, by keeping the coupler $\hat{c}$ in its ground state so that $\langle\hat{c}^\dagger \hat{c}\rangle = 0$, we can approximate $\cos\hat{\varphi}_c \approx 1$:
\begin{align*}
&\hat{H}_{\rm RAR,ZZ}/\hbar \approx \sum_{o = \{a, b,c\}}\omega_o \hat{o}^\dagger\hat{o} \\&\quad + 2E_J \epsilon(t)\Big[\cos(\hat{\varphi}_a)\cos(\hat{\varphi}_b)
 - \sin(\hat{\varphi}_a)\sin(\hat{\varphi}_b)\Big].\numberthis
\end{align*}
Again, none of the cosine terms in the above expression will be rotating at the frequency of the pump and will thus rotate away, leaving:
\begin{align}
\hat{H}_{\rm RAR,ZZ}/\hbar &\approx \sum_{o = \{a, b,c\}}\omega_o \hat{o}^\dagger\hat{o} - 2E_J \epsilon(t) \sin{(\hat{\varphi}_a)\sin(\hat{\varphi}_b)}.
\end{align}
In the rotating frame of all three normal modes, the above Hamiltonian becomes:
\begin{align*}
\tilde{\hat{H}}_{\rm RAR,ZZ}/\hbar &\approx -2E_J \frac{\epsilon_p}{2}\left(e^{i\omega_p t} + e^{-i\omega_p t}\right)\\
&\quad \times \varphi_a\left(\hat{a}e^{-i\omega_a t} + \hat{a}^\dagger e^{i\omega_a t}\right)\\
&\quad \times\varphi_b\left(\hat{b} e^{-i\omega_b t} + \hat{b}^\dagger e^{i\omega_b t}\right) + \cdots\numberthis
\end{align*}
Finally, taking the above Hamiltonian only to leading order in the ZPFs since $\varphi_j \ll 1$ and removing fast-rotating terms, we are left with a beam-splitter Hamiltonian:
\begin{align}
\tilde{\hat{H}}_{\rm RAR,ZZ}/\hbar \approx \underbrace{-E_J \epsilon_p \varphi_a\varphi_b}_{g_{\rm ab}}\left(\hat{a}^\dagger\hat{b} + \hat{b}^\dagger\hat{a}\right).
\end{align}
\end{appendices}

\end{document}